\documentclass[twocolumn,twocolappendix]{aastex631}

\usepackage{amsmath}

\usepackage{subcaption}
\usepackage{upgreek}
\usepackage{makecell}
\usepackage{lipsum}
\graphicspath{{./}{figures/}}

\received{...}
\revised{...}
\accepted{...}
\submitjournal{ApJ}
\usepackage{makecell}

\def\gtorder{\mathrel{\raise.3ex\hbox{$>$}\mkern-14mu\lower0.6ex\hbox{$\sim$}}}
\def\ltorder{\mathrel{\raise.3ex\hbox{$<$}\mkern-14mu\lower0.6ex\hbox{$\sim$}}}

\shorttitle{The Radius of PSR~J0740+6620}
\shortauthors{Dittmann, Miller, Lamb, et al.}

\begin{document}

\title{A More Precise Measurement of the Radius of PSR~J0740+6620 Using Updated NICER Data}

\correspondingauthor{Alexander J. Dittmann}
\email{dittmann@umd.edu}

\author[0000-0001-6157-6722]{Alexander J. Dittmann}
\affil{Department of Astronomy and Joint Space-Science Institute, University of Maryland, College Park, MD 20742-2421, USA}
\affil{Theoretical Division, Los Alamos National Laboratory, Los Alamos, NM 87545, USA}

\author[0000-0002-2666-728X]{M. Coleman Miller}
\affil{Department of Astronomy and Joint Space-Science Institute, University of Maryland, College Park, MD 20742-2421, USA}

\author[0000-0002-3862-7402]{Frederick K. Lamb}
\affil{Illinois Center for Advanced Studies of the Universe and Department of Physics, University of Illinois at Urbana-Champaign, 1110 West Green Street, Urbana, IL 61801-3080, USA}
\affil{Department of Astronomy, University of Illinois at Urbana-Champaign, 1002 West Green Street, Urbana, IL 61801-3074, USA}

\author[0000-0002-3097-942X]{Isiah Holt}
\affil{Department of Astronomy and Joint Space-Science Institute, University of Maryland, College Park, MD 20742-2421, USA}
\affil{X-Ray Astrophysics Laboratory, NASA Goddard Space Flight Center, Code 662, Greenbelt, MD 20771, USA}

\author[0000-0003-2759-1368]{Cecilia Chirenti}
\affil{Department of Astronomy and Joint Space-Science Institute, University of Maryland, College Park, MD 20742-2421, USA}
\affil{Astroparticle Physics Laboratory NASA/GSFC, Greenbelt, MD 20771, USA}
\affil{Center for Research and Exploration in Space Science and Technology, NASA/GSFC, Greenbelt, MD 20771, USA}
\affil{Center for Mathematics, Computation, and Cognition, UFABC, Santo Andr\'{e}, SP 09210-170, Brazil}

\author[0000-0002-4013-5650]{Michael T.~Wolff}
\affil{Space Science Division, U.S. Naval Research Laboratory, Washington, DC 20375-5352, USA}

\author[0000-0002-9870-2742]{Slavko Bogdanov}
\affil{Columbia Astrophysics Laboratory, Columbia University, 550 West 120th Street, New York, NY 10027, USA}

\author[0000-0002-6449-106X]{Sebastien~Guillot}
\affil{Institut de Recherche en Astrophysique et Plan\'{e}tologie, UPS-OMP, CNRS, CNES, 9 avenue du Colonel Roche, BP 44346, F-31028 Toulouse Cedex 4, France}

\author[0000-0002-6089-6836]{Wynn C.~G. Ho}
\affiliation{Department of Physics and Astronomy, Haverford College, 370 Lancaster Avenue, Haverford, PA 19041, USA}

\author[0000-0003-4357-0575]{Sharon M. Morsink}
\affiliation{Department of Physics, University of Alberta, Edmonton, AB T6G 2E1, Canada}

\author{Zaven Arzoumanian}
\affil{X-Ray Astrophysics Laboratory, NASA Goddard Space Flight Center, Code 662, Greenbelt, MD 20771, USA}

\author[0000-0001-7115-2819]{Keith C. Gendreau}
\affil{X-Ray Astrophysics Laboratory, NASA Goddard Space Flight Center, Code 662, Greenbelt, MD 20771, USA}

\begin{abstract}
PSR~J0740+6620 is the neutron star with the highest precisely determined mass, inferred from radio observations to be $2.08\pm0.07\,\rm M_\odot$. Measurements of its radius therefore hold promise to constrain the properties of the cold, catalyzed, high-density matter in neutron star cores. Previously, \citet{2021ApJ...918L..28M} and \citet{2021ApJ...918L..27R} reported measurements of the radius of PSR~J0740+6620 based on Neutron Star Interior Composition Explorer (NICER) observations accumulated through 17 April 2020, and an exploratory analysis utilizing NICER background estimates and a data set accumulated through 28 December 2021 was presented in \citet{salmi2022}. 
Here we report an updated radius measurement, derived by fitting models of X-ray emission from the neutron star surface to NICER data accumulated through 21 April 2022, totaling $\sim1.1$ Ms additional exposure compared to the data set analyzed in \citet{2021ApJ...918L..28M} and \citet{2021ApJ...918L..27R}, and to data from X-ray Multi-Mirror (XMM-Newton) observations. 
We find that the equatorial circumferential radius of PSR~J0740+6620 is $12.92_{-1.13}^{+2.09}$~km (68\% credibility), a fractional uncertainty $\sim83\%$ the width of that reported in \citet{2021ApJ...918L..28M}, in line with statistical expectations given 
the additional data. If we were to require the radius to be less than 16~km, as was done in \citet{salmi2023filler}, then our 68\% credible region would become $R=12.76^{+1.49}_{-1.02}$~km, which is close to the headline result of \citet{salmi2023filler}. 
Our updated measurements, along with other laboratory and astrophysical constraints, imply a slightly softer equation of state than that inferred from our previous measurements.
\end{abstract}


\section{Introduction} \label{sec:intro}
The cores of neutron stars consist of cold matter that is believed to be catalyzed to its ground state at a few times nuclear saturation density (saturation corresponds to a baryon number density $n_s\approx0.16\,\rm{fm^{-3}}$, equivalent to a mass density $\rho_s\approx2.7\times10^{14}\,\rm{g\,cm^{-3}}$). Laboratories are unable to replicate the extremely high densities, and likely very high neutron-proton asymmetries, present in neutron star cores. Thus, observations of neutron stars provide unique insights about the nature of dense matter. Notably, the densities in neutron star cores partially bridge the gap between the regime $\rho\lesssim\rho_s$ probed by nuclear experimentation \citep[e.g.,]{2002Sci...298.1592D,PhysRevLett.126.172502}
and the regime $\rho\gg\rho_s$ where perturbative quantum chromodynamics is currently able to make predictions about the nature of nuclear matter \citep[e.g.,][]{2015PhRvL.114c1103B,2016PhRvD..94j6008H,2017JHEP...11..031E}.

In recent years, observations of neutron stars have added considerably to our knowledge of the equation of state (EOS: pressure as a function of energy density) of cold matter - matter in which the thermal energies of the particles are much smaller than their Fermi energies - at high densities \citep[e.g.,][]{1997ApJ...490L..91P,2005ApJ...619..483B,2013arXiv1312.0029M,2016ApJ...820...28O,2017A&A...608A..31N}. 

Three neutron stars have precisely measured masses $\sim 2\,M_\odot$: PSR~J1614-2230, with {$M=1.937\pm0.014\,M_\odot$} \citep{2010Natur.467.1081D,2016ApJ...832..167F,2018ApJ...859...47A,2023ApJ...951L...9A}; PSR~J0348+0432, with $M=2.01\pm0.04\,M_\odot$ \citep{2013Sci...340..448A}; and PSR~J0740+6620, with $M=2.08\pm0.07\,M_\odot$ \citep{2020NatAs...4...72C,2021ApJ...915L..12F}, where the quoted uncertainties denote the 68\% credible regions. Observations of such massive neutron stars can rule out equations of state that predict appreciably lower maximum stable masses.
Additionally, observations of the gravitational wave event GW170817 have provided constraints on the tidal deformability of inspiraling neutron stars \citep[e.g.,][]{2017PhRvL.119p1101A,2018PhRvL.121p1101A,2018PhRvL.121i1102D,2020CQGra..37d5006A}, ruling out equations of state that would cause neutron stars to have relatively high radii at a given mass.
Observations of the kilonovae following binary neutron star mergers may also place upper limits on the maximum stable mass of nonrotating neutron stars \citep[e.g.,][]{2017ApJ...850L..19M,2018ApJ...852L..25R,2021ApJ...922...14P}.  

The Neutron Star Interior Composition Explorer (NICER) can make phase- and energy-resolved measurements of the thermal X-ray pulses produced by some rotating neutron stars. These can be used to constrain the masses and radii of these stars, and thus the neutron star EOS \citep[e.g.,][]{2016RvMP...88b1001W,2016ApJ...832...92O}. The first mass and radius constraints derived from modeling NICER X-ray pulse profiles were reported in two independent analyses of the energy-resolved  pulse profile of PSR~J0030+0451: \citet{2019ApJ...887L..24M} measured its mass and equatorial radius to be $1.44_{-0.14}^{+0.15}\,M_\odot$ and $13.02_{-1.06}^{+1.24}\,{\rm km}$, respectively, while \citet{2019ApJ...887L..21R} measured its mass and radius to be $1.34_{-0.16}^{+0.15}\,M_\odot$ and $12.71_{-1.19}^{+1.14}\,{\rm km}$. 
Later, a pair of independent analyses of NICER and XMM-Newton data constrained the radius of PSR~J0740+6620 to be $13.7_{-1.5}^{+2.6}\,{\rm km}$ \citep{2021ApJ...918L..28M} and $12.39_{-0.98}^{+1.30},{\rm km}$ \citep{2021ApJ...918L..27R},\footnote{See, for example, Section 4.6 of \citet{2021ApJ...918L..28M} for a summary of the differences between these two works. Subsequent analyses of the PSR~J0740+6620 data resolved some of the differences between these estimates, as noted in Section 3.3 of \citet{salmi2022}.} providing unique insight into the EOS, given the higher-density material present in the core of such a massive neutron star. The incorporation of XMM-Newton observations is particularly useful when analyzing NICER data on PSR~J0740+6620, given the faintness of the source and the crowded field of view, because the imaging data provided by XMM-Newton constrain the phase-averaged stellar flux and the non-stellar astrophysical background \citep{2021ApJ...918L..26W}. Models of the spurious X-ray counts caused by particle interactions with the NICER detectors \citep[e.g.,][]{2022AJ....163..130R} can also provide valuable information about the actual astrophysical flux, and can provide similar constraints in practice to the inclusion of XMM-Newton data \citep[e.g.,][]{salmi2022}.

Here we present a new measurement of the equatorial radius of PSR~J0740+6620, incorporating previous NICER and XMM-Newton observations as well as an additional $\sim 1.1$ Ms of NICER data. We outline our data selection and processing procedures in Section~\ref{sec:data} and describe our methodology in Section~\ref{sec:methods}. The results of our X-ray data analysis are presented in Section~\ref{sec:results} and  their implications are discussed in Section~\ref{sec:disc}. We summarize our conclusions in Section~\ref{sec:conclusions} and report our full posteriors in Appendices~\ref{app:NO} and~\ref{app:NX}. The posterior samples from our analyses are available on Zenodo.\footnote{ \href{https://zenodo.org/records/10215109}{https://zenodo.org/records/10215109} \citep{dittmann_2024_10215109}.}

\section{Data}\label{sec:data}
The present analysis of the pulse profile of PSR~J0740+6620 uses NICER event data collected between 21 September 2018 (ObsID 1031020101) and 21 April 2022 (ObsID 5031020445). These NICER data add roughly two years of data to the set presented in \citet{2021ApJ...918L..26W} and analyzed in \citet{2021ApJ...918L..28M} and \citet{2021ApJ...918L..27R}. The net exposure time of the resulting data set (after application of the filtering procedure described below) was 2733.81 ks, in comparison to the 1602.68 ks net exposure time of the observations analyzed in \citet{2021ApJ...918L..28M} and \citet{2021ApJ...918L..27R}.

We utilized NICER event data processed using the xti20210707 calibration release and HEASoft version 6.30.1 to extract events \citep{2014ascl.soft08004N}. NICER event data were filtered to minimize the background signal resulting from high-energy particles interacting with the NICER X-ray timing instrument (XTI). We rejected events obtained at low cosmic-ray cut-off rigidities (COR\_SAX $<$ 2). We excluded all events from NICER XTI detector 34 because it exhibited anomalously high levels of noise compared to the others, and in some time intervals we excluded data recorded by detector 14 for the same reason. We did not explicitly exclude data based on the angle between NICER's pointing direction and the Sun. Rather, we limited our event extractions to time intervals where the maximum ``undershoot rate" (the detector reset rate) was below a set threshold. This limited contamination by the accumulation of optical photons (``optical loading"), which leads to spectral broadening of low-energy electronic noise that would otherwise be confined below $\approx0.2$ keV, and negatively affects both the detector gain and spectral resolution.
Holding all other data reduction choices constant, we compared the significances with which pulsations were detected (as quantified by the $Z^2_2$ test; see \citealt{1983A&A...128..245B}) for data selected by requiring maximum undershoot rate limits of 50, 100, 150, and 200~cps (counts per second). The 100~cps undershoot rate limit  produced the most significant detection and was therefore used in our subsequent analysis. Unlike in \citet{2021ApJ...918L..26W}, we did not employ the good time interval (GTI) sorting method of \citet{2019ApJ...887L..27G}, which might have biased the resulting radius measurements by up to $\sim10\%$ \citep[under simplified assumptions;][]{2022ApJ...927..195E}, but in practice by $\ltorder1\%$ \citep[see also][]{salmi2022}. 
Because our data selection procedure is based on the optical-photon-induced detector reset rate, we expect any biases potentially introduced by our consideration of pulsation detection significance to be even smaller than the $\lesssim1\%$ bias in previous analyses.

We analyzed data from NICER pulse invariant (PI) channels 30 through 123 inclusive, which correspond to a calibrated photon energy range of 0.3 to 1.24 keV. We folded the waveform observed in each NICER energy channel on the spin period of the pulsar to produce a pulse profile at each energy for subsequent analysis. 
In our analyses incorporating XMM-Newton data,  we used the same XMM-Newton data sets and channel ranges as \citet{2021ApJ...918L..28M}: channels 57 through 299 (inclusive) for the European Photon Imaging Camera (EPIC) pn instrument and channels 20 through 99 (inclusive) of the EPIC MOS1 and MOS2 instruments.

\section{Methods}\label{sec:methods} 
Our approach to modeling the X-ray data is similar in most aspects to the procedures outlined in \citet{2019ApJ...887L..24M} and \citet{2021ApJ...918L..28M}. We review our models of emission from the stellar surface in Section \ref{sec:surface} and our pulse profile models in Section \ref{sec:modeling}.
We then describe our parameter estimation procedure in Section \ref{sec:BayesWept}, which is generally similar to that described in \citet{2021ApJ...918L..28M}.

\subsection{Modeling X-Ray Emission from the Stellar Surface}\label{sec:surface}
PSR~J0740+6620 emits soft X-rays that are modulated on the rotational period of the neutron star \citep{2021ApJ...918L..26W}. The accepted model for soft X-ray emission in old, non-accreting pulsars such as PSR~J0740+6620 is that particles with large ($\gg100$) Lorentz factors,\footnote{If particles with Lorentz factors $\lesssim100$ contribute significantly to atmospheric heating, the resulting emission pattern may be relatively limb-brightened compared to predictions in the deep-heating approximation \citep{2019ApJ...872..162B,2020A&A...641A..15S,2023arXiv230809319S}. The resulting reduction in the pulsed emission fraction could confound the influence of spacetime curvature; underestimating the degree of shallow heating is therefore expected to cause the underestimation of $R/M$.} 
produced by magnetospheric pair cascades \citep[e.g.,][]{2015ApJ...810..144T,2022ApJ...933L..37T}, penetrate deep into the neutron star's atmosphere, where their interaction with the atmosphere heats it \citep{1974RvMP...46..815T,2001ApJ...556..987H,2002ApJ...568..862H,2021ApJ...914L..15B}.
The dependence of the specific intensity of the radiation emerging from the atmosphere on the angle between the direction of propagation and the local  normal to the surface depends on the atmospheric composition, which we expect to be pure hydrogen or, less probably, pure helium (see Section \ref{sec:atms_comp}); the ionization state (see \ref{sec:ionization}); and the strength and orientation of the stellar magnetic field, which we assume has a negligible effect on the specific intensity emerging from the atmosphere of PSR~J0740+6620 (see \ref{sec:B_surface}). 

\subsubsection{Upper Atmosphere Composition}\label{sec:atms_comp}
Calculations in \citet{1980ApJ...235..534A} suggest that the strong surface gravity of neutron stars causes the lightest elements present in their atmospheres to rise to the surface within seconds or minutes. Because PSR~J0740+6620 is expected to have accreted a substantial quantity of matter from its binary companion to reach its current spin frequency, the prevalence of hydrogen in the outer layers of the envelopes of stars like its companion suggests that PSR~J0740+6620's upper atmosphere is likely to consist of pure hydrogen \citep[see, e.g.,][]{1992ApJ...399..634B,2019MNRAS.484..974W,2020MNRAS.493.4936W}. Although some neutron stars in X-ray binaries are thought to have accreted little or no hydrogen \citep[see, e.g., 4U 1820-30,][]{2002ApJ...566.1045S}, such binaries have very short periods \citep[685 s for 4U 1820-30,][]{1987ApJ...312L..17S}, implying that the neutron star and its companion are very close to one another and that the companion's envelope has been completely stripped of hydrogen. The orbital period of the system containing PSR~J0740+6620 is 4.77 days \citep{2020NatAs...4...72C}, suggesting that the pulsar's companion had a significant hydrogen envelope when its envelope was accreting onto PSR~J0740+6620. There is some evidence that the companion of PSR~J0740+6620 may now be a cool helium-atmosphere white dwarf \citep[e.g.,][]{2019MNRAS.485.3715B,2020MNRAS.495.2509E}. Even if little hydrogen was accreted onto PSR~J0740+6620, spallation may have created enough hydrogen to dominate the atmospheric composition \citep{1992ApJ...384..143B,2004ApJ...605..830C,2019ApJ...887..100R}. For the sake of completeness, we performed analyses using helium \citep{2009Natur.462...71H} as well as hydrogen model atmospheres, but consider them less likely to accurately represent the surface composition of PSR~J0740+6620 a priori. 

\subsubsection{Ionization Fraction}\label{sec:ionization}
The analyses of PSR~J0740+6620 presented in \citet{2021ApJ...918L..28M}, \citet{2021ApJ...918L..27R}, and \citet{salmi2022} inferred that the emitting regions of the stellar surface had temperatures $k_bT_{\rm eff}\gtrsim0.07\, {\rm keV}$. In spite of these high temperatures, atmospheric densities may be sufficient to create an appreciable neutral hydrogen fraction via recombination. We therefore performed two sets of analyses using tabulated atmospheric data generated by the NSX code, one assuming complete ionization \citep{2001MNRAS.327.1081H} and a second allowing for partial ionization \citep{2009Natur.462...71H}.\footnote{See \citet{2005MNRAS.360..458B} for the relevant opacity tables.} 
We report results from both sets of analyses, although we consider the fully ionized models to be more reliable because of systematic uncertainties in the partially ionized hydrogen atmosphere models due to the incomplete coverage of the temperature, density, and energy ranges relevant to neutron star atmospheres in the  opacity tables that are currently available \citep{2021ApJ...914L..15B}. 

\subsubsection{Surface Magnetic Fields}\label{sec:B_surface}
Sufficiently strong magnetic fields can significantly affect radiative transfer of X-rays through neutron star atmospheres \citep[e.g.,][]{2003MNRAS.338..233H,2003ApJ...599.1293H}. 
Using the measured spin period ($P=0.00289\,{\rm s}$) and period derivative ($\dot{P}=1.219\times10^{-20}$) of PSR~J0740+6620 \citep{2020NatAs...4...72C} along with Equation (12) of \citet{2006ApJ...643.1139C} assuming $\alpha=0$ (i.e., that the closed field line region extends to the light cylinder) and a magnetic field inclination of $\pi/2$, we estimate a surface magnetic field strength of $B\approx3\times10^8\,{\rm G}$ if the field configuration is a centered dipole. \citet{2021ApJ...918L..28M} inferred that the centers of emitting regions were separated on the surface by $\approx 2$ radians in the best-fit model and were thus not antipodal. The chord length between the two spots was therefore $\approx\sin{(1)}\approx0.84$ times the diameter, suggesting a surface magnetic field strength $\approx0.84^{-3}\approx1.7$ times larger than that of a centered dipole, i.e., a surface magnetic field strength $B\approx5\times10^8\,{\rm G}$. 
This field strength would have only a small effect on atomic structure \citep{2001RvMP...73..629L,2014PhyU...57..735P} and the corresponding electron cyclotron energy is only $\approx6\,{\rm eV}$, much smaller than the typical thermal energies we infer for particles in the atmosphere of PSR~J0740+6620. 
Therefore, we use nonmagnetic atmosphere models throughout the present work. However, we caution that we have not rigorously tested this assumption, which would require the construction of grids of model atmospheres with magnetic fields covering the range $B\sim10^9-10^{10}\,{\rm G}$, an especially challenging task due to the comparable strengths of Coulomb and magnetic effects.

\subsubsection{X-ray Propagation from Surface to Detector}
We model the propagation of light in the vicinity of the pulsar using the oblate Schwarzschild approximation \citep[e.g.,][]{1998ApJ...499L..37M,2014ApJ...791...78A,2019ApJ...887L..26B,2021PhRvD.103f3038S}, which takes into account the oblate shape and Doppler boosts resulting from the rapidity of the spins of millisecond pulsars, but treats the exterior spacetime as Schwarzschild; the oblate Schwarzschild approximation typically results in fractional systematic errors $\ltorder 0.1\%$ for pulsars rotating at frequencies of a few hundred Hz \citep{2019ApJ...887L..26B,2021PhRvD.103f3038S}. We account for the absorption of X-rays in the interstellar medium using the \texttt{TBabs} model \citep{2000ApJ...542..914W}. We convert our model photon spectra reaching the telescope to model count spectra, as would be registered by the NICER detectors, using a version of the NICER response matrix constructed to match that used over the course of a set of GTIs comprising each data set (see Section \ref{sec:data}). The XMM-Newton data are treated similarly, using response matrices generated by the XMM-Newton Scientific Analysis
System tools “arfgen” and “rmfgen” relevant to the individual CCDs where the pulsar source image fell during the observations for the EPIC-pn and EPIC-MOS instruments \citep[][]{2004ASPC..314..759G,2014ascl.soft04004S}.

\subsection{Pulse Profile Modeling}\label{sec:modeling}
Our approach to modeling the NICER pulse profile of PSR~J0740+6620 is the same as that described in \citet{2021ApJ...918L..28M}, which built upon the approach discussed in \citet{2019ApJ...887L..24M}. Detailed descriptions of our pulse profile generation procedure and tests of its accuracy are given in \citet{2019ApJ...887L..26B} and \citet{2021ApJ...914L..15B}. The X-ray pulse profile of PSR~J0740+6620 has two clear peaks separated by $\sim0.4$ in phase \citep{2021ApJ...918L..26W}, and is therefore incompatible with emission from a single circular emitting region of uniform temperature (hereafter we use the term ``spot'' to refer to X-ray-emitting regions of uniform temperature). While modeling the pulse profile of PSR~J0030+0451 necessitated oval \citep{2019ApJ...887L..24M} or crescent \citep{2019ApJ...887L..21R} spots, \citet{2021ApJ...918L..28M} and \citet{2021ApJ...918L..27R} found that emission from two circular spots, although not antipodal ones, was sufficient to describe the X-ray pulse profile of PSR~J0740+6620, with no evidence for more complicated spot geometries; however, we have not rigorously explored this possibility with the present data set.

\subsubsection{Model Parameters}\label{sec:parameters}

\begin{deluxetable*}{ccc}
\tablehead{
\colhead{Parameter} & \colhead{Definition} & \colhead{Assumed Prior}
}
\startdata
$c^2R_e/(GM)$ & Inverse stellar compactness at equator& 3.2 to 8.0 \\
$M$ & Gravitational mass & $\exp{[-(M-2.08\,M_\odot)^2(0.09\,M_\odot)^{-2}/2]}$\\
$\theta_{\rm c,1}$ & Colatitude of spot 1 center & 0 to $\pi$ radians \\
$\Delta\theta_1$ & Angular radius of spot 1 & 0 to 3 radians \\
$kT_{\rm eff,1}$ & Effective temperature of spot 1 & 0.011 to 0.5 keV\\
$\theta_{\rm c,2}$ & Colatitude of spot 2 center & 0 to $\pi$ radians \\
$\Delta\theta_2$ & Angular radius of spot 2 & 0 to 3 radians \\
$kT_{\rm eff,2}$ & Effective temperature of spot 2 & 0.011 to 0.5 keV\\
$\Delta\phi$ & Longitudinal offset between spots 1 and 2 & 0 to 1 cycles\\
$\theta_{\rm obs}$ & Observer inclination to stellar rotation axis & 1.44 to 1.62 radians \\
$N_H$ & Neutral H column density & 0 to $2\times10^{21}\,{\rm cm^{-2}}$\\
$D$ & Distance & $\exp{[-(D-1.136\,{\rm kpc})^2(0.20\,{\rm kpc})^{-2}/2]}$, $D\geq1.136\,{\rm kpc}$\\
    &          & $\exp{[-(D-1.136\,{\rm kpc})^2(0.18\,{\rm kpc})^{-2}/2]}$, $D\leq1.136\,{\rm kpc}$
\enddata
\caption{The primary model parameters, and their corresponding priors, for the pulse profile models considered in this work. Except where noted, the prior is uniform across the given range and zero elsewhere. We place no additional restrictions on the potential values for each parameter.}
\label{tab:parameters}
\end{deluxetable*}

Having specialized our models to a pair of circular spots, we construct them as follows: each spot ($i=1,2$) is characterized by an effective temperature ($kT_{{\rm eff},i}$), where $k$ is Boltzmann's constant; an angular radius ($\Delta \theta_i$); and a colatitude ($\theta_{{\rm c},i}$). We allow the two spots to have an arbitrary phase offset, measured in rotational cycles from one spot center to the other ($\Delta \phi$). Furthermore, we characterize the star using its gravitational mass ($M$) and inverse stellar compactness ($c^2R_e/(GM)$), where $R_e$ is its equatorial circumferential radius. We also model the observer inclination to the pulsar spin axis ($\theta_{\rm obs}$), the neutral hydrogen column density between NICER and PSR~J0740+6620 ($N_H$), and the distance from NICER to PSR~J0740+6620 ($D$). 
We allow arbitrary overlaps or lack thereof between spots. This is accomplished by labeling the spots ``1'' and ``2,'' and assigning any overlapping regions the effective temperature of the lower-numbered spot. This allows, for example, the creation of a single crescent-shaped spot by covering a hotter spot with another so cool that the latter emits virtually no radiation.

The full set of parameters used in our analyses is listed in Table~\ref{tab:parameters}, along with the priors we assume for each parameter. For most parameters we use a flat prior, defined as having a uniform nonzero probability density between an upper and lower bound (inclusive) and zero probability density outside of those bounds. 
Because PSR~J0740+6620 is in a binary system, we are able to incorporate additional informative priors. Specifically, \citet{2021ApJ...915L..12F} reported based on Shapiro delay measurements that the mass of PSR~J0740+6620 was $2.08_{-0.069}^{+0.072}\,M_\odot$, that the distance to the pulsar was $1.136_{-0.152}^{+0.174}\,\rm{kpc}$, and that the angle between our line of sight and the orbital axis of the system was $87.5^\circ\pm0.17^\circ$, all at 68\% credibility. 

We use these radio-timing-derived measurements to place priors on the mass, distance, and observer inclination in our analysis. We implement a Gaussian prior on the gravitational mass of the pulsar, with a median of $2.08\,M_\odot$ and standard deviation of $0.09\,M_\odot$, where we have linearly added an estimate of the systematic error in the Shapiro delay measurement (E.~Fonseca, personal communication). We implement an asymmetric quasi-Gaussian distance prior, which has a probability distribution that falls off at different rates above and below the mode, according to the functional form listed in Table~\ref{tab:parameters}. This distribution is based on the results presented in \citet{2021ApJ...915L..12F} and includes an additional, linearly added estimate of the systematic error of 0.03~kpc (E.~Fonseca, personal communication).

Radio observations are able to constrain the angle between the observer's line of sight and the orbital axis of the binary system, but our analysis depends on the angle between the observer's line of sight and the rotational axis of PSR~J0740+6620. Accretion from their companion stars is thought to be the primary mechanism for the spin-up of millisecond pulsars such as PSR~J0740+6620 \citep[e.g.,][]{1991PhR...203....1B}. Accordingly, the spin axis of the pulsar should gradually align with the orbital axis of the system. However, it is unlikely that the alignment is perfect, and thus the pulsar spin axis may be tilted a few degrees with respect to the orbital axis of the binary system. Thus, we have adopted a flat prior on the angle between the observer's line of sight and pulsar spin axis centered at $87.5^\circ$, which is the best estimate of the orbital inclination of the binary, with a width of $5^\circ$.\footnote{\citet{2021ApJ...918L..28M} found that the inferred radius of PSR~J0740+6620 was insensitive to the observer inclination within this range, and in an exploratory analysis we found that broadening the inclination prior further also did not affect our radius measurement.}

\subsubsection{Pulse Profile Models}\label{sec:waveformModels}
Given values for the set of model parameters described in Section \ref{sec:parameters}, a stellar rotation period (which is known to high precision from radio observations, e.g., \citealt{2020NatAs...4...72C,2021ApJ...915L..12F}), a neutron star atmosphere model, and a specified observation time, we can generate a spots-only pulse profile. However to compare these model pulse profiles with data, it is necessary to account for non-spot (or `background') contributions to the observed X-ray counts, which are not modulated on the spin period of the pulsar. To compare with NICER data it is also necessary to select a starting rotational phase for the pulse profile. 
We are then able to calculate the log likelihood of the available data sets given a particular model. When we incorporate information from various XMM-Newton instruments, the overall log likelihood is simply the sum of the log likelihoods for the data given the model for each data set, i.e.,
\begin{equation}\label{eq:loglsum}
\begin{split}
\log{\mathcal{L}_{\rm tot}}=\log{\mathcal{L}_{\rm NICER}}+\log{\mathcal{L}_{\rm XMM-pn}}\\+\log{\mathcal{L}_{\rm XMM-MOS1}}+\log{\mathcal{L}_{\rm XMM-MOS2}}
\end{split}
\end{equation}
for analyses of NICER data along with XMM-Newton imaging data from the pn, MOS1, and MOS2 instruments, or $\log{\mathcal{L}_{\rm tot}}=\log{\mathcal{L}_{\rm NICER}}$ for analyses of NICER data alone. The straightforward summation of the terms in Equation (\ref{eq:loglsum}) assumes that the different terms are uncorrelated, which is valid for data from NICER and the different instruments mounted on XMM-Newton \citep{2001A&A...365L..27T,2001A&A...365L..18S}.

We compute the log likelihood corresponding to each data set by summing the Poisson log likelihoods of the data given the model for each energy channel ($E_j$) and phase ($\phi_k$). Given an observed number of counts $d_{jk}$, which is a non-negative integer, and a predicted number of counts $m_{jk}$, which is a positive real number, the Poisson likelihood of the data given the model is $m_{jk}^{d_{jk}}e^{-m_{jk}}/d_{jk}!$. However, the factor $1/d_{jk}!$ is shared by all models, and can thus be neglected. The log likelihood we use is therefore $\log{\mathcal{L}}=\sum_{\phi_k}\sum_{E_j}(d_{jk}\log{m_{jk}}-m_{jk})$, where NICER data are analyzed over 32 phases and XMM-Newton data that we use are resolved only in energy, with a single time-averaged phase. 

As previously mentioned, observed pulsar X-ray pulse profiles may include contributions not modulated on the spin period of the pulsar, from sources such as particle background, optical loading, and both resolved and unresolved sources in the telescope field of view, among others. 
We analytically marginalize over a phase-independent background parameter for each NICER energy channel, as described in section 3.4 of \citet{2019ApJ...887L..24M}. 
Concerning the XMM-Newton background, we take previously measured XMM-Newton blank-sky background spectra to be a Poisson realization of the actual XMM-Newton background, and calculate the probability of the data given the spot counts and the distribution of possible background counts, as described in Section 3.4.2 of \citet{2021ApJ...918L..28M}.

\subsection{Bayesian Parameter Estimation}\label{sec:BayesWept}
We adopt a hybrid sampling approach that utilizes an initial suite of nested sampling \citep{2004AIPC..735..395S} analyses followed by a suite of Markov chain Monte Carlo (MCMC) analyses for the purpose of refining our posterior inferences. To wit, each of the initial nested sampling analyses, which we carry out using the MultiNest package and its Python bindings \citep{2009MNRAS.398.1601F,2016ascl.soft06005B}, \footnote{\url{https://github.com/farhanferoz/MultiNest},\url{https://github.com/JohannesBuchner/PyMultiNest}} is able to sample the parameter space globally, identify multiple modes if they exist, and approximate the Bayesian evidence for each mode. The evidence ($\mathcal{Z}$), which is the average of the likelihood $\mathcal{L}$ over the normalized prior $\uppi$, is given by the following integral over the model parameters $\mathbf{\theta}$
\begin{equation}
\mathcal{Z} = \int \mathcal{L}(\mathbf{\theta})\uppi(\mathbf{\theta})d\mathbf{\theta},
\end{equation}
and can be used to compare models, where the model with the highest evidence is preferred.

Basic nested sampling algorithms such as MultiNest, while often relatively fast, produce a very limited number of posterior samples during each analysis and provide neither theoretical nor practical convergence guarantees.\footnote{Other nested sampling packages \emph{do} allow users to target posterior rather than evidence accuracy, and allow for further sampling after the convergence of an initial nested sampling of the parameter space \citep[e.g.,][]{2020MNRAS.493.3132S,2021JOSS....6.3001B}.} Therefore, following each nested sampling analysis we used the posterior probability distribution inferred from that analysis to initialize an MCMC analysis, which we continued until the posterior distribution appeared to be stationary and we were able to obtain approximately $10^6$ independent posterior samples from the stationary distribution.

\subsubsection{Nested Sampling Analyses}\label{sec:nested}
We began our analysis using the publicly available MultiNest algorithm \citep{2009MNRAS.398.1601F,2016ascl.soft06005B}, which begins by randomly sampling a number of points (a user-specified number of ``live points,'' $N_{\rm live}$) from the prior. 
Subsequently, MultiNest sequentially replaces the lowest-likelihood point with a higher-likelihood point. Proposals for these higher-likelihood points are drawn from within the region bounded by an approximation of the isolikelihood surface corresponding to the likelihood of the point to be replaced.
A larger number of live points will lead to more thorough sampling of the parameter space, all other parameters being equal. MultiNest constructs approximate isolikelihood surfaces using multiple hyper-ellipsoids constructed to envelop the set of live points. If these minimal bounding hyper-ellipsoids contain a volume smaller than the expectation value of the remaining prior volume, the bounding ellipsoids are expanded until the overlap-corrected enclosed volume matches the expected remaining prior volume \citep[][Section 5.2]{2009MNRAS.398.1601F}.
However, if the true isolikelihood surface is not described accurately by the (possibly expanded) hyper-ellipsoids encompassing a given set of live points, MultiNest can draw samples from a biased region of parameter space, resulting in biased estimates of the posterior and Bayesian evidence \citep[e.g.,][]{2016S&C....26..383B,2020AJ....159...73N,2023StSur..17..169B,2023MNRAS.521.1184L,2024arXiv240416928D}. The MultiNest algorithm attempts to ameliorate this shortcoming through an `efficiency' parameter, which is the inverse of a factor used to further increase the hypervolume of the MultiNest bounding ellipsoids (for example, an efficiency value of $\epsilon=0.1$ multiplies by 10 the target volume for the expanded hyper-ellipsoids). However, this procedure is still not guaranteed to encompass the intended isolikelihood surface.\footnote{For example, \citet{2023MNRAS.521.1184L} found that a value $\epsilon\leq10^{-3}$ was necessary for MultiNest to produce unbiased Bayesian evidence estimates when performing cosmological inferences, and that $\epsilon\leq10^{-2}$ was necessary to recover the true value (within MultiNest's error estimates) of a simple high-dimensional Gaussian likelihood. See \citet{2024arXiv240416928D} for a thorough investigation into the accuracy of MultiNest's posterior and evidence estimates.}

Previous analyses of NICER data have found that MultiNest rarely produces converged posterior estimates. For example, \citet{2021ApJ...918L..28M} reported that a MultiNest analysis of an earlier PSR~J0740+6620 data set using $N_{\rm live}=1000$ and $\epsilon=0.01$ systematically underestimated the stellar radius and its uncertainty. Analyzing the same data set, \citet{2021ApJ...918L..27R} found that the settings $\epsilon=0.1$ and $N_{\rm live}=4\times10^4$ underestimated the width of the inferred radius posterior compared to a run with the same efficiency and twice as many live points. Moreover, \citet{2021ApJ...918L..28M} showed, by performing MultiNest inferences and comparing the number of parameters inferred to fall within the $\pm(1,2,3)\sigma$ credible intervals with statistical expectations, that MultiNest systematically underestimated the width of credible intervals, although less significantly when more live points and lower sampling efficiencies were used.
Analyzing NICER data for PSR~J0030+0451, \citet{2024ApJ...961...62V} found that analyses using  $\epsilon=0.3$ and $N_{\rm live}=10^3$ often overestimated the median stellar radius and underestimated the width of the posterior distribution compared to higher-resolution analyses using $\epsilon=0.1,~N_{\rm live}=10^3$ and $\epsilon=0.3,~N_{\rm live}=10^4$; in some cases the posteriors estimated by lower-resolution analyses strongly excluded the median from higher-resolution analyses \citep{2024ApJ...961...62V}. Despite these shortcomings, MultiNest is in our experience able to provide a suitable starting point for subsequent MCMC analyses using orders of magnitude fewer computational resources than would initializing an MCMC analysis using uniform samples drawn from the prior.

Based on these considerations and a series of convergence tests, our default MultiNest settings were $\epsilon=0.01$ and $N_{\rm live}=4096$. In each analysis we enabled MultiNest to evaluate multiple modes. Because our parameter inferences do not rely on the capacity of MultiNest to estimate posteriors with high accuracy, these settings were sufficient for our preliminary nested sampling analysis. 

\subsubsection{Markov Chain Monte Carlo Analyses}
Following the completion of each nested sampling analysis, we used the posterior probability distribution estimates it produced to draw initial walker positions for an MCMC analysis using the \texttt{emcee} package \citep{2013PASP..125..306F} and the affine-invariant `stretch' proposal of \citet{2010CAMCS...5...65G}.\footnote{\url{https://github.com/dfm/emcee}} Because the proposal distribution utilized used in this analysis satisfies detailed balance, the distribution of walker positions will converge to and provide samples from the stationary posterior distribution. In principle the convergence of each chain of walkers to the stationary distribution may take an arbitrarily long time if the walkers are initialized very far from equilibrium, which is why we use a prelimiary nested sampling analysis to generate initial walker positions that more accurately approximate the posterior distribution.

Each analysis used 4096 walkers, for which we drew initial positions by re-sampling a Gaussian kernel density estimate of the corresponding MultiNest-derived posterior distributions. We performed MCMC sampling until we had collected $\sim10^6$ effective samples (in practice $\sim 10^7$ samples, because our analyses had proposal acceptance fractions of $\sim 0.1$ and autocorrelation times of $\sim10$ iterations). We judged the sampling to be converged when over the aforementioned $\sim10^7$ MCMC iterations we observed no secular variation in the 1st, 16th, 50th, 84th, or 99th percentiles. 

\section{Pulse Profile Modeling Results}\label{sec:results}

\begin{figure*}
     \centering
     \begin{subfigure}[b]{0.45\textwidth}
         \centering
         \includegraphics[width=\textwidth]{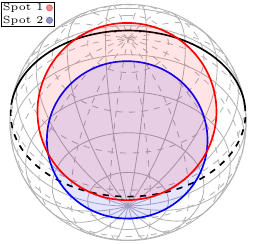}
         \caption{Lunate spot configuration}
         \label{fig:lunate}
     \end{subfigure}
     \hfill
     \begin{subfigure}[b]{0.45\textwidth}
         \centering
         \includegraphics[width=\textwidth]{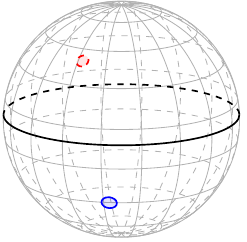}
         \caption{Dual-spot configuration}
         \label{fig:time2dual}
     \end{subfigure}
     \hfill
    \caption{Examples of the two spot configurations identified during our analysis of the updated data set for PSR~J0740+6620 able to produce good fits to the NICER data. The black circle denotes the colatitude of the observer line of sight. Dashed lines indicate features on the side of each sphere opposite the observer. To reiterate, in our models regions that overlap emit with the temperature of the `lowest-number' spot; accordingly, emission from the configuration shown in panel (a), for which the temperature of `spot 1' tends towards the lower end of our prior, effectively comes from the protruding crescent of `spot 2.'}\label{fig:configurations}
\end{figure*}
We first present in Section \ref{sec:modes} the different modes identified during the initial stage of our analysis, assess the thoroughness of that preliminary nested sampling stage, and report approximate evidences for each atmosphere model and data set analyzed. Subsequently, we report in Section \ref{sec:massRadius} the mass and radius inferences from each data set and atmosphere model combination. We assess the goodness of fit of our models in Section \ref{sec:adequacy}, and compare our results with those of \citet{salmi2023filler} in Section \ref{sec:diffs}.

\subsection{Inferred Emitting Region Geometries}\label{sec:modes}
\setlength{\tabcolsep}{2.6pt}
\begin{deluxetable}{ccccc}
\tablehead{
\colhead{Dataset} & \colhead{Atmosphere} & \colhead{Lunate spot} & \colhead{Dual spot} & \colhead{$\Delta \log{Z}$} }
\startdata
NICER & {$\rm H_{full}$} & $\checkmark$ & - & - \\
NICER & {$\rm H_{partial}$} & $\checkmark$ & $\checkmark$  &  $-0.11$ \\
NICER & {$\rm He_{full}$} & $\checkmark$ & - & $1.33$ \\ \hline
NICER+XMM & {$\rm H_{full}$} & - & $\checkmark$ & - \\
NICER+XMM & {$\rm H_{partial}$} & - & $\checkmark$  & $1.0$ \\
NICER+XMM & {$\rm He_{full}$} & - & $\checkmark$  & $2.08$ \\
\enddata
\caption{We present the inferred spot configuration(s) and log Bayes factors for each data set and each atmosphere model considered in this work. In the second column, subscripts indicate whether each atmosphere model assumed full ionization or allowed for partial ionization. Check marks ($\checkmark$) indicate which spot configurations provided good fits to each data set under the assumptions of a particular atmosphere model.
For each data set (NICER and NICER+XMM), we report the evidence for each model (including all identified modes)
relative to the evidence obtained using that data set assuming the fully ionized hydrogen atmosphere, which we consider to be the most astrophysically relevant atmosphere (see Section 3).}
\label{tab:modes}
\end{deluxetable}
\citet{2021ApJ...918L..28M} identified two emitting region geometries that were capable of reproducing the observed pulse profile and spectrum of PSR~J0740+6620: one was a crescent (or lunate) configuration formed by an extremely cool circular spot eclipsing a hotter spot; the other consisted of two fairly small spots with similar temperatures, well-separated in azimuth. Both spot configurations are also allowed by the priors used in this work (Table~\ref{tab:parameters}). 

Figure~\ref{fig:configurations} shows ilustrative schematic examples of both configurations. \citet{2021ApJ...918L..28M} found that the configuration with two fairly small spots was favored by a Bayes factor of $\sim 3000$ over the lunate configuration. When only the updated NICER data are considered, we find that either spot configuration can provide good fits, depending on the assumed atmospheric composition and whether or not we also consider the XMM-Newton data. 

Table~\ref{tab:modes} shows the spot configurations that provided good fits to each data set under each assumed atmosphere model. We also include in Table \ref{tab:modes}, for each data set, the log evidence for each model relative to the fit assuming a fully ionized  hydrogen atmosphere to the same data set.\footnote{Throughout this work we use the natural logarithm unless otherwise denoted by a subscript.} For example, the logarithm of the Bayesian evidence for our fully ionized  helium fit to only the NICER data was $\approx1.33$ larger than the logarithm of the Bayesian evidence for the fully ionized hydrogen fit to the same data set. Given the sensitivity of the evidence to priors, and the potential for appreciable systematic errors in evidences estimated by MultiNest, no atmosphere model is significantly favored over another.  

When only the available NICER data are considered, the preferred mode depends on the assumed atmospheric composition and conditions: while the fully ionized  hydrogen and helium atmosphere models overwhelmingly favor the lunate configuration, the partially ionized hydrogen atmosphere models favor both configurations. However, when the XMM data---which constrain the stellar count rate and non-stellar background---are  included, the data prefer models with two smaller, near-antipodal spots.

\begin{figure*}
 \includegraphics[width=\linewidth]{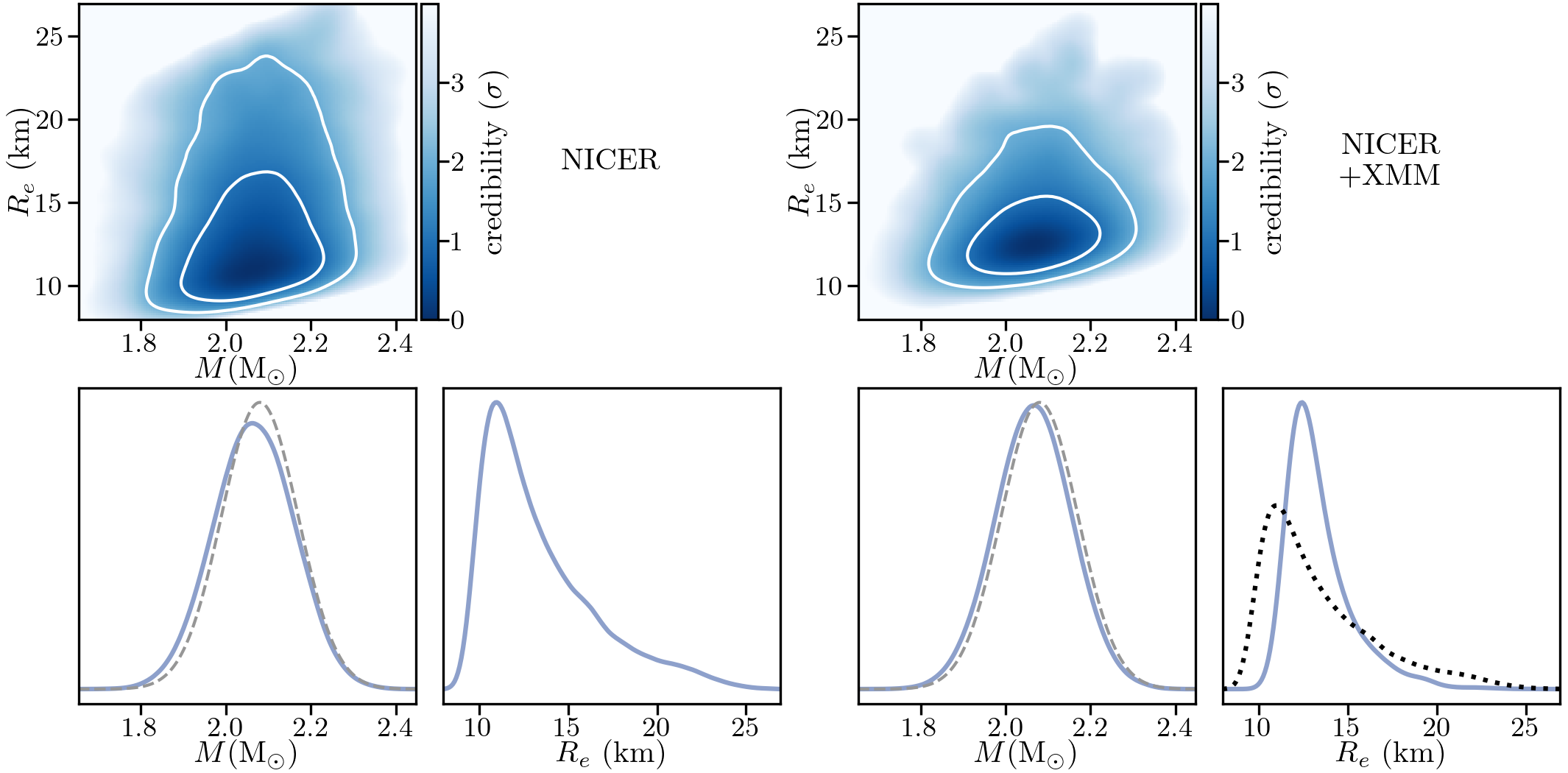}
 \caption{A comparison of the $M$, $R_e$, and joint $M-R_e$ posterior probability distributions for PSR~J0740+6620 inferred by analyzing the NICER data only (left panels) and the NICER and XMM-Newton data analyzed jointly (right panels).
 The results shown here assume a fully ionized, purely hydrogen atmosphere, which we consider the most likely atmosphere for the reasons explained in Section~\ref{sec:surface}. The solid curve in each bottom-left panel shows the one-dimensional probability distribution for the stellar mass inferred from these data sets; the mass prior derived from \citet{2021ApJ...915L..12F} is shown by the dashed gray line. The color shading in the two-dimensional $M$--$R_e$ plots illustrates the shapes of the minimum-area credible regions; the overplotted white lines show the $1\sigma$ and $2\sigma$ contours outside which $\sim31.7\%$ and $\sim4.6\%$ of the posterior mass of each distribution lies. The solid curve in the  bottom-right panel on the left shows the probability density of the equatorial radius inferred by analyzing only the NICER data. The solid curve in the bottom-right panel on the right shows the probability density inferred by jointly analyzing the NICER and XMM-Newton data; the black dotted curve in this panel shows the probability density inferred by analyzing the NICER data alone.}
 \label{fig:MRboth}
\end{figure*}

Before attempting to sample any of the posteriors more thoroughly, we first verified that our initial nested sampling analyses were thorough enough to identify the most significant mode, and that each did not find one mode or the other only by chance. 
The first test to which we subjected our nested sampling results was to evaluate the best fit from an analysis that identified only one mode using the data and atmospheric model of analyses which found other modes. 
Taking the best fits from our fully ionized hydrogen atmosphere analyses as an example, when we apply the (lunate) best fit parameters from our analysis of the NICER data alone to the joint NICER and XMM-Newton data set, we find a \textit{log} likelihood $\sim400$ lower than for the (dual-spot) best joint fit to the NICER and XMM-Newton data.
Similarly, the log likelihood of the best fit found analyzing the NICER data alone assuming a helium atmosphere was $\sim 600$ lower than the log likelihood of the best fit found in the joint analysis of NICER and XMM-Newton data. This analysis thus suggests that the XMM-Newton data strongly disfavor the lunate spot configuration. For comparison, the model that best fits the NICER data only, using the partially ionized hydrogen atmosphere tables, which is the smaller near-antipodal spot configuration, gives a NICER+XMM log likelihood only $\sim35$ lower than the best fit for this model found in the joint NICER and XMM-Newton analysis. 

Based on the comparisons above, and additional tests discussed in Appendix \ref{app:sampling}, we believe that our MultiNest analyses have sampled parameter space thoroughly enough to identify the relevant modes in each analysis. Because we are able to sample each posterior further using MCMC techniques, we do not need to rely on the often-questionable convergence of the posterior estimates reported by MultiNest. Instead, we require only that the different modes identified by MultiNest be assigned roughly the correct weights.\footnote{In principle, various MCMC implementations are able to properly sample [delete: from] multi-modal posteriors, including by allowing walkers to move from one mode to another. Although our requirement that MultiNest be able to adequately identify each mode certainly accelerated our subsequent MCMC analysis, that requirement was not strictly necessary.} In practice, the two modes present in $\Delta\phi$ for the smaller-spot configurations (see Appendix \ref{app:NX}), which are a result of the near reflection symmetry due to the observer's inclination being so close to the rotational equator of the star, have very little effect on the radius or mass posteriors. Thus, the only capability of MultiNest on which we rely is its ability to differentiate between the lunate and near-antipodal configuration for each data set and atmosphere model, a capability our analyses have demonstrated. 

\subsection{Mass and Radius Constraints}\label{sec:massRadius}
After we completed our preliminary nested sampling analyses, we initialized a series of MCMC analyses to thoroughly sample each posterior. We discuss the results in this section. 

\begin{figure}
 \includegraphics[width=\linewidth]{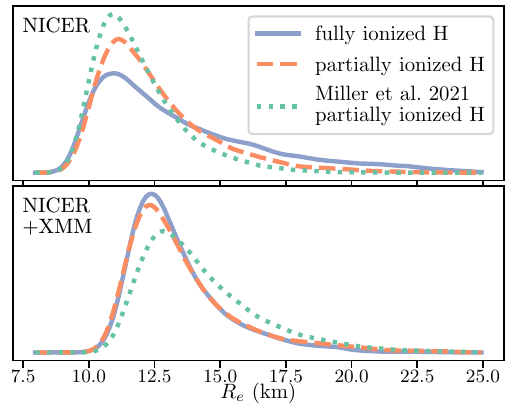}
 \caption{Radius posteriors obtained when only the NICER data are used (top panel) and when the NICER and XMM-Newton data are both used (bottom panel). The solid blue lines show the posterior for a fully ionized hydrogen atmosphere whereas the dashed orange lines are for a partially ionized hydrogen atmosphere. The dotted green lines show the posterior reported by \citet{2021ApJ...918L..28M}, who assumed a partially ionized hydrogen atmosphere. The results obtained here using only the NICER data depend on the atmospheric model, because different spot patterns are favored by the different models (see Figure~\ref{fig:configurations}). However, when both the NICER and XMM-Newton data are used, the posterior is almost independent of the atmospheric model assumed.}
 \label{fig:allRadii}
\end{figure}

Unsurprisingly, given the faintness of PSR~J0740+6620, we are able to infer little about the stellar mass, although we systematically infer masses very slightly lower than the median value of our prior (e.g., $2.06\,M_\odot$ rather than $2.08\,M_\odot$). We present the credible regions for the equatorial radius of PSR~J0740+6620 derived from each of our analyses in Table~\ref{tab:radii}, the posterior probability distributions of the mass and equatorial radius derived from our analyses assuming a fully ionized hydrogen atmosphere in Figure~\ref{fig:MRboth}, and a comparison of the present constraints on the equatorial radius to those derived in \citet{2021ApJ...918L..28M} in Figure~\ref{fig:allRadii}.

We find that, when the constraints provided by XMM-Newton data are included, the radii we infer from analyses utilizing differing neutron star atmosphere models are consistent with one another: the $-1\sigma$ values range only from $11.76\, \rm km$ to $11.82\, \rm km$ and the $+1\sigma$ values range from $15.01\, \rm km$ to $15.52\, \rm km$. In contrast, when we analyze the NICER data alone, we find that the constraints on the stellar radius vary much more between our analyses. This is largely due to the faintness of PSR~J0740+6620 and the presence of other sources within the field of view of NICER when it is observing PSR~J0740+6620, as shown in \citet{2021ApJ...918L..26W}. Because the number of phase-dependent counts in the pulse profile of PSR~J0740+6620 is much smaller than the number of phase-independent counts, the NICER data alone allow a wide range of emitting region geometries and values of the stellar compactness, which limits the precision of the resulting constraints on the radius of the pulsar. When the XMM-Newton imaging data are included, the stellar and background fluxes are better constrained.

\begin{deluxetable}{ccccccc}
\caption{Summary of Inferred Equatorial Radii}
\tablehead{
\colhead{Dataset} & \colhead{\makecell{Atmosphere\\ Model}} & \colhead{$-2\sigma$} & \colhead{$-1\sigma$} & \colhead{median} & \colhead{$+1\sigma$} & \colhead{$+2\sigma$} 
}
\startdata
NICER&$\rm{H_{full}}$&9.60 & 10.54 & 12.48 & 16.54 & 21.84 \\
NICER&$\rm{H_{partial}}$&9.72 & 10.59 & 11.95 & 14.46 & 18.32 \\
NICER& $\rm{He_{full}}$ &9.62 & 10.56 & 12.72 & 17.39 & 22.55 \\
\hline
NICER+XMM&$\rm{H_{full}}$&10.99 & 11.79 & 12.92 & 15.01 & 18.57 \\
NICER+XMM&$\rm{H_{partial}}$&10.94 & 11.76 & 12.99 & 15.36 & 19.75 \\
NICER+XMM&$\rm He_{full}$&10.98 & 11.82 & 13.07 & 15.52 & 20.50 \\
\enddata
\tablecomments{A comparison of the $-2\sigma$, $-1\sigma$, median, $+1\sigma$, and $+2\sigma$ constraints on the equatorial radius of PSR~J0740+6620 (measured in kilometers), using different data sets and atmospheric models. When we include only the NICER data, the inferred stellar radius and especially its upper limit depend on the model atmosphere assumed. However, when we also include the XMM-Newton data, which constrain the spectra of the stellar flux and the non-stellar background, we find more consistent results across different assumed model atmospheres.}
\label{tab:radii}
\end{deluxetable}

A general trend illustrated in Table~\ref{tab:radii} and Figure~\ref{fig:allRadii} is that in each analysis, when the XMM-Newton data are included the lower limits on the stellar radius increase. Because the XMM-Newton observations constrain the number of non-stellar counts in the star's pulse profile the inferred fractional modulation in the profile increases when the XMM-Newton data are included, placing a strong upper bound on the stellar compactness. The XMM-Newton data also constrain the total flux from the stellar surface and its spectrum, and when we assume the atmosphere is fully ionized hydrogen or helium we find that, compared to fitting the NICER data alone, including the XMM-Newton data places more stringent upper limits on the stellar radius. As shown in Appendices \ref{app:NO} and \ref{app:NX}, the maximum-likelihood radius is also slightly reduced. When we instead assume the atmosphere is partially ionized hydrogen, we find that the upper limits on the radius are more lenient when the XMM-Newton data are included, as previously found in \citet{2021ApJ...918L..28M}, which used these same model atmospheres. 

If we compare our results to those presented in \citet{2021ApJ...918L..28M}, temporarily restricting ourselves to the model atmospheres used in that work, we find a $68\%$ credible interval for the equatorial radius that is narrower (only $83\%$ as wide) when the NICER and XMM-Newton data are jointly fit. On the other hand, when we fit the NICER data alone we find no improvement over \citet{2021ApJ...918L..28M}. We have not rigorously determined why this is the case. One clear difference is that our current analyses have identified support for lunate modes in our NICER-only analyses, which tend to support larger-radius configurations (see Tables \ref{tab:modes} and \ref{tab:radii}, and Appendices \ref{app:NO} and \ref{app:NX}). \citet{2021ApJ...918L..28M} did not find evidence for these modes. We find that including the XMM-Newton data excludes models with those modes, possibly due to the information about the total stellar X-ray flux provided by these data. This may explain why the analyses that include the XMM-Newton data find lower upper limits on the equatorial radius. A possible but less likely explanation is the different data selection choices made in this work.

We find that fits that assume a fully ionized helium atmosphere infer slightly larger values of the stellar radius than the corresponding fits that assume either a partial or fully ionized hydrogen atmosphere. Similar though much stronger trends have been noted in analyses of the PSR~J0030+0451 NICER data \citep{2019ApJ...887L..24M,2023arXiv230809319S}. These trends are likely due to differences between the emergent spectra and the beaming patterns produced by the two model atmospheres, which are larger for neutron stars that emit X-rays from larger fractions of their surfaces. The previous analyses of the NICER data on PSR~J0740+6620 assuming an ionized helium atmosphere that were reported in \citet{2021ApJ...918L..27R} and \citet{2023arXiv230809319S} did not show any appreciable differences between the results for the helium and hydrogen atmospheres used in these works. However, both works assumed an \textit{a priori} upper limit on the stellar radius of $16\,\rm km$, which reduced the variation of the upper limits they derived from their posteriors.

\subsection{Adequacy of the Fits}\label{sec:adequacy}
\begin{figure}
 \includegraphics[width=\linewidth]{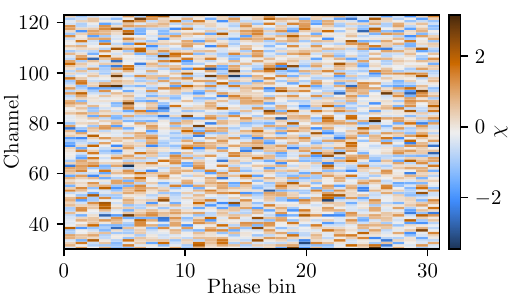}
 \caption{The value of $\chi_i=(m_i-d_i)/\sqrt{m_i}$ (where $m_i$ is the predicted number of model counts and $d_i$ is the number of observed counts in phase-channel bin $i$) for each of the 3008 NICER phase-energy bins considered in this study (32 bins in rotational phase and 94 energy channels), for the energy-resolved pulse profile model that best fits the NICER and XMM-Newton data, assuming a fully ionized hydrogen atmosphere. No patterns in the values of $\chi$ are evident, as expected of a good fit.  For this fit, $\chi^2/{\rm dof}=2818.8/2901$, where $\chi^2=\sum \chi^2_i$. The probability of finding $\chi^2\geq 2818.8$ for 2901 degrees of freedom is $\sim86$\% if the model is correct.}
 \label{fig:residualsNXf}
\end{figure}

As in \citet{2019ApJ...887L..24M} and \citet{2021ApJ...918L..28M}, we performed several tests to assess the adequacy of the fits we obtained. Here we consider only the maximum-likelihood parameters from our joint analysis of the NICER and XMM-Newton data, assuming a fully ionized hydrogen atmosphere. 

We first performed a standard $\chi^2$ test comparing our best-fit phase- and energy-resolved pulse profile model the NICER data and found the ratio of the resulting $\chi^2$ to the number of degrees of freedom of 2818.8/2901. The probability of finding a value of $\chi^2$ this high or higher for this many degrees of freedom using the correct model is 86\%. Figure~\ref{fig:residualsNXf} shows the resulting signed residuals in each phase and energy bin. There is no evidence for clustering or systematic trends. Although this test cannot show that the model being used is correct, this value of $\chi^2$ shows that this model is not obviously deficient. 

\begin{figure}
 \includegraphics[width=\linewidth]{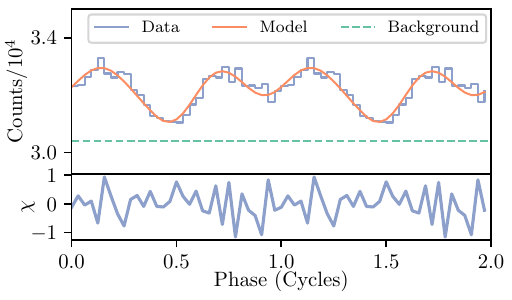}
 \caption{Top: A comparison of the bolometric NICER pulse profile (in 32 bins, plotted in blue) to the bolometric pulse profile from our best fit to the combined NICER and XMM-Newton data set, shown in orange. The dashed green line plots the unmodulated background in our best fit, which was added to the pulse profile generated by the two hot spots, as described in Section \ref{sec:waveformModels}. Bottom: bolometric residuals, $\chi$, as a function of phase. For our best fit we find a bolometric $\chi^2$/dof of 28.8/27, which has a probability of 37\% if the model is correct.}
 \label{fig:bolometricNXf}
\end{figure}

We have also checked the ability of our analysis to reproduce the bolometric pulse profile. This is a nontrivial test, because our Bayesian analysis fits an energy-resolved pulse profile model to the energy- and phase-resolved NICER data. Because the $\chi^2$ statistic is nonlinear, it is possible to significantly over- or under-estimate the total flux at a particular phase, even though the pulse-phase--energy-channel residuals are negligible, causing the residuals obtained by comparing the model bolometric pulse profile to the observed bolometric profile to be significant. We find that the $\chi^2$ per degree of freedom obtained by comparing this bolometric pulse profile model to the observed bolometric pulse profile is $28.8/27$. The probability of finding a value of $\chi^2$ this high or higher given the correct model is $\sim37\%$.\footnote{Because it is not clear a priori how many of our model parameters influence the bolometric pulse profile, we used synthetic data tests to determine that the effective number of parameters is $\approx 5$.  These tests were suggested by Serena Vinciguerra during discussions within the NICER Light Curves working group.} We show the bolometric fit in Figure~\ref{fig:bolometricNXf}, which also shows the size of the background relative to the X-ray emission from the pulsar. 

\begin{figure}
 \includegraphics[width=\linewidth]{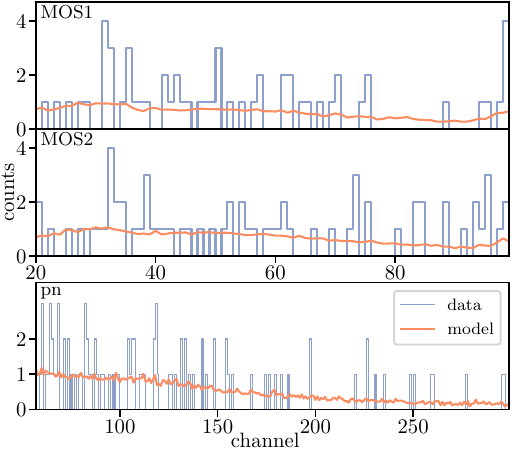}
 \caption{Comparison of the observed energy spectra predicted by our model of a fully ionized hydrogen atmosphere that best fits the combined NICER and XMM-Newton data with the spectra observed by the MOS1, MOS2, and pn detectors on XMM-Newton in the energy channels used in our analysis. Each panel shows the spectrum observed by the indicated XMM-Newton detector in blue as a histogram and the spectrum predicted by the best-fit model in orange. The model provides acceptable fits to all three observed spectra.}
 \label{fig:compareXMM}
\end{figure}

Figure~\ref{fig:compareXMM} compares the observed energy spectra predicted by our model of a fully ionized hydrogen atmosphere that best fits the combined NICER and XMM-Newton data with the spectra observed by the MOS1, MOS2, and pn detectors on XMM-Newton in the energy channels used in our analysis. Because very few counts were recorded in each energy channel, we evaluated the quality of the fit by generating a set of $10^5$ synthetic data sets by Poisson sampling the spectrum predicted by our best-fit model in each channel of each instrument. We then computed the log likelihood for each synthetic data set and instrument, and compared it to the log likelihood of the actual data given our model. We found the total log likelihood of the real data was at the 78th percentile, that the log likelihood of the real pn data alone was at the 99th percentile, and that the log likelihoods of the real data for MOS1 and MOS2 were at the 15th and 4th percentiles, respectively. The overall fit to the XMM data is therefore good, the fit to the pn data is anomalously good, and the fits to the MOS1 and MOS2 data are relatively bad. \citet{2021ApJ...918L..28M} observed the same trend, although to a lesser extent.

\subsection{Differences Between Our Results and Those of Salmi et al.(2024)}\label{sec:diffs}
An independent analysis of the data analyzed in this paper is presented in \citet{salmi2023filler}, which reports a radius measurement of $12.49^{+1.28}_{-0.88}$ km (68\% credibility). This radius measurement overlaps with our measurement of $12.92_{-1.13}^{+2.09}$~km, particularly at the $-1\sigma$ point in the posterior distribution, which is especially important for constraining the dense matter EOS, because larger radii are disfavored by the current gravitational wave observations of binary neutron star mergers \citep[e.g.,][]{2017PhRvL.119p1101A,PhysRevLett.121.091102}.
However, substantial discrepancies remain between the median values and uncertainties in these two measurements. 
As we demonstrate below, these discrepancies are due largely to the different methods used to sample the posterior, the different priors assumed for the equatorial radius (\citealt{salmi2023filler} imposed a hard upper limit of 16~km on the equatorial radius), and the different methods used to model the XMM-Newton data.\footnote{Although we have imposed priors on the stellar mass and the inclination angle between the stellar rotation axis and observer line of sight that are broader than those imposed by \citet{salmi2023filler} in order to account for systematic errors in the Shapiro delay mass measurement and the possibility of a slight spin-orbit misalignment, we have determined that these priors make negligible contributions to the differences between our results and those of \citet{salmi2023filler}. Given that we have found that the new data analyzed here place tighter upper limits on the stellar radius, the upper limit of $16\,{\rm km}$ placed on the radius by \citet{salmi2023filler} should have a smaller effect than it had in previous analyses.} For reference, the results of both our preliminary nested sampling analyses and our converged MCMC analyses are provided in Appendix \ref{app:sampling}.

\subsubsection{Analyses of the NICER Data Alone}
For our initial, preliminary analysis of the data we used MultiNest, which \citet{salmi2023filler} use exclusively for their parameter estimation. We can therefore make a nearly apples-to-apples comparison of some of our preliminary results with their final results in some cases. As we now discuss, our results are consistent with those of \citet{salmi2023filler} when we use the same sampling methodology, consider only the NICER data, and employ the same radius prior.

Consider first analyses that assume a fully ionized hydrogen atmosphere. Our preliminary, MultiNest analysis of only the NICER data gave an estimate for the equatorial radius of $12.15^{+2.3}_{-1.51}~{\rm km}$; the corresponding absolute and fractional widths of the $\pm 1 \sigma$ interval are $3.81$~km and 31\%. The final radius estimate quoted in \citet{salmi2023filler} for this same case is slightly smaller, namely, $11.29^{+1.13}_{-0.81}$~km, while the absolute and fractional widths of the $\pm 1 \sigma$ interval are almost a factor of two smaller, namely, 1.94~km and 17\%. If we now discard all posterior samples with $R_e>16\,\rm km$ to mimic the radius prior assumed in \citet{salmi2023filler}, our radius estimate becomes $11.98^{+1.9}_{-1.4}$~km, and the absolute and fractional widths of the $\pm 1 \sigma$ interval become 3.28~km and 27\%. We note, however, that the lunate spot mode was enormously favored over the dual-spot configuration for this data set and atmosphere model in our analysis, and that the analysis of \citet{salmi2023filler} was precluded from finding this mode by their constraint that each spot have an angular radius less than $\pi/2$ and that the spots were not allowed to overlap.

Consider now the preliminary, MultiNest results we obtained assuming a partially ionized hydrogen atmosphere and the final results \citet{salmi2023filler} obtained assuming a fully ionized hydrogen atmosphere. If we again discard from our analysis all samples with $R_e>16$ km, our radius estimate becomes $11.47^{+1.2}_{-0.91}$~km, and the absolute and fractional widths of the $\pm 1 \sigma$ interval become 3.28~km and 27\%. If we then also discard all samples with spot angular radii larger than $0.4$ radians, which effectively eliminates the lunate spot mode, we obtain a radius estimate of $11.36^{+0.97}_{-0.81}$~km and the absolute and fractional widths of the $\pm 1 \sigma$ interval become 1.78~km and 16\%, in agreement with the result reported by \citet{salmi2023filler}.

\subsubsection{Analyses of the NICER and XMM-Newton Data}
Our preliminary joint analysis of the NICER and XMM-Newton data using MultiNest and assuming a fully ionized hydrogen atmosphere gave a radius estimate of $12.70^{+1.48}_{-0.97}$~km. Using a prior as above to exclude radii greater than 16~km reduced the estimated radius to $12.66^{+1.37}_{-0.94}$~km. Because the lower bounds on the radius provide the most information about the EOS, it is useful to compare the value of the radius at the $-1\sigma$ point in the posterior distribution obtained by various analyses. In our preliminary analysis we obtained, whereas \citet{salmi2023filler} found 11.61~km when analyzing the same data. Based on the above comparison of analyses that use only the NICER data, this difference, as well as some of the difference in the reported widths of the credible regions, appears to be due largely to differences in the treatment of the XMM-Newton data. As one example, we treated the distribution of blank-sky counts in each energy channel as a realization of a Poisson process, whereas \citet{salmi2023filler} followed \citet{2021ApJ...918L..27R} in assigning a flat prior for the distribution of blank-sky counts in each channel with bounds chosen to be a few times larger than the statistical uncertainty in the number of blank-sky counts, clipped to maintain positivity.
Although notable, the differences in the assumed radius prior and in the treatments of the XMM-Newton data are insufficient to explain the difference between the width of our radius posteriors and those of \citet{salmi2023filler}.

Another significant difference between our analysis and that of \citet{salmi2023filler} is the statistical sampling method employed. Although MultiNest is often suitable for exploring the parameter space, numerous studies have shown that it can be unreliable when used for parameter estimation, as discussed in Section~\ref{sec:nested}. Appendix~\ref{app:sampling} shows that using MultiNest for parameter estimation tends to underestimate the widths of radius credible regions, and that these are typically biased towards low radii. 

To derive converged results, we performed MCMC sampling, using our initial MultiNest results only to select the initial positions of the walkers. Over the course of the MCMC sampling, the radius credible interval expanded from the initial, MultiNest-derived interval of $12.70^{+1.48}_{-0.97}$~km to $12.92^{+2.09}_{-1.13}$ km.\footnote{Figure~3 of \citet{2021ApJ...918L..28M} provides an illustrative example of the expansion of the initial, MultiNest-determined credible regions over the course of more thorough MCMC sampling.} In section 4.3 of \citet{salmi2023filler}, the authors report an additional MultiNest analysis that used a lower sampling efficiency ($\epsilon=10^{-3}$) than the efficiency they used to obtain the headline result featured in the abstract of their paper. A lower efficiency allows MultiNest to better sample the parameter space. The additional analysis gave a radius of $12.55^{+1.37}_{-0.92}$, in better agreement with our result than the headline result, despite the different treatments of the XMM data and radius prior in these two analyses. 

The headline result of \citet{salmi2023filler} ($R_e\approx12.49^{+1.28}_{-0.88}$ km) is only very slightly narrower than the value reported in \citet{2021ApJ...918L..27R} ($R_e\approx12.39^{+1.30}_{-0.98}$ km), despite the inclusion of about 50\% more data.\footnote{This is not quite a fair comparison because the headline results of \citet{2021ApJ...918L..27R} used an unrealistically broad prior on the relative calibration between NICER and XMM-Newton, which strongly affected the inferred radius. Section 4.1 of \citet{salmi2023filler} presents a more detailed comparison to \citet{2021ApJ...918L..27R}. It is also worth noting that the headline results of \citet{salmi2023filler} used $\epsilon=0.01$, whereas \citet{2021ApJ...918L..27R} used $\epsilon=0.1$} Adopting instead the results presented in Section~4.3 of \citet{salmi2023filler}, which used a MultiNest sampling efficiency that is expected to better sample the parameter space, their radius estimate using the present data is $12.55^{+1.37}_{-0.92}$~km, very slightly broader than the result reported in \citet{2021ApJ...918L..27R}. In contrast, our radius estimate improved from $13.71^{+2.61}_{-1.50}$ km in \citet{2021ApJ...918L..28M} to $12.92^{+2.09}_{-1.13}$ km in the present work. Overall, the differences between the results obtained in different analyses have diminished as NICER has accumulated more data. 

\section{Discussion}\label{sec:disc}
As discussed in Section \ref{sec:data}, we have analyzed an additional $1.1$~Ms of NICER data on PSR~J0740+6620, in addition to the NICER and XMM-Newton data analyzed in \citet{2021ApJ...918L..28M}.
Analysis of these NICER and XMM-Newton data improves the precision of the constraint on the radius of this neutron star from $13.71^{+2.61}_{-1.50}$~km to $12.92^{+2.09}_{-1.13}$~km, reducing the fractional uncertainty in the radius from $\sim30\%$ to $\sim25\%$. This more precise measurement of the radius incrementally improves the constraints on the EOS of dense matter. We expect the EOS to be further constrained by analysis of the additional data  on PSR~J0740+6620 that is currently being collected. Although measurements of the radius of PSR~J0740+6620 made using NICER still depend somewhat on the procedure used to analyze the data (see Section \ref{sec:diffs}), the consequences of using different procedures have become less significant as the amount of NICER data has increased.

\subsection{Equation of State Constraints}

Our current radius measurement is more precise than that of \citet{2021ApJ...918L..28M}, which was based on the data collected through 2020 (see \citet{2021ApJ...918L..28M}). For example, the value of the equatorial radius at the $+1\sigma$ point in the posterior distribution has decreased from 16.32~km to 14.34~km, and the radius width at $\pm 1\sigma$ is now 3.22~km, which is 78\% of the 4.12~km width reported in \citet{2021ApJ...918L..28M}.  This improvement is almost exactly what is expected under the assumption that the radius uncertainty scales as the inverse square root of the exposure time, because $\sqrt{1602.68~{\rm ks}/2733.81~{\rm ks}}\approx 0.77$.

However, this decrease has a relatively minor effect on our EOS inferences, because in making these inferences we also include other results that bear on the properties of cold, catalyzed matter at densities above the saturation density of nuclear matter.  
In particular, the upper bound placed on the tidal deformability of neutron stars with moderate masses by the observations of GW170817 \citep{2017PhRvL.119p1101A,2018PhRvL.121p1101A,PhysRevLett.121.091102} disfavors large stellar radii, so equatorial radii $\gtrsim 14$~km already had low posterior weight. 
Our new constraints on the radius of PSR~J0740+6620 can be viewed as being in better agreement with current tidal deformability measurements.

In this section we update our EOS constraints based on our new measurement of the radius of PSR~J0740+6620.  Our method is described in detail elsewhere \citep{2019ApJ...887L..24M,2020ApJ...888...12M,2021ApJ...918L..28M}.  In brief:
\begin{enumerate}

    \item We assume that we know the EOS below a threshold density, which we take to be half the saturation density of nuclear matter, because this is roughly the density of the transition between the solid crust and the fluid core \citep{2013ApJ...773...11H}.  We use the QHC19 EOS of \citet{2019ApJ...885...42B} below this density, but the mass, radius, tidal deformability, and other stellar properties we derive are insensitive to the low-density EOS.

    \item We then extend the EOS to higher densities.  There are many frameworks for such extensions, but for the most direct comparison with previous work we use the Gaussian process approach that was introduced in this context by \citet{2019PhRvD..99h4049L}.  We use the same set of 100,000 realizations of a high-density EOS that were used in \citet{2021ApJ...918L..28M}, so any differences between the results we obtain here and our previous results are due to the new measurement rather than to different samples in the EOS space. Our particular Gaussian process sampling of possible equations of state does not explicitly take into account the possibility of phase transitions, but initial studies suggest that phase transitions (or more generally, complex variations of the sound speed with density) are neither favored nor disfavored by current data \citep{2023PhRvD.108d3013E,2023JPhCS2536a2006M,2023arXiv230902345M}. As discussed in \citet{2021ApJ...918L..28M}, this prior on the sound speed tends towards the speed of light at very high densities.

    \item The EOS constraints presented here also impose a Gaussian prior on the nuclear symmetry energy at nuclear saturation density $S = 32 \pm 2$~MeV, \citep{2012PhRvC..86a5803T}, take into account the existence of the three high-mass pulsars \citep{2013Sci...340..448A,2018ApJ...859...47A,2021ApJ...915L..12F}, and make use of the tidal deformability posteriors from the gravitational wave observations of GW170817 \citep{2017PhRvL.119p1101A,2018PhRvL.121p1101A,2018PhRvL.121i1102D} and GW190425 \citep{2020ApJ...892L...3A}.
    
\end{enumerate}

\begin{figure}
 \includegraphics[width=\linewidth]{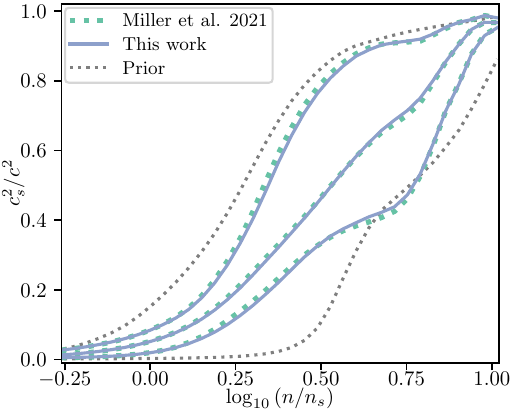}
 \caption{The constraints we infer on the square of the sound speed in dense matter using the results obtained in this paper. At a given baryonic number density (shown here in units of the nuclear saturation density $n_s=0.16~\rm fm^{-3}$), the lower line shows the 5th percentile of the squared sound speed, the upper line shows the 95th percentile, and the middle line shows the median value of the squared sound speed. The dotted green lines show the constraints inferred by \citet{2021ApJ...918L..28M} whereas the solid blue lines show the constraints obtained using the results presented in this paper. The dotted black lines show the prior 5th and 95th percentiles of our EOS samples. We find a slightly lower sound speed at densities below $n\lesssim3n_s\,(\log_{10}{n/n_s}\lesssim0.5)$ and slightly tighter overall constraints on the sound speed in the density range from $n\sim 3-5n_s~(\log_{10}{n/n_s}\sim 0.5-0.7)$ than those found by \citet{2021ApJ...918L..28M}.}
 \label{fig:eos}
\end{figure}

\begin{figure}
 \includegraphics[width=\linewidth]{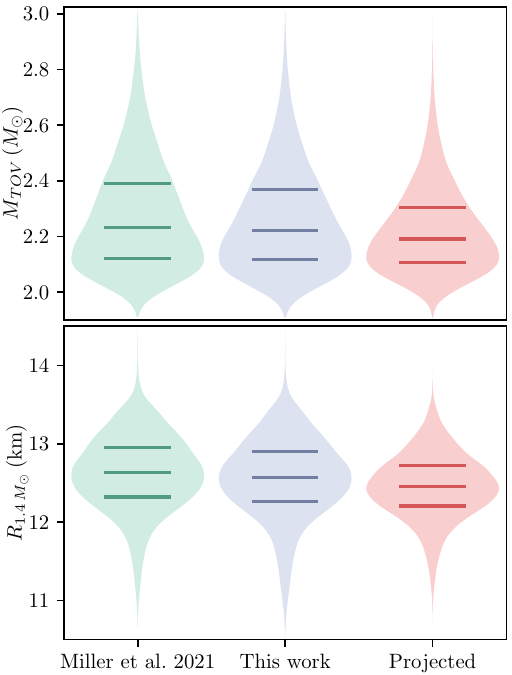}
 \caption{The posterior distributions (shaded regions) and the 25th, 50th, and 75th percentiles of these distributions (horizontal lines) for the maximum stable gravitational mass of a nonrotating neutron star (top panel) and the circumferential radius of a nonrotating $1.4\,M_\odot$ neutron star (bottom panel) inferred from the current constraints on the EOS of neutron star matter. The results reported by \citet{2021ApJ...918L..28M} are shown in green, the constraints based on the updated measurements presented in this work are shown in blue, and the projected constraints that might follow from analysis of an additional $\sim2.7\, {\rm Ms}$ of NICER data on PSR~J0740+6620 are shown in red.} \label{fig:EOSRM}
\end{figure}

\begin{deluxetable}{ccccc}
\caption{Updated Maximum Mass and Fiducial Radius}
\tablehead{
\colhead{Quantity} & \colhead{Data set} &  \colhead{$-1\sigma$} & \colhead{median} & \colhead{$+1\sigma$} 
}
\startdata
$M_{\rm TOV}(M_\odot)$&Miller et al. (2021)&2.08&2.23&2.47 \\
$M_{\rm TOV}(M_\odot)$&This work&2.08 &2.22 &2.44 \\
$M_{\rm TOV}(M_\odot)$&Projected &2.08 &2.19 &2.37 \\\hline
$R_e(1.4~M_\odot)({\rm km})$&Miller et al. (2021)&12.17 &12.63 &13.11 \\
$R_e(1.4~M_\odot)({\rm km})$&This work&12.09 &12.57 &13.06 \\
$R_e(1.4~M_\odot)({\rm km})$&Projected&12.07 &12.46 &12.85 \\
\enddata
\tablecomments{Comparison of the $-1\sigma$, median, and $+1\sigma$ points in the posterior distributions of the maximum gravitational mass of a nonrotating neutron star ($M_{\rm TOV}$) and the equatorial circumferential radius of a fiducial nonrotating $1.4~M_\odot$ neutron star ($R_e$) inferred by \citet{2021ApJ...918L..28M} using the PSR~J0740+6620 data then available and the values we infer here after including an additional $1.1$ Ms of NICER data. The improvements achieved by using the larger present data set are relatively small because other information, notably that provided by the gravitational waves produced by the double neutron star coalescence event GW170817, plays an important role in both analyses and by itself disfavors the high radii that are more disfavored by the present NICER data set than by the earlier NICER data set. The table also shows a projection of the constraints on $M_{\rm TOV}$ and $R_e(1.4\,M_\odot)$ that could potentially be achieved using twice the presently analyzed NICER data on PSR~J0740+6620.}
\label{tab:maxmass}
\end{deluxetable}

Figure~\ref{fig:eos} shows how the constraints on the square of the sound speed in dense matter are revised when the additional NICER results obtained in this paper are included in the analysis. The details of this analysis and the priors we used can be found in Section~5.3 and Figure~13 of \citet{2021ApJ...918L..28M}. Table~\ref{tab:maxmass} and Figure~\ref{fig:EOSRM} show how the inferred maximum gravitational mass of a nonrotating neutron star ($M_{\rm TOV}$) and the circumferential radius of a nonrotating neutron star of fiducial gravitational mass $1.4~M_\odot$ are revised when our current results are included. All the inferences in Table~\ref{tab:maxmass} include the current measurements of the masses of the three high-mass pulsars, the gravitational wave-inferred limits on the tidal deformability of neutron stars, and the current experimental constraints on the nuclear symmetry energy (see section~5.3 of \citealt{2021ApJ...918L..28M} for details).

We find evidence for a slightly softer EOS and consequent reduction of the maximum mass and fiducial radius, but the changes are relatively small. The high values for the radius of PSR~J0740+6620 that are more disfavored by the present NICER data than by the previous data are also disfavored by the gravitational wave observations. However, as the slight improvements in Figure~\ref{fig:eos} suggest, \textit{the NICER data has begun to exclude high stellar radii independent of the constraints provided by gravitational wave data.} We have included, in Table~\ref{tab:maxmass} and Figure \ref{fig:EOSRM}, a forecast of how the constraints on the equatorial radius of a 1.4 solar mass neutron star and the maximum mass of a nonrotating neutron star might tighten using twice the present data on PSR~J0740+6620, or equivalent measurements of another pulsar. However, we caution that the precise values of the constraints on the equatorial radius of a 1.4 solar mass neutron star 
and the maximum mass of a nonrotating neutron star depend more strongly on the details of a given EOS parameterization than do the EOS constraints themselves, as shown in Table 4 of \citet{2021ApJ...918L..28M}.
Thus, additional NICER observations will continue to improve our knowledge of the properties of dense matter; for example, our projections suggest that doubling the current exposure time on PSR J0740 would by itself reduce the uncertainty of the radius of a fiducial $1.4\,M_\odot$ pulsar to $\sim80\%$ of its present value, at least based on the set of EOS that we currently consider.

\subsection{Future Prospects}
The X-ray faintness of PSR~J0740+6620 makes measurement of its radius using NICER challenging. Based on the data analyzed here, which yield an uncertainty in the radius of $\pm\sim12\%$, and assuming the uncertainty is proportional to the inverse of the square root of the observing duration, more than 12~Ms of NICER data would be required to achieve an uncertainty of $\pm\sim5\%$. Although it is unlikely that NICER will be able to collect such a large quantity of data on PSR~J0740+6620, even including the data from $\sim2-3$~Ms of additional observations could appreciably improve the constraints on the equation of state of dense matter, as Table~\ref{tab:maxmass} shows.

Future X-ray timing missions such as STROBE-X \citep{2019arXiv190303035R}, which would have an effective area $\sim2~\rm m^2$ at 1.5 keV, or eXTP \citep{2019SCPMA..6229502Z} should be able to provide estimates of the radii of even faint pulsars like PSR~J0740+6620 with a precision of a few percent or better. Given the paucity of counts observed by XMM-Newton and the apparent discrepancy between the XMM-Newton pn and MOS1/MOS2 data (see Section~\ref{sec:adequacy}), additional imaging observations of PSR~J0740+6620 may further improve our knowledge of the flux from the pulsar and the background flux, and therefore decrease the limit on the radius of PSR~J0740+6620. 
\pagebreak
\section{Conclusions}\label{sec:conclusions}
The present NICER observations, which include approximately 1.1~Ms of data in addition to the $\sim\,1.6$~Ms of data that was previously available, have improved the constraints on the radius of PSR~J0740+6620 from $13.71^{+2.61}_{-1.50}$~km \citep{2021ApJ...918L..28M} to $12.92_{-1.13}^{+2.09}$~km. These additional data have also reduced the effects of using different analysis methods. Being able to  obtain results that are insensitive to the analysis methods used is particularly important when analyzing data on faint sources such as PSR~J0740+6620 \citep[see, e.g., Section \ref{sec:diffs} and][]{salmi2023filler}. Although the improvements in estimates of the mass and radius of PSR~J0740+6620 presented here are incremental, these new estimates provide further information about the properties of matter at supranuclear densities. They imply that the EOS of cold, dense matter is slightly softer than was implied by the previously available data. Future NICER observations will provide even more information about the EOS.

\section*{Software}
emcee \citep{2013PASP..125..306F}, MultiNest \citep{2009MNRAS.398.1601F}, matplotlib \citep{4160265}, numpy \citep{5725236}, scipy \citep{2020SciPy-NMeth}, PyMultiNest \citep{2016ascl.soft06005B}, PGF/Ti$k$Z \citep{tantau:2013a}, PINT \citep{2019ascl.soft02007L,2021ApJ...911...45L}, XMM-Newton SAS \citep[][]{2004ASPC..314..759G,2014ascl.soft04004S}, HEASoft \citep{2014ascl.soft08004N}

\section*{Acknowledgments}

A.J.D. and M.C.M. were supported in part by NASA ADAP grants 80NSSC21K0649 and 80NSSC20K0288. Part of this work was performed at the Aspen Center for Physics, which is supported by U.S. National Science Foundation grant PHY-1607611.  Some of the resources used in this work were provided by the NASA High-End Computing (HEC) Program through the NASA Center for Climate Simulation (NCCS) at Goddard Space Flight Center. A.J.D. gratefully acknowledges the support of LANL/LDRD under project number 20220087DR. The LA-UR number is LA-UR-24-20120.
The authors acknowledge the University of Maryland supercomputing resources (http://hpcc.umd.edu) that were made available for conducting the research reported in this paper, as well as the YORP and ASTRA clusters administered by the Center for Theory and Computation within the University of Maryland Department of Astronomy. We are particularly thankful to Tuomo Salmi and Serena Vinciguerra for useful discussions.

S.B.~acknowledges funding from NASA grants 80NSSC20K0275 and 80NSSC22K0728.
S.G. acknowledges the support of the Centre National d'Etudes Spatiales (CNES). 
W.C.G.H. acknowledges support through grant 80NSSC23K0078 from NASA.
S.M. acknowledges support from NSERC Discovery Grant RGPIN-2019-06077.
Portions of this work performed at the Naval Research Laboratory were supported by NASA.

We acknowledge extensive use of NASA’s Astrophysics Data System (ADS) Bibliographic Services and the ArXiv.
\facilities{NICER, XMM-Newton}
\bibliographystyle{aasjournal}
\bibliography{references}

\begin{thebibliography}{}
\expandafter\ifx\csname natexlab\endcsname\relax\def\natexlab#1{#1}\fi
\providecommand{\url}[1]{\href{#1}{#1}}
\providecommand{\dodoi}[1]{doi:~\href{http://doi.org/#1}{\nolinkurl{#1}}}
\providecommand{\doeprint}[1]{\href{http://ascl.net/#1}{\nolinkurl{http://ascl.net/#1}}}
\providecommand{\doarXiv}[1]{\href{https://arxiv.org/abs/#1}{\nolinkurl{https://arxiv.org/abs/#1}}}

\bibitem[{{Abbott} {et~al.}(2017){Abbott}, {Abbott}, {Abbott}, {Acernese},
  {Ackley}, {Adams}, {Adams}, {Addesso}, {Adhikari}, {Adya}, {Affeldt},
  {Afrough}, {Agarwal}, {Agathos}, {Agatsuma}, {Aggarwal}, {Aguiar}, {Aiello},
  {Ain}, {Ajith}, {Allen}, {Allen}, {Allocca}, {Altin}, {Amato}, {Ananyeva},
  {Anderson}, {Anderson}, {Angelova}, {Antier}, {Appert}, {Arai}, {Araya},
  {Areeda}, {Arnaud}, {Arun}, {Ascenzi}, {Ashton}, {Ast}, {Aston}, {Astone},
  {Atallah}, {Aufmuth}, {Aulbert}, {AultONeal}, {Austin}, {Avila-Alvarez},
  {Babak}, {Bacon}, {Bader}, {Bae}, {Bailes}, {Baker}, {Baldaccini},
  {Ballardin}, {Ballmer}, {Banagiri}, {Barayoga}, {Barclay}, {Barish},
  {Barker}, {Barkett}, {Barone}, {Barr}, {Barsotti}, {Barsuglia}, {Barta},
  {Barthelmy}, {Bartlett}, {Bartos}, {Bassiri}, {Basti}, {Batch}, {Bawaj},
  {Bayley}, {Bazzan}, {B{\'e}csy}, {Beer}, {Bejger}, {Belahcene}, {Bell},
  {Berger}, {Bergmann}, {Bernuzzi}, {Bero}, {Berry}, {Bersanetti}, {Bertolini},
  {Betzwieser}, {Bhagwat}, {Bhandare}, {Bilenko}, {Billingsley}, {Billman},
  {Birch}, {Birney}, {Birnholtz}, {Biscans}, {Biscoveanu}, {Bisht}, {Bitossi},
  {Biwer}, {Bizouard}, {Blackburn}, {Blackman}, {Blair}, {Blair}, {Blair},
  {Bloemen}, {Bock}, {Bode}, {Boer}, {Bogaert}, {Bohe}, {Bondu}, {Bonilla},
  {Bonnand}, {Boom}, {Bork}, {Boschi}, {Bose}, {Bossie}, {Bouffanais}, {Bozzi},
  {Bradaschia}, {Brady}, {Branchesi}, {Brau}, {Briant}, {Brillet}, {Brinkmann},
  {Brisson}, {Brockill}, {Broida}, {Brooks}, {Brown}, {Brown}, {Brunett},
  {Buchanan}, {Buikema}, {Bulik}, {Bulten}, {Buonanno}, {Buskulic}, {Buy},
  {Byer}, {Cabero}, {Cadonati}, {Cagnoli}, {Cahillane}, {Calder{\'o}n
  Bustillo}, {Callister}, {Calloni}, {Camp}, {Canepa}, {Canizares}, {Cannon},
  {Cao}, {Cao}, {Capano}, {Capocasa}, {Carbognani}, {Caride}, {Carney},
  {Carullo}, {Casanueva Diaz}, {Casentini}, {Caudill}, {Cavagli{\`a}},
  {Cavalier}, {Cavalieri}, {Cella}, {Cepeda}, {Cerd{\'a}-Dur{\'a}n},
  {Cerretani}, {Cesarini}, {Chamberlin}, {Chan}, {Chao}, {Charlton}, {Chase},
  {Chassande-Mottin}, {Chatterjee}, {Chatziioannou}, {Cheeseboro}, {Chen},
  {Chen}, {Chen}, {Cheng}, {Chia}, {Chincarini}, {Chiummo}, {Chmiel}, {Cho},
  {Cho}, {Chow}, {Christensen}, {Chu}, {Chua}, {Chua}, {Chung}, {Chung},
  {Ciani}, {Ciolfi}, {Cirelli}, {Cirone}, {Clara}, {Clark}, {Clearwater},
  {Cleva}, {Cocchieri}, {Coccia}, {Cohadon}, {Cohen}, {Colla}, {Collette},
  {Cominsky}, {Constancio}, {Conti}, {Cooper}, {Corban}, {Corbitt},
  {Cordero-Carri{\'o}n}, {Corley}, {Cornish}, {Corsi}, {Cortese}, {Costa},
  {Coughlin}, {Coughlin}, {Coulon}, {Countryman}, {Couvares}, {Covas}, {Cowan},
  {Coward}, {Cowart}, {Coyne}, {Coyne}, {Creighton}, {Creighton}, {Cripe},
  {Crowder}, {Cullen}, {Cumming}, {Cunningham}, {Cuoco}, {Dal Canton},
  {D{\'a}lya}, {Danilishin}, {D'Antonio}, {Danzmann}, {Dasgupta}, {Da Silva
  Costa}, {Dattilo}, {Dave}, {Davier}, {Davis}, {Daw}, {Day}, {De}, {DeBra},
  {Degallaix}, {De Laurentis}, {Del{\'e}glise}, {Del Pozzo}, {Demos}, {Denker},
  {Dent}, {De Pietri}, {Dergachev}, {De Rosa}, {DeRosa}, {De Rossi}, {DeSalvo},
  {de Varona}, {Devenson}, {Dhurandhar}, {D{\'\i}az}, {Dietrich}, {Di Fiore},
  {Di Giovanni}, {Di Girolamo}, {Di Lieto}, {Di Pace}, {Di Palma}, {Di Renzo},
  {Doctor}, {Dolique}, {Donovan}, {Dooley}, {Doravari}, {Dorrington},
  {Douglas}, {Dovale {\'A}lvarez}, {Downes}, {Drago}, {Dreissigacker},
  {Driggers}, {Du}, {Ducrot}, {Dudi}, {Dupej}, {Dwyer}, {Edo}, {Edwards},
  {Effler}, {Eggenstein}, {Ehrens}, {Eichholz}, {Eikenberry}, {Eisenstein},
  {Essick}, {Estevez}, {Etienne}, {Etzel}, {Evans}, {Evans}, {Factourovich},
  {Fafone}, {Fair}, {Fairhurst}, {Fan}, {Farinon}, {Farr}, {Farr},
  {Fauchon-Jones}, {Favata}, {Fays}, {Fee}, {Fehrmann}, {Feicht}, {Fejer},
  {Fernandez-Galiana}, {Ferrante}, {Ferreira}, {Ferrini}, {Fidecaro},
  {Finstad}, {Fiori}, {Fiorucci}, {Fishbach}, {Fisher}, {Fitz-Axen},
  {Flaminio}, {Fletcher}, {Fong}, {Font}, {Forsyth}, {Forsyth}, {Fournier},
  {Frasca}, {Frasconi}, {Frei}, {Freise}, {Frey}, {Frey}, {Fries}, {Fritschel},
  {Frolov}, {Fulda}, {Fyffe}, {Gabbard}, {Gadre}, {Gaebel}, {Gair},
  {Gammaitoni}, {Ganija}, {Gaonkar}, {Garcia-Quiros}, {Garufi}, {Gateley},
  {Gaudio}, {Gaur}, {Gayathri}, {Gehrels}, {Gemme}, {Genin}, {Gennai},
  {George}, {George}, {Gergely}, {Germain}, {Ghonge}, {Ghosh}, {Ghosh},
  {Ghosh}, {Giaime}, {Giardina}, {Giazotto}, {Gill}, {Glover}, {Goetz},
  {Goetz}, {Gomes}, {Goncharov}, {Gonz{\'a}lez}, {Gonzalez Castro},
  {Gopakumar}, {Gorodetsky}, {Gossan}, {Gosselin}, {Gouaty}, {Grado}, {Graef},
  {Granata}, {Grant}, {Gras}, {Gray}, {Greco}, {Green}, {Gretarsson}, {Groot},
  {Grote}, {Grunewald}, {Gruning}, {Guidi}, {Guo}, {Gupta}, {Gupta}, {Gushwa},
  {Gustafson}, {Gustafson}, {Halim}, {Hall}, {Hall}, {Hamilton}, {Hammond},
  {Haney}, {Hanke}, {Hanks}, {Hanna}, {Hannam}, {Hannuksela}, {Hanson},
  {Hardwick}, {Harms}, {Harry}, {Harry}, {Hart}, {Haster}, {Haughian}, {Healy},
  {Heidmann}, {Heintze}, {Heitmann}, {Hello}, {Hemming}, {Hendry}, {Heng},
  {Hennig}, {Heptonstall}, {Heurs}, {Hild}, {Hinderer}, {Ho}, {Hoak}, {Hofman},
  {Holt}, {Holz}, {Hopkins}, {Horst}, {Hough}, {Houston}, {Howell}, {Hreibi},
  {Hu}, {Huerta}, {Huet}, {Hughey}, {Husa}, {Huttner}, {Huynh-Dinh}, {Indik},
  {Inta}, {Intini}, {Isa}, {Isac}, {Isi}, {Iyer}, {Izumi}, {Jacqmin}, {Jani},
  {Jaranowski}, {Jawahar}, {Jim{\'e}nez-Forteza}, {Johnson},
  {Johnson-McDaniel}, {Jones}, {Jones}, {Jonker}, {Ju}, {Junker}, {Kalaghatgi},
  {Kalogera}, {Kamai}, {Kandhasamy}, {Kang}, {Kanner}, {Kapadia}, {Karki},
  {Karvinen}, {Kasprzack}, {Kastaun}, {Katolik}, {Katsavounidis}, {Katzman},
  {Kaufer}, {Kawabe}, {K{\'e}f{\'e}lian}, {Keitel}, {Kemball}, {Kennedy},
  {Kent}, {Key}, {Khalili}, {Khan}, {Khan}, {Khan}, {Khazanov}, {Kijbunchoo},
  {Kim}, {Kim}, {Kim}, {Kim}, {Kim}, {Kim}, {Kimbrell}, {King}, {King},
  {Kinley-Hanlon}, {Kirchhoff}, {Kissel}, {Kleybolte}, {Klimenko}, {Knowles},
  {Koch}, {Koehlenbeck}, {Koley}, {Kondrashov}, {Kontos}, {Korobko}, {Korth},
  {Kowalska}, {Kozak}, {Kr{\"a}mer}, {Kringel}, {Krishnan}, {Kr{\'o}lak},
  {Kuehn}, {Kumar}, {Kumar}, {Kumar}, {Kuo}, {Kutynia}, {Kwang}, {Lackey},
  {Lai}, {Landry}, {Lang}, {Lange}, {Lantz}, {Lanza}, {Larson},
  {Lartaux-Vollard}, {Lasky}, {Laxen}, {Lazzarini}, {Lazzaro}, {Leaci},
  {Leavey}, {Lee}, {Lee}, {Lee}, {Lee}, {Lee}, {Lehmann}, {Lenon}, {Leon},
  {Leonardi}, {Leroy}, {Letendre}, {Levin}, {Li}, {Linker}, {Littenberg},
  {Liu}, {Liu}, {Lo}, {Lockerbie}, {London}, {Lord}, {Lorenzini}, {Loriette},
  {Lormand}, {Losurdo}, {Lough}, {Lousto}, {Lovelace}, {L{\"u}ck}, {Lumaca},
  {Lundgren}, {Lynch}, {Ma}, {Macas}, {Macfoy}, {Machenschalk}, {MacInnis},
  {Macleod}, {Maga{\~n}a Hernandez}, {Maga{\~n}a-Sandoval}, {Maga{\~n}a
  Zertuche}, {Magee}, {Majorana}, {Maksimovic}, {Man}, {Mandic}, {Mangano},
  {Mansell}, {Manske}, {Mantovani}, {Marchesoni}, {Marion}, {M{\'a}rka},
  {M{\'a}rka}, {Markakis}, {Markosyan}, {Markowitz}, {Maros}, {Marquina},
  {Marsh}, {Martelli}, {Martellini}, {Martin}, {Martin}, {Martynov}, {Marx},
  {Mason}, {Massera}, {Masserot}, {Massinger}, {Masso-Reid}, {Mastrogiovanni},
  {Matas}, {Matichard}, {Matone}, {Mavalvala}, {Mazumder}, {McCarthy},
  {McClelland}, {McCormick}, {McCuller}, {McGuire}, {McIntyre}, {McIver},
  {McManus}, {McNeill}, {McRae}, {McWilliams}, {Meacher}, {Meadors}, {Mehmet},
  {Meidam}, {Mejuto-Villa}, {Melatos}, {Mendell}, {Mercer}, {Merilh},
  {Merzougui}, {Meshkov}, {Messenger}, {Messick}, {Metzdorff}, {Meyers},
  {Miao}, {Michel}, {Middleton}, {Mikhailov}, {Milano}, {Miller}, {Miller},
  {Miller}, {Millhouse}, {Milovich-Goff}, {Minazzoli}, {Minenkov}, {Ming},
  {Mishra}, {Mitra}, {Mitrofanov}, {Mitselmakher}, {Mittleman}, {Moffa},
  {Moggi}, {Mogushi}, {Mohan}, {Mohapatra}, {Molina}, {Montani}, {Moore},
  {Moraru}, {Moreno}, {Morisaki}, {Morriss}, {Mours}, {Mow-Lowry}, {Mueller},
  {Muir}, {Mukherjee}, {Mukherjee}, {Mukherjee}, {Mukund}, {Mullavey}, {Munch},
  {Mu{\~n}iz}, {Muratore}, {Murray}, {Nagar}, {Napier}, {Nardecchia},
  {Naticchioni}, {Nayak}, {Neilson}, {Nelemans}, {Nelson}, {Nery}, {Neunzert},
  {Nevin}, {Newport}, {Newton}, {Ng}, {Nguyen}, {Nguyen}, {Nichols}, {Nielsen},
  {Nissanke}, {Nitz}, {Noack}, {Nocera}, {Nolting}, {North}, {Nuttall},
  {Oberling}, {O'Dea}, {Ogin}, {Oh}, {Oh}, {Ohme}, {Okada}, {Oliver},
  {Oppermann}, {Oram}, {O'Reilly}, {Ormiston}, {Ortega}, {O'Shaughnessy},
  {Ossokine}, {Ottaway}, {Overmier}, {Owen}, {Pace}, {Page}, {Page}, {Pai},
  {Pai}, {Palamos}, {Palashov}, {Palomba}, {Pal-Singh}, {Pan}, {Pan}, {Pang},
  {Pang}, {Pankow}, {Pannarale}, {Pant}, {Paoletti}, {Paoli}, {Papa}, {Parida},
  {Parker}, {Pascucci}, {Pasqualetti}, {Passaquieti}, {Passuello}, {Patil},
  {Patricelli}, {Pearlstone}, {Pedraza}, {Pedurand}, {Pekowsky}, {Pele},
  {Penn}, {Perez}, {Perreca}, {Perri}, {Pfeiffer}, {Phelps}, {Piccinni},
  {Pichot}, {Piergiovanni}, {Pierro}, {Pillant}, {Pinard}, {Pinto}, {Pirello},
  {Pitkin}, {Poe}, {Poggiani}, {Popolizio}, {Porter}, {Post}, {Powell},
  {Prasad}, {Pratt}, {Pratten}, {Predoi}, {Prestegard}, {Prijatelj},
  {Principe}, {Privitera}, {Prix}, {Prodi}, {Prokhorov}, {Puncken}, {Punturo},
  {Puppo}, {P{\"u}rrer}, {Qi}, {Quetschke}, {Quintero}, {Quitzow-James},
  {Raab}, {Rabeling}, {Radkins}, {Raffai}, {Raja}, {Rajan}, {Rajbhandari},
  {Rakhmanov}, {Ramirez}, {Ramos-Buades}, {Rapagnani}, {Raymond}, {Razzano},
  {Read}, {Regimbau}, {Rei}, {Reid}, {Reitze}, {Ren}, {Reyes}, {Ricci},
  {Ricker}, {Rieger}, {Riles}, {Rizzo}, {Robertson}, {Robie}, {Robinet},
  {Rocchi}, {Rolland}, {Rollins}, {Roma}, {Romano}, {Romano}, {Romel}, {Romie},
  {Rosi{\'n}ska}, {Ross}, {Rowan}, {R{\"u}diger}, {Ruggi}, {Rutins}, {Ryan},
  {Sachdev}, {Sadecki}, {Sadeghian}, {Sakellariadou}, {Salconi}, {Saleem},
  {Salemi}, {Samajdar}, {Sammut}, {Sampson}, {Sanchez}, {Sanchez},
  {Sanchis-Gual}, {Sandberg}, {Sanders}, {Sassolas}, {Sathyaprakash},
  {Saulson}, {Sauter}, {Savage}, {Sawadsky}, {Schale}, {Scheel}, {Scheuer},
  {Schmidt}, {Schmidt}, {Schnabel}, {Schofield}, {Sch{\"o}nbeck}, {Schreiber},
  {Schuette}, {Schulte}, {Schutz}, {Schwalbe}, {Scott}, {Scott}, {Seidel},
  {Sellers}, {Sengupta}, {Sentenac}, {Sequino}, {Sergeev}, {Shaddock},
  {Shaffer}, {Shah}, {Shahriar}, {Shaner}, {Shao}, {Shapiro}, {Shawhan},
  {Sheperd}, {Shoemaker}, {Shoemaker}, {Siellez}, {Siemens}, {Sieniawska},
  {Sigg}, {Silva}, {Singer}, {Singh}, {Singhal}, {Sintes}, {Slagmolen},
  {Smith}, {Smith}, {Smith}, {Somala}, {Son}, {Sonnenberg}, {Sorazu},
  {Sorrentino}, {Souradeep}, {Spencer}, {Srivastava}, {Staats}, {Staley},
  {Steinke}, {Steinlechner}, {Steinlechner}, {Steinmeyer}, {Stevenson},
  {Stone}, {Stops}, {Strain}, {Stratta}, {Strigin}, {Strunk}, {Sturani},
  {Stuver}, {Summerscales}, {Sun}, {Sunil}, {Suresh}, {Sutton}, {Swinkels},
  {Szczepa{\'n}czyk}, {Tacca}, {Tait}, {Talbot}, {Talukder}, {Tanner},
  {T{\'a}pai}, {Taracchini}, {Tasson}, {Taylor}, {Taylor}, {Tewari}, {Theeg},
  {Thies}, {Thomas}, {Thomas}, {Thomas}, {Thorne}, {Thorne}, {Thrane},
  {Tiwari}, {Tiwari}, {Tokmakov}, {Toland}, {Tonelli}, {Tornasi},
  {Torres-Forn{\'e}}, {Torrie}, {T{\"o}yr{\"a}}, {Travasso}, {Traylor},
  {Trinastic}, {Tringali}, {Trozzo}, {Tsang}, {Tse}, {Tso}, {Tsukada}, {Tsuna},
  {Tuyenbayev}, {Ueno}, {Ugolini}, {Unnikrishnan}, {Urban}, {Usman},
  {Vahlbruch}, {Vajente}, {Valdes}, {Vallisneri}, {van Bakel}, {van Beuzekom},
  {van den Brand}, {Van Den Broeck}, {Vander-Hyde}, {van der Schaaf}, {van
  Heijningen}, {van Veggel}, {Vardaro}, {Varma}, {Vass}, {Vas{\'u}th},
  {Vecchio}, {Vedovato}, {Veitch}, {Veitch}, {Venkateswara}, {Venugopalan},
  {Verkindt}, {Vetrano}, {Vicer{\'e}}, {Viets}, {Vinciguerra}, {Vine}, {Vinet},
  {Vitale}, {Vo}, {Vocca}, {Vorvick}, {Vyatchanin}, {Wade}, {Wade}, {Wade},
  {Walet}, {Walker}, {Wallace}, {Walsh}, {Wang}, {Wang}, {Wang}, {Wang},
  {Wang}, {Ward}, {Warner}, {Was}, {Watchi}, {Weaver}, {Wei}, {Weinert},
  {Weinstein}, {Weiss}, {Wen}, {Wessel}, {We{\ss}els}, {Westerweck},
  {Westphal}, {Wette}, {Whelan}, {Whitcomb}, {Whiting}, {Whittle}, {Wilken},
  {Williams}, {Williams}, {Williamson}, {Willis}, {Willke}, {Wimmer},
  {Winkler}, {Wipf}, {Wittel}, {Woan}, {Woehler}, {Wofford}, {Wong}, {Worden},
  {Wright}, {Wu}, {Wysocki}, {Xiao}, {Yamamoto}, {Yancey}, {Yang}, {Yap},
  {Yazback}, {Yu}, {Yu}, {Yvert}, {Zadro{\.Z}ny}, {Zanolin}, {Zelenova},
  {Zendri}, {Zevin}, {Zhang}, {Zhang}, {Zhang}, {Zhang}, {Zhao}, {Zhou},
  {Zhou}, {Zhu}, {Zhu}, {Zimmerman}, {Zucker}, {Zweizig}, {LIGO Scientific
  Collaboration}, \& {Virgo Collaboration}}]{2017PhRvL.119p1101A}
{Abbott}, B.~P., {Abbott}, R., {Abbott}, T.~D., {et~al.} 2017, \prl, 119,
  161101, \dodoi{10.1103/PhysRevLett.119.161101}

\bibitem[{{Abbott} {et~al.}(2018){Abbott}, {Abbott}, {Abbott}, {Acernese},
  {Ackley}, {Adams}, {Adams}, {Addesso}, {Adhikari}, {Adya}, {Affeldt},
  {Agarwal}, {Agathos}, {Agatsuma}, {Aggarwal}, {Aguiar}, {Aiello}, {Ain},
  {Ajith}, {Allen}, {Allen}, {Allocca}, {Aloy}, {Altin}, {Amato}, {Ananyeva},
  {Anderson}, {Anderson}, {Angelova}, {Antier}, {Appert}, {Arai}, {Araya},
  {Areeda}, {Ar{\`e}ne}, {Arnaud}, {Arun}, {Ascenzi}, {Ashton}, {Ast}, {Aston},
  {Astone}, {Atallah}, {Aubin}, {Aufmuth}, {Aulbert}, {AultONeal}, {Austin},
  {Avila-Alvarez}, {Babak}, {Bacon}, {Badaracco}, {Bader}, {Bae}, {Baker},
  {Baldaccini}, {Ballardin}, {Ballmer}, {Banagiri}, {Barayoga}, {Barclay},
  {Barish}, {Barker}, {Barkett}, {Barnum}, {Barone}, {Barr}, {Barsotti},
  {Barsuglia}, {Barta}, {Bartlett}, {Bartos}, {Bassiri}, {Basti}, {Batch},
  {Bawaj}, {Bayley}, {Bazzan}, {B{\'e}csy}, {Beer}, {Bejger}, {Belahcene},
  {Bell}, {Beniwal}, {Bensch}, {Berger}, {Bergmann}, {Bernuzzi}, {Bero},
  {Berry}, {Bersanetti}, {Bertolini}, {Betzwieser}, {Bhandare}, {Bilenko},
  {Bilgili}, {Billingsley}, {Billman}, {Birch}, {Birney}, {Birnholtz},
  {Biscans}, {Biscoveanu}, {Bisht}, {Bitossi}, {Bizouard}, {Blackburn},
  {Blackman}, {Blair}, {Blair}, {Blair}, {Bloemen}, {Bock}, {Bode}, {Boer},
  {Boetzel}, {Bogaert}, {Bohe}, {Bondu}, {Bonilla}, {Bonnand}, {Booker},
  {Boom}, {Booth}, {Bork}, {Boschi}, {Bose}, {Bossie}, {Bossilkov}, {Bosveld},
  {Bouffanais}, {Bozzi}, {Bradaschia}, {Brady}, {Bramley}, {Branchesi}, {Brau},
  {Briant}, {Brighenti}, {Brillet}, {Brinkmann}, {Brisson}, {Brockill},
  {Brooks}, {Brown}, {Brunett}, {Buchanan}, {Buikema}, {Bulik}, {Bulten},
  {Buonanno}, {Buskulic}, {Buy}, {Byer}, {Cabero}, {Cadonati}, {Cagnoli},
  {Cahillane}, {Calder{\'o}n Bustillo}, {Callister}, {Calloni}, {Camp},
  {Canepa}, {Canizares}, {Cannon}, {Cao}, {Cao}, {Capano}, {Capocasa},
  {Carbognani}, {Caride}, {Carney}, {Carullo}, {Casanueva Diaz}, {Casentini},
  {Caudill}, {Cavagli{\`a}}, {Cavalier}, {Cavalieri}, {Cella}, {Cepeda},
  {Cerd{\'a}-Dur{\'a}n}, {Cerretani}, {Cesarini}, {Chaibi}, {Chamberlin},
  {Chan}, {Chao}, {Charlton}, {Chase}, {Chassande-Mottin}, {Chatterjee},
  {Chatziioannou}, {Cheeseboro}, {Chen}, {Chen}, {Chen}, {Cheng}, {Chia},
  {Chincarini}, {Chiummo}, {Chmiel}, {Cho}, {Cho}, {Chow}, {Christensen},
  {Chu}, {Chua}, {Chua}, {Chung}, {Chung}, {Ciani}, {Ciobanu}, {Ciolfi},
  {Cipriano}, {Cirelli}, {Cirone}, {Clara}, {Clark}, {Clearwater}, {Cleva},
  {Cocchieri}, {Coccia}, {Cohadon}, {Cohen}, {Colla}, {Collette}, {Collins},
  {Cominsky}, {Constancio}, {Conti}, {Cooper}, {Corban}, {Corbitt},
  {Cordero-Carri{\'o}n}, {Corley}, {Cornish}, {Corsi}, {Cortese}, {Costa},
  {Cotesta}, {Coughlin}, {Coughlin}, {Coulon}, {Countryman}, {Couvares},
  {Covas}, {Cowan}, {Coward}, {Cowart}, {Coyne}, {Coyne}, {Creighton},
  {Creighton}, {Cripe}, {Crowder}, {Cullen}, {Cumming}, {Cunningham}, {Cuoco},
  {Canton}, {D{\'a}lya}, {Danilishin}, {D'Antonio}, {Danzmann}, {Dasgupta}, {Da
  Silva Costa}, {Dattilo}, {Dave}, {Davier}, {Davis}, {Daw}, {Day}, {DeBra},
  {Deenadayalan}, {Degallaix}, {De Laurentis}, {Del{\'e}glise}, {Del Pozzo},
  {Demos}, {Denker}, {Dent}, {De Pietri}, {Derby}, {Dergachev}, {De Rosa}, {De
  Rossi}, {DeSalvo}, {de Varona}, {Dhurandhar}, {D{\'\i}az}, {Dietrich}, {Di
  Fiore}, {Di Giovanni}, {Di Girolamo}, {Di Lieto}, {Ding}, {Di Pace}, {Di
  Palma}, {Di Renzo}, {Dmitriev}, {Doctor}, {Dolique}, {Donovan}, {Dooley},
  {Doravari}, {Dorrington}, {Dovale {\'A}lvarez}, {Downes}, {Drago},
  {Dreissigacker}, {Driggers}, {Du}, {Dupej}, {Dwyer}, {Easter}, {Edo},
  {Edwards}, {Effler}, {Eggenstein}, {Ehrens}, {Eichholz}, {Eikenberry},
  {Eisenmann}, {Eisenstein}, {Essick}, {Estelles}, {Estevez}, {Etienne},
  {Etzel}, {Evans}, {Evans}, {Fafone}, {Fair}, {Fairhurst}, {Fan}, {Farinon},
  {Farr}, {Farr}, {Fauchon-Jones}, {Favata}, {Fays}, {Fee}, {Fehrmann},
  {Feicht}, {Fejer}, {Feng}, {Fernandez-Galiana}, {Ferrante}, {Ferreira},
  {Ferrini}, {Fidecaro}, {Fiori}, {Fiorucci}, {Fishbach}, {Fisher}, {Fishner},
  {Fitz-Axen}, {Flaminio}, {Fletcher}, {Fong}, {Font}, {Forsyth}, {Forsyth},
  {Fournier}, {Frasca}, {Frasconi}, {Frei}, {Freise}, {Frey}, {Frey},
  {Fritschel}, {Frolov}, {Fulda}, {Fyffe}, {Gabbard}, {Gadre}, {Gaebel},
  {Gair}, {Gammaitoni}, {Ganija}, {Gaonkar}, {Garcia},
  {Garc{\'\i}a-Quir{\'o}s}, {Garufi}, {Gateley}, {Gaudio}, {Gaur}, {Gayathri},
  {Gemme}, {Genin}, {Gennai}, {George}, {George}, {Gergely}, {Germain},
  {Ghonge}, {Ghosh}, {Ghosh}, {Ghosh}, {Giacomazzo}, {Giaime}, {Giardina},
  {Giazotto}, {Gill}, {Giordano}, {Glover}, {Goetz}, {Goetz}, {Goncharov},
  {Gonz{\'a}lez}, {Gonzalez Castro}, {Gopakumar}, {Gorodetsky}, {Gossan},
  {Gosselin}, {Gouaty}, {Grado}, {Graef}, {Granata}, {Grant}, {Gras}, {Gray},
  {Greco}, {Green}, {Green}, {Gretarsson}, {Groot}, {Grote}, {Grunewald},
  {Gruning}, {Guidi}, {Gulati}, {Guo}, {Gupta}, {Gupta}, {Gushwa}, {Gustafson},
  {Gustafson}, {Halim}, {Hall}, {Hall}, {Hamilton}, {Hamilton}, {Hammond},
  {Haney}, {Hanke}, {Hanks}, {Hanna}, {Hannam}, {Hannuksela}, {Hanson},
  {Hardwick}, {Harms}, {Harry}, {Harry}, {Hart}, {Haster}, {Haughian}, {Healy},
  {Heidmann}, {Heintze}, {Heitmann}, {Hello}, {Hemming}, {Hendry}, {Heng},
  {Hennig}, {Heptonstall}, {Hernandez}, {Heurs}, {Hild}, {Hinderer}, {Ho},
  {Hoak}, {Hochheim}, {Hofman}, {Holland}, {Holt}, {Holz}, {Hopkins}, {Horst},
  {Hough}, {Houston}, {Howell}, {Hreibi}, {Huerta}, {Huet}, {Hughey}, {Hulko},
  {Husa}, {Huttner}, {Huynh-Dinh}, {Iess}, {Indik}, {Ingram}, {Inta}, {Intini},
  {Irwin}, {Isa}, {Isac}, {Isi}, {Iyer}, {Izumi}, {Jacqmin}, {Jani},
  {Jaranowski}, {Johnson}, {Johnson}, {Jones}, {Jones}, {Jonker}, {Ju},
  {Junker}, {Kalaghatgi}, {Kalogera}, {Kamai}, {Kandhasamy}, {Kang}, {Kanner},
  {Kapadia}, {Karki}, {Karvinen}, {Kasprzack}, {Katolik}, {Katsanevas},
  {Katsavounidis}, {Katzman}, {Kaufer}, {Kawabe}, {Keerthana},
  {K{\'e}f{\'e}lian}, {Keitel}, {Kemball}, {Kennedy}, {Key}, {Khalili},
  {Khamesra}, {Khan}, {Khan}, {Khan}, {Khan}, {Khazanov}, {Kijbunchoo}, {Kim},
  {Kim}, {Kim}, {Kim}, {Kim}, {Kim}, {King}, {King}, {Kinley-Hanlon},
  {Kirchhoff}, {Kissel}, {Kleybolte}, {Klimenko}, {Knowles}, {Koch},
  {Koehlenbeck}, {Koley}, {Kondrashov}, {Kontos}, {Korobko}, {Korth},
  {Kowalska}, {Kozak}, {Kr{\"a}mer}, {Kringel}, {Krishnan}, {Kr{\'o}lak},
  {Kuehn}, {Kumar}, {Kumar}, {Kumar}, {Kuo}, {Kutynia}, {Kwang}, {Lackey},
  {Lai}, {Landry}, {Landry}, {Lang}, {Lange}, {Lantz}, {Lanza},
  {Lartaux-Vollard}, {Lasky}, {Laxen}, {Lazzarini}, {Lazzaro}, {Leaci},
  {Leavey}, {Lee}, {Lee}, {Lee}, {Lee}, {Lee}, {Lehmann}, {Lenon}, {Leonardi},
  {Leroy}, {Letendre}, {Levin}, {Li}, {Li}, {Li}, {Linker}, {Littenberg},
  {Liu}, {Liu}, {Lo}, {Lockerbie}, {London}, {Longo}, {Lorenzini}, {Loriette},
  {Lormand}, {Losurdo}, {Lough}, {Lousto}, {Lovelace}, {L{\"u}ck}, {Lumaca},
  {Lundgren}, {Lynch}, {Ma}, {Macas}, {Macfoy}, {Machenschalk}, {MacInnis},
  {Macleod}, {Maga{\~n}a Hernandez}, {Maga{\~n}a-Sandoval}, {Maga{\~n}a
  Zertuche}, {Magee}, {Majorana}, {Maksimovic}, {Man}, {Mandic}, {Mangano},
  {Mansell}, {Manske}, {Mantovani}, {Marchesoni}, {Marion}, {M{\'a}rka},
  {M{\'a}rka}, {Markakis}, {Markosyan}, {Markowitz}, {Maros}, {Marquina},
  {Martelli}, {Martellini}, {Martin}, {Martin}, {Martynov}, {Mason}, {Massera},
  {Masserot}, {Massinger}, {Masso-Reid}, {Mastrogiovanni}, {Matas},
  {Matichard}, {Matone}, {Mavalvala}, {Mazumder}, {McCann}, {McCarthy},
  {McClelland}, {McCormick}, {McCuller}, {McGuire}, {McIver}, {McManus},
  {McRae}, {McWilliams}, {Meacher}, {Meadors}, {Mehmet}, {Meidam},
  {Mejuto-Villa}, {Melatos}, {Mendell}, {Mendoza-Gandara}, {Mercer}, {Mereni},
  {Merilh}, {Merzougui}, {Meshkov}, {Messenger}, {Messick}, {Metzdorff},
  {Meyers}, {Miao}, {Michel}, {Middleton}, {Mikhailov}, {Milano}, {Miller},
  {Miller}, {Miller}, {Miller}, {Millhouse}, {Mills}, {Milovich-Goff},
  {Minazzoli}, {Minenkov}, {Ming}, {Mishra}, {Mitra}, {Mitrofanov},
  {Mitselmakher}, {Mittleman}, {Moffa}, {Mogushi}, {Mohan}, {Mohapatra},
  {Montani}, {Moore}, {Moraru}, {Moreno}, {Morisaki}, {Mours}, {Mow-Lowry},
  {Mueller}, {Muir}, {Mukherjee}, {Mukherjee}, {Mukherjee}, {Mukund},
  {Mullavey}, {Munch}, {Mu{\~n}iz}, {Muratore}, {Murray}, {Nagar}, {Napier},
  {Nardecchia}, {Naticchioni}, {Nayak}, {Neilson}, {Nelemans}, {Nelson},
  {Nery}, {Neunzert}, {Nevin}, {Newport}, {Ng}, {Ng}, {Nguyen}, {Nguyen},
  {Nichols}, {Nielsen}, {Nissanke}, {Nitz}, {Nocera}, {Nolting}, {North},
  {Nuttall}, {Obergaulinger}, {Oberling}, {O'Brien}, {O'Dea}, {Ogin}, {Oh},
  {Oh}, {Ohme}, {Ohta}, {Okada}, {Oliver}, {Oppermann}, {Oram}, {O'Reilly},
  {Ormiston}, {Ortega}, {O'Shaughnessy}, {Ossokine}, {Ottaway}, {Overmier},
  {Owen}, {Pace}, {Pagano}, {Page}, {Page}, {Pai}, {Pai}, {Palamos},
  {Palashov}, {Palomba}, {Pal-Singh}, {Pan}, {Pan}, {Pang}, {Pang}, {Pankow},
  {Pannarale}, {Pant}, {Paoletti}, {Paoli}, {Papa}, {Parida}, {Parker},
  {Pascucci}, {Pasqualetti}, {Passaquieti}, {Passuello}, {Patil}, {Patricelli},
  {Pearlstone}, {Pedersen}, {Pedraza}, {Pedurand}, {Pekowsky}, {Pele}, {Penn},
  {Perego}, {Perez}, {Perreca}, {Perri}, {Pfeiffer}, {Phelps}, {Phukon},
  {Piccinni}, {Pichot}, {Piergiovanni}, {Pierro}, {Pillant}, {Pinard}, {Pinto},
  {Pirello}, {Pitkin}, {Poggiani}, {Popolizio}, {Porter}, {Possenti}, {Post},
  {Powell}, {Prasad}, {Pratt}, {Pratten}, {Predoi}, {Prestegard}, {Principe},
  {Privitera}, {Prodi}, {Prokhorov}, {Puncken}, {Punturo}, {Puppo},
  {P{\"u}rrer}, {Qi}, {Quetschke}, {Quintero}, {Quitzow-James}, {Raab},
  {Rabeling}, {Radkins}, {Raffai}, {Raja}, {Rajan}, {Rajbhandari}, {Rakhmanov},
  {Ramirez}, {Ramos-Buades}, {Rana}, {Rapagnani}, {Raymond}, {Razzano}, {Read},
  {Regimbau}, {Rei}, {Reid}, {Reitze}, {Ren}, {Ricci}, {Ricker},
  {Riemenschneider}, {Riles}, {Rizzo}, {Robertson}, {Robie}, {Robinet},
  {Robson}, {Rocchi}, {Rolland}, {Rollins}, {Roma}, {Romano}, {Romel}, {Romie},
  {Rosi{\'n}ska}, {Ross}, {Rowan}, {R{\"u}diger}, {Ruggi}, {Rutins}, {Ryan},
  {Sachdev}, {Sadecki}, {Sakellariadou}, {Salconi}, {Saleem}, {Salemi},
  {Samajdar}, {Sammut}, {Sampson}, {Sanchez}, {Sanchez}, {Sanchis-Gual},
  {Sandberg}, {Sanders}, {Sarin}, {Sassolas}, {Sathyaprakash}, {Saulson},
  {Sauter}, {Savage}, {Sawadsky}, {Schale}, {Scheel}, {Scheuer}, {Schmidt},
  {Schnabel}, {Schofield}, {Sch{\"o}nbeck}, {Schreiber}, {Schuette}, {Schulte},
  {Schutz}, {Schwalbe}, {Scott}, {Scott}, {Seidel}, {Sellers}, {Sengupta},
  {Sentenac}, {Sequino}, {Sergeev}, {Setyawati}, {Shaddock}, {Shaffer}, {Shah},
  {Shahriar}, {Shaner}, {Shao}, {Shapiro}, {Shawhan}, {Shen}, {Shoemaker},
  {Shoemaker}, {Siellez}, {Siemens}, {Sieniawska}, {Sigg}, {Silva}, {Singer},
  {Singh}, {Singhal}, {Sintes}, {Slagmolen}, {Slaven-Blair}, {Smith}, {Smith},
  {Smith}, {Somala}, {Son}, {Sorazu}, {Sorrentino}, {Souradeep}, {Spencer},
  {Srivastava}, {Staats}, {Steinke}, {Steinlechner}, {Steinlechner},
  {Steinmeyer}, {Steltner}, {Stevenson}, {Stocks}, {Stone}, {Stops}, {Strain},
  {Stratta}, {Strigin}, {Strunk}, {Sturani}, {Stuver}, {Summerscales}, {Sun},
  {Sunil}, {Suresh}, {Sutton}, {Swinkels}, {Szczepa{\'n}czyk}, {Tacca}, {Tait},
  {Talbot}, {Talukder}, {Tanner}, {T{\'a}pai}, {Taracchini}, {Tasson},
  {Taylor}, {Taylor}, {Tewari}, {Theeg}, {Thies}, {Thomas}, {Thomas}, {Thomas},
  {Thorne}, {Thrane}, {Tiwari}, {Tiwari}, {Tokmakov}, {Toland}, {Tonelli},
  {Tornasi}, {Torres-Forn{\'e}}, {Torrie}, {T{\"o}yr{\"a}}, {Travasso},
  {Traylor}, {Trinastic}, {Tringali}, {Trovato}, {Trozzo}, {Tsang}, {Tse},
  {Tso}, {Tsuna}, {Tsukada}, {Tuyenbayev}, {Ueno}, {Ugolini}, {Urban}, {Usman},
  {Vahlbruch}, {Vajente}, {Valdes}, {van Bakel}, {van Beuzekom}, {van den
  Brand}, {Van Den Broeck}, {Vander-Hyde}, {van der Schaaf}, {van Heijningen},
  {van Veggel}, {Vardaro}, {Varma}, {Vass}, {Vas{\'u}th}, {Vecchio},
  {Vedovato}, {Veitch}, {Veitch}, {Venkateswara}, {Venugopalan}, {Verkindt},
  {Vetrano}, {Vicer{\'e}}, {Viets}, {Vinciguerra}, {Vine}, {Vinet}, {Vitale},
  {Vo}, {Vocca}, {Vorvick}, {Vyatchanin}, {Wade}, {Wade}, {Wade}, {Walet},
  {Walker}, {Wallace}, {Walsh}, {Wang}, {Wang}, {Wang}, {Wang}, {Wang}, {Ward},
  {Warner}, {Was}, {Watchi}, {Weaver}, {Wei}, {Weinert}, {Weinstein}, {Weiss},
  {Wellmann}, {Wen}, {Wessel}, {We{\ss}els}, {Westerweck}, {Wette}, {Whelan},
  {Whiting}, {Whittle}, {Wilken}, {Williams}, {Williams}, {Williamson},
  {Willis}, {Willke}, {Wimmer}, {Winkler}, {Wipf}, {Wittel}, {Woan}, {Woehler},
  {Wofford}, {Wong}, {Worden}, {Wright}, {Wu}, {Wysocki}, {Xiao}, {Yam},
  {Yamamoto}, {Yancey}, {Yang}, {Yap}, {Yazback}, {Yu}, {Yu}, {Yvert},
  {Zadro{\.Z}ny}, {Zanolin}, {Zelenova}, {Zendri}, {Zevin}, {Zhang}, {Zhang},
  {Zhang}, {Zhang}, {Zhang}, {Zhao}, {Zhou}, {Zhou}, {Zhu}, {Zhu}, {Zimmerman},
  {Zlochower}, {Zucker}, {Zweizig}, {LIGO Scientific Collaboration}, \& {Virgo
  Collaboration}}]{2018PhRvL.121p1101A}
---. 2018, \prl, 121, 161101, \dodoi{10.1103/PhysRevLett.121.161101}

\bibitem[{{Abbott} {et~al.}(2020{\natexlab{a}}){Abbott}, {Abbott}, {Abbott},
  {Abraham}, {Acernese}, {Ackley}, {Adams}, {Adya}, {Affeldt}, {Agathos},
  {Agatsuma}, {Aggarwal}, {Aguiar}, {Aiello}, {Ain}, {Ajith}, {Allen},
  {Allocca}, {Aloy}, {Altin}, {Amato}, {Anand}, {Ananyeva}, {Anderson},
  {Anderson}, {Angelova}, {Antier}, {Appert}, {Arai}, {Araya}, {Areeda},
  {Ar{\`e}ne}, {Arnaud}, {Aronson}, {Arun}, {Ascenzi}, {Ashton}, {Aston},
  {Astone}, {Aubin}, {Aufmuth}, {AultONeal}, {Austin}, {Avendano},
  {Avila-Alvarez}, {Babak}, {Bacon}, {Badaracco}, {Bader}, {Bae}, {Baird},
  {Baker}, {Baldaccini}, {Ballardin}, {Ballmer}, {Bals}, {Banagiri},
  {Barayoga}, {Barbieri}, {Barclay}, {Barish}, {Barker}, {Barkett}, {Barnum},
  {Barone}, {Barr}, {Barsotti}, {Barsuglia}, {Barta}, {Bartlett}, {Bartos},
  {Bassiri}, {Basti}, {Bawaj}, {Bayley}, {Bazzan}, {B{\'e}csy}, {Bejger},
  {Belahcene}, {Bell}, {Beniwal}, {Benjamin}, {Berger}, {Bergmann}, {Bernuzzi},
  {Berry}, {Bersanetti}, {Bertolini}, {Betzwieser}, {Bhandare}, {Bidler},
  {Biggs}, {Bilenko}, {Bilgili}, {Billingsley}, {Birney}, {Birnholtz},
  {Biscans}, {Bischi}, {Biscoveanu}, {Bisht}, {Bitossi}, {Bizouard},
  {Blackburn}, {Blackman}, {Blair}, {Blair}, {Blair}, {Bloemen}, {Bobba},
  {Bode}, {Boer}, {Boetzel}, {Bogaert}, {Bondu}, {Bonnand}, {Booker}, {Boom},
  {Bork}, {Boschi}, {Bose}, {Bossilkov}, {Bosveld}, {Bouffanais}, {Bozzi},
  {Bradaschia}, {Brady}, {Bramley}, {Branchesi}, {Brau}, {Breschi}, {Briant},
  {Briggs}, {Brighenti}, {Brillet}, {Brinkmann}, {Brockill}, {Brooks},
  {Brooks}, {Brown}, {Brunett}, {Buikema}, {Bulik}, {Bulten}, {Buonanno},
  {Buskulic}, {Buy}, {Byer}, {Cabero}, {Cadonati}, {Cagnoli}, {Cahillane},
  {Calder{\'o}n Bustillo}, {Callister}, {Calloni}, {Camp}, {Campbell},
  {Canepa}, {Cannon}, {Cao}, {Cao}, {Carapella}, {Carbognani}, {Caride},
  {Carney}, {Carullo}, {Casanueva Diaz}, {Casentini}, {Caudill},
  {Cavagli{\`a}}, {Cavalier}, {Cavalieri}, {Cella}, {Cerd{\'a}-Dur{\'a}n},
  {Cesarini}, {Chaibi}, {Chakravarti}, {Chamberlin}, {Chan}, {Chao},
  {Charlton}, {Chase}, {Chassande-Mottin}, {Chatterjee}, {Chaturvedi},
  {Chatziioannou}, {Cheeseboro}, {Chen}, {Chen}, {Chen}, {Cheng}, {Cheong},
  {Chia}, {Chiadini}, {Chincarini}, {Chiummo}, {Cho}, {Cho}, {Cho},
  {Christensen}, {Chu}, {Chua}, {Chung}, {Chung}, {Ciani}, {Cie{\'s}lar},
  {Ciobanu}, {Ciolfi}, {Cipriano}, {Cirone}, {Clara}, {Clark}, {Clearwater},
  {Cleva}, {Coccia}, {Cohadon}, {Cohen}, {Colleoni}, {Collette}, {Collins},
  {Colpi}, {Cominsky}, {Constancio}, {Conti}, {Cooper}, {Corban}, {Corbitt},
  {Cordero-Carri{\'o}n}, {Corezzi}, {Corley}, {Cornish}, {Corre}, {Corsi},
  {Cortese}, {Costa}, {Cotesta}, {Coughlin}, {Coughlin}, {Coulon},
  {Countryman}, {Couvares}, {Covas}, {Cowan}, {Coward}, {Cowart}, {Coyne},
  {Coyne}, {Creighton}, {Creighton}, {Cripe}, {Croquette}, {Crowder}, {Cullen},
  {Cumming}, {Cunningham}, {Cuoco}, {Dal Canton}, {D{\'a}lya}, {D'Angelo},
  {Danilishin}, {D'Antonio}, {Danzmann}, {Dasgupta}, {Da Silva Costa},
  {Datrier}, {Dattilo}, {Dave}, {Davier}, {Davis}, {Daw}, {DeBra},
  {Deenadayalan}, {Degallaix}, {De Laurentis}, {Del{\'e}glise}, {Del Pozzo},
  {DeMarchi}, {Demos}, {Dent}, {De Pietri}, {De Rosa}, {De Rossi}, {DeSalvo},
  {de Varona}, {Dhurandhar}, {D{\'\i}az}, {Dietrich}, {Di Fiore}, {DiFronzo},
  {Di Giorgio}, {Di Giovanni}, {Di Giovanni}, {Di Girolamo}, {Di Lieto},
  {Ding}, {Di Pace}, {Di Palma}, {Di Renzo}, {Divakarla}, {Dmitriev}, {Doctor},
  {Donovan}, {Dooley}, {Doravari}, {Dorrington}, {Downes}, {Drago}, {Driggers},
  {Du}, {Ducoin}, {Dupej}, {Durante}, {Dwyer}, {Easter}, {Eddolls}, {Edo},
  {Effler}, {Ehrens}, {Eichholz}, {Eikenberry}, {Eisenmann}, {Eisenstein},
  {Errico}, {Essick}, {Estelles}, {Estevez}, {Etienne}, {Etzel}, {Evans},
  {Evans}, {Fafone}, {Fairhurst}, {Fan}, {Farinon}, {Farr}, {Farr},
  {Fauchon-Jones}, {Favata}, {Fays}, {Fazio}, {Fee}, {Feicht}, {Fejer}, {Feng},
  {Fernandez-Galiana}, {Ferrante}, {Ferreira}, {Ferreira}, {Fidecaro}, {Fiori},
  {Fiorucci}, {Fishbach}, {Fisher}, {Fishner}, {Fittipaldi}, {Fitz-Axen},
  {Fiumara}, {Flaminio}, {Fletcher}, {Floden}, {Flynn}, {Fong}, {Font},
  {Forsyth}, {Fournier}, {Frasca}, {Frasconi}, {Frei}, {Freise}, {Frey},
  {Frey}, {Fritschel}, {Frolov}, {Fronz{\`e}}, {Fulda}, {Fyffe}, {Gabbard},
  {Gadre}, {Gaebel}, {Gair}, {Gammaitoni}, {Gaonkar}, {Garc{\'\i}a-Quir{\'o}s},
  {Garufi}, {Gateley}, {Gaudio}, {Gaur}, {Gayathri}, {Gemme}, {Genin},
  {Gennai}, {George}, {George}, {Gergely}, {Ghonge}, {Ghosh}, {Ghosh}, {Ghosh},
  {Giacomazzo}, {Giaime}, {Giardina}, {Gibson}, {Gill}, {Glover}, {Gniesmer},
  {Godwin}, {Goetz}, {Goetz}, {Goncharov}, {Gonz{\'a}lez}, {Gonzalez Castro},
  {Gopakumar}, {Gossan}, {Gosselin}, {Gouaty}, {Grace}, {Grado}, {Granata},
  {Grant}, {Gras}, {Grassia}, {Gray}, {Gray}, {Greco}, {Green}, {Green},
  {Gretarsson}, {Grimaldi}, {Grimm}, {Groot}, {Grote}, {Grunewald}, {Gruning},
  {Guidi}, {Gulati}, {Guo}, {Gupta}, {Gupta}, {Gupta}, {Gustafson},
  {Gustafson}, {Haegel}, {Halim}, {Hall}, {Hall}, {Hamilton}, {Hammond},
  {Haney}, {Hanke}, {Hanks}, {Hanna}, {Hannam}, {Hannuksela}, {Hansen},
  {Hanson}, {Harder}, {Hardwick}, {Haris}, {Harms}, {Harry}, {Harry},
  {Hasskew}, {Haster}, {Haughian}, {Hayes}, {Healy}, {Heidmann}, {Heintze},
  {Heitmann}, {Hellman}, {Hello}, {Hemming}, {Hendry}, {Heng}, {Hennig},
  {Hernandez Vivanco}, {Heurs}, {Hild}, {Hinderer}, {Ho}, {Hochheim}, {Hofman},
  {Holgado}, {Holland}, {Holt}, {Holz}, {Hopkins}, {Horst}, {Hough}, {Howell},
  {Hoy}, {Huang}, {H{\"u}bner}, {Huerta}, {Huet}, {Hughey}, {Hui}, {Husa},
  {Huttner}, {Huynh-Dinh}, {Idzkowski}, {Iess}, {Inchauspe}, {Ingram}, {Inta},
  {Intini}, {Irwin}, {Isa}, {Isac}, {Isi}, {Iyer}, {Jacqmin}, {Jadhav}, {Jani},
  {Janthalur}, {Jaranowski}, {Jariwala}, {Jenkins}, {Jiang}, {Johnson},
  {Jones}, {Jones}, {Jones}, {Jones}, {Jonker}, {Ju}, {Junker}, {Kalaghatgi},
  {Kalogera}, {Kamai}, {Kandhasamy}, {Kang}, {Kanner}, {Kapadia}, {Karki},
  {Kashyap}, {Kasprzack}, {Kastaun}, {Katsanevas}, {Katsavounidis}, {Katzman},
  {Kaufer}, {Kawabe}, {Keerthana}, {K{\'e}f{\'e}lian}, {Keitel}, {Kennedy},
  {Key}, {Khalili}, {Khan}, {Khan}, {Khazanov}, {Khetan}, {Khursheed},
  {Kijbunchoo}, {Kim}, {Kim}, {Kim}, {Kim}, {Kim}, {Kim}, {Kimball}, {King},
  {Kinley-Hanlon}, {Kirchhoff}, {Kissel}, {Kleybolte}, {Klika}, {Klimenko},
  {Knowles}, {Koch}, {Koehlenbeck}, {Koekoek}, {Koley}, {Kondrashov}, {Kontos},
  {Koper}, {Korobko}, {Korth}, {Kovalam}, {Kozak}, {Kr{\"a}mer}, {Kringel},
  {Krishnendu}, {Kr{\'o}lak}, {Krupinski}, {Kuehn}, {Kumar}, {Kumar}, {Kumar},
  {Kumar}, {Kuo}, {Kutynia}, {Kwang}, {Lackey}, {Laghi}, {Lai}, {Lam},
  {Landry}, {Landry}, {Lane}, {Lang}, {Lange}, {Lantz}, {Lanza},
  {Lartaux-Vollard}, {Lasky}, {Laxen}, {Lazzarini}, {Lazzaro}, {Leaci},
  {Leavey}, {Lecoeuche}, {Lee}, {Lee}, {Lee}, {Lee}, {Lee}, {Lee}, {Lehmann},
  {Lenon}, {Leroy}, {Letendre}, {Levin}, {Li}, {Li}, {Li}, {Li}, {Li}, {Lin},
  {Linde}, {Linker}, {Littenberg}, {Liu}, {Liu}, {Llorens-Monteagudo}, {Lo},
  {London}, {Longo}, {Lorenzini}, {Loriette}, {Lormand}, {Losurdo}, {Lough},
  {Lousto}, {Lovelace}, {Lower}, {L{\"u}ck}, {Lumaca}, {Lundgren}, {Lynch},
  {Ma}, {Macas}, {Macfoy}, {MacInnis}, {Macleod}, {Macquet}, {Maga{\~n}a
  Hernandez}, {Maga{\~n}a-Sandoval}, {Magee}, {Majorana}, {Maksimovic},
  {Malik}, {Man}, {Mandic}, {Mangano}, {Mansell}, {Manske}, {Mantovani},
  {Mapelli}, {Marchesoni}, {Marion}, {M{\'a}rka}, {M{\'a}rka}, {Markakis},
  {Markosyan}, {Markowitz}, {Maros}, {Marquina}, {Marsat}, {Martelli},
  {Martin}, {Martin}, {Martinez}, {Martynov}, {Masalehdan}, {Mason}, {Massera},
  {Masserot}, {Massinger}, {Masso-Reid}, {Mastrogiovanni}, {Matas},
  {Matichard}, {Matone}, {Mavalvala}, {McCann}, {McCarthy}, {McClelland},
  {McCormick}, {McCuller}, {McGuire}, {McIsaac}, {McIver}, {McManus}, {McRae},
  {McWilliams}, {Meacher}, {Meadors}, {Mehmet}, {Mehta}, {Meidam}, {Mejuto
  Villa}, {Melatos}, {Mendell}, {Mercer}, {Mereni}, {Merfeld}, {Merilh},
  {Merzougui}, {Meshkov}, {Messenger}, {Messick}, {Messina}, {Metzdorff},
  {Meyers}, {Meylahn}, {Miani}, {Miao}, {Michel}, {Middleton}, {Milano},
  {Miller}, {Millhouse}, {Mills}, {Milovich-Goff}, {Minazzoli}, {Minenkov},
  {Mishkin}, {Mishra}, {Mistry}, {Mitra}, {Mitrofanov}, {Mitselmakher},
  {Mittleman}, {Mo}, {Moffa}, {Mogushi}, {Mohapatra}, {Molina-Ruiz}, {Mondin},
  {Montani}, {Moore}, {Moraru}, {Morawski}, {Moreno}, {Morisaki}, {Mours},
  {Mow-Lowry}, {Muciaccia}, {Mukherjee}, {Mukherjee}, {Mukherjee}, {Mukherjee},
  {Mukund}, {Mullavey}, {Munch}, {Mu{\~n}iz}, {Muratore}, {Murray}, {Nagar},
  {Nardecchia}, {Naticchioni}, {Nayak}, {Neil}, {Neilson}, {Nelemans},
  {Nelson}, {Nery}, {Neunzert}, {Nevin}, {Ng}, {Ng}, {Nguyen}, {Nguyen},
  {Nichols}, {Nichols}, {Nissanke}, {Nocera}, {North}, {Nuttall},
  {Obergaulinger}, {Oberling}, {O'Brien}, {Oganesyan}, {Ogin}, {Oh}, {Oh},
  {Ohme}, {Ohta}, {Okada}, {Oliver}, {Oppermann}, {Oram}, {O'Reilly},
  {Ormiston}, {Ortega}, {O'Shaughnessy}, {Ossokine}, {Ottaway}, {Overmier},
  {Owen}, {Pace}, {Pagano}, {Page}, {Pagliaroli}, {Pai}, {Pai}, {Palamos},
  {Palashov}, {Palomba}, {Pan}, {Panda}, {Pang}, {Pankow}, {Pannarale}, {Pant},
  {Paoletti}, {Paoli}, {Parida}, {Parker}, {Pascucci}, {Pasqualetti},
  {Passaquieti}, {Passuello}, {Patil}, {Patricelli}, {Payne}, {Pearlstone},
  {Pechsiri}, {Pedersen}, {Pedraza}, {Pedurand}, {Pele}, {Penn}, {Perego},
  {Perez}, {P{\'e}rigois}, {Perreca}, {Petermann}, {Pfeiffer}, {Phelps},
  {Phukon}, {Piccinni}, {Pichot}, {Piergiovanni}, {Pierro}, {Pillant},
  {Pinard}, {Pinto}, {Pirello}, {Pitkin}, {Plastino}, {Poggiani}, {Pong},
  {Ponrathnam}, {Popolizio}, {Porter}, {Powell}, {Prajapati}, {Prasad},
  {Prasai}, {Prasanna}, {Pratten}, {Prestegard}, {Principe}, {Prodi},
  {Prokhorov}, {Punturo}, {Puppo}, {P{\"u}rrer}, {Qi}, {Quetschke}, {Quinonez},
  {Raab}, {Raaijmakers}, {Radkins}, {Radulesco}, {Raffai}, {Raja}, {Rajan},
  {Rajbhandari}, {Rakhmanov}, {Ramirez}, {Ramos-Buades}, {Rana}, {Rao},
  {Rapagnani}, {Raymond}, {Razzano}, {Read}, {Regimbau}, {Rei}, {Reid},
  {Reitze}, {Rettegno}, {Ricci}, {Richardson}, {Richardson}, {Ricker},
  {Riemenschneider}, {Riles}, {Rizzo}, {Robertson}, {Robinet}, {Rocchi},
  {Rolland}, {Rollins}, {Roma}, {Romanelli}, {Romano}, {Romel}, {Romie},
  {Rose}, {Rose}, {Rose}, {Rosi{\'n}ska}, {Rosofsky}, {Ross}, {Rowan},
  {R{\"u}diger}, {Ruggi}, {Rutins}, {Ryan}, {Sachdev}, {Sadecki},
  {Sakellariadou}, {Salafia}, {Salconi}, {Saleem}, {Samajdar}, {Sammut},
  {Sanchez}, {Sanchez}, {Sanchis-Gual}, {Sanders}, {Santiago}, {Santos},
  {Sarin}, {Sassolas}, {Sathyaprakash}, {Sauter}, {Savage}, {Schale}, {Scheel},
  {Scheuer}, {Schmidt}, {Schnabel}, {Schofield}, {Sch{\"o}nbeck}, {Schreiber},
  {Schulte}, {Schutz}, {Scott}, {Scott}, {Seidel}, {Sellers}, {Sengupta},
  {Sennett}, {Sentenac}, {Sequino}, {Sergeev}, {Setyawati}, {Shaddock},
  {Shaffer}, {Shahriar}, {Shaner}, {Sharma}, {Sharma}, {Shawhan}, {Shen},
  {Shink}, {Shoemaker}, {Shoemaker}, {Shukla}, {ShyamSundar}, {Siellez},
  {Sieniawska}, {Sigg}, {Singer}, {Singh}, {Singh}, {Singhal}, {Sintes},
  {Sitmukhambetov}, {Skliris}, {Slagmolen}, {Slaven-Blair}, {Smith}, {Smith},
  {Somala}, {Son}, {Soni}, {Sorazu}, {Sorrentino}, {Souradeep}, {Sowell},
  {Spencer}, {Spera}, {Srivastava}, {Srivastava}, {Staats}, {Stachie},
  {Standke}, {Steer}, {Steinke}, {Steinlechner}, {Steinlechner}, {Steinmeyer},
  {Stevenson}, {Stocks}, {Stone}, {Stops}, {Strain}, {Stratta}, {Strigin},
  {Strunk}, {Sturani}, {Stuver}, {Sudhir}, {Summerscales}, {Sun}, {Sunil},
  {Sur}, {Suresh}, {Sutton}, {Swinkels}, {Szczepa{\'n}czyk}, {Tacca}, {Tait},
  {Talbot}, {Tanner}, {Tao}, {T{\'a}pai}, {Tapia}, {Tasson}, {Taylor},
  {Tenorio}, {Terkowski}, {Thomas}, {Thomas}, {Thondapu}, {Thorne}, {Thrane},
  {Tiwari}, {Tiwari}, {Tiwari}, {Toland}, {Tonelli}, {Tornasi},
  {Torres-Forn{\'e}}, {Torrie}, {T{\"o}yr{\"a}}, {Travasso}, {Traylor},
  {Tringali}, {Tripathee}, {Trovato}, {Trozzo}, {Tsang}, {Tse}, {Tso},
  {Tsukada}, {Tsuna}, {Tsutsui}, {Tuyenbayev}, {Ueno}, {Ugolini},
  {Unnikrishnan}, {Urban}, {Usman}, {Vahlbruch}, {Vajente}, {Valdes},
  {Valentini}, {van Bakel}, {van Beuzekom}, {van den Brand}, {Van Den Broeck},
  {Van der-Hyde}, {van der Schaaf}, {Van Heijningen}, {van Veggel}, {Vardaro},
  {Varma}, {Vass}, {Vas{\'u}th}, {Vecchio}, {Vedovato}, {Veitch}, {Veitch},
  {Venkateswara}, {Venugopalan}, {Verkindt}, {Vetrano}, {Vicer{\'e}}, {Viets},
  {Vinciguerra}, {Vine}, {Vinet}, {Vitale}, {Vo}, {Vocca}, {Vorvick},
  {Vyatchanin}, {Wade}, {Wade}, {Wade}, {Walet}, {Walker}, {Wallace}, {Walsh},
  {Wang}, {Wang}, {Wang}, {Wang}, {Wang}, {Ward}, {Warden}, {Warner}, {Was},
  {Watchi}, {Weaver}, {Wei}, {Weinert}, {Weinstein}, {Weiss}, {Wellmann},
  {Wen}, {Wessel}, {We{\ss}els}, {Westhouse}, {Wette}, {Whelan}, {Whiting},
  {Whittle}, {Wilken}, {Williams}, {Williamson}, {Willis}, {Willke}, {Winkler},
  {Wipf}, {Wittel}, {Woan}, {Woehler}, {Wofford}, {Wright}, {Wu}, {Wysocki},
  {Xiao}, {Xu}, {Yamamoto}, {Yancey}, {Yang}, {Yang}, {Yang}, {Yap}, {Yazback},
  {Yeeles}, {Yu}, {Yu}, {Yuen}, {Zadro{\.z}ny}, {Zadro{\.z}ny}, {Zanolin},
  {Zappa}, {Zelenova}, {Zendri}, {Zevin}, {Zhang}, {Zhang}, {Zhang}, {Zhao},
  {Zhao}, {Zhou}, {Zhou}, {Zhu}, {Zimmerman}, {Zlochower}, {Zucker}, {Zweizig},
  {(The LIGO Scientific Collaboration}, \& {Virgo
  Collaboration)}}]{2020CQGra..37d5006A}
---. 2020{\natexlab{a}}, Classical and Quantum Gravity, 37, 045006,
  \dodoi{10.1088/1361-6382/ab5f7c}

\bibitem[{{Abbott} {et~al.}(2020{\natexlab{b}}){Abbott}, {Abbott}, {Abbott},
  {Abraham}, {Acernese}, {Ackley}, {Adams}, {Adhikari}, {Adya}, {Affeldt},
  {Agathos}, {Agatsuma}, {Aggarwal}, {Aguiar}, {Aiello}, {Ain}, {Ajith},
  {Allen}, {Allocca}, {Aloy}, {Altin}, {Amato}, {Anand}, {Ananyeva},
  {Anderson}, {Anderson}, {Angelova}, {Antier}, {Appert}, {Arai}, {Araya},
  {Areeda}, {Ar{\`e}ne}, {Arnaud}, {Aronson}, {Arun}, {Ascenzi}, {Ashton},
  {Aston}, {Astone}, {Aubin}, {Aufmuth}, {AultONeal}, {Austin}, {Avendano},
  {Avila-Alvarez}, {Babak}, {Bacon}, {Badaracco}, {Bader}, {Bae}, {Baird},
  {Baker}, {Baldaccini}, {Ballardin}, {Ballmer}, {Bals}, {Banagiri},
  {Barayoga}, {Barbieri}, {Barclay}, {Barish}, {Barker}, {Barkett}, {Barnum},
  {Barone}, {Barr}, {Barsotti}, {Barsuglia}, {Barta}, {Bartlett}, {Bartos},
  {Bassiri}, {Basti}, {Bawaj}, {Bayley}, {Baylor}, {Bazzan}, {B{\'e}csy},
  {Bejger}, {Belahcene}, {Bell}, {Beniwal}, {Benjamin}, {Berger}, {Bergmann},
  {Bernuzzi}, {Berry}, {Bersanetti}, {Bertolini}, {Betzwieser}, {Bhandare},
  {Bidler}, {Biggs}, {Bilenko}, {Bilgili}, {Billingsley}, {Birney},
  {Birnholtz}, {Biscans}, {Bischi}, {Biscoveanu}, {Bisht}, {Bitossi},
  {Bizouard}, {Blackburn}, {Blackman}, {Blair}, {Blair}, {Blair}, {Bloemen},
  {Bobba}, {Bode}, {Boer}, {Boetzel}, {Bogaert}, {Bondu}, {Bonnand}, {Booker},
  {Boom}, {Bork}, {Boschi}, {Bose}, {Bossilkov}, {Bosveld}, {Bouffanais},
  {Bozzi}, {Bradaschia}, {Brady}, {Bramley}, {Branchesi}, {Brau}, {Breschi},
  {Briant}, {Briggs}, {Brighenti}, {Brillet}, {Brinkmann}, {Brockill},
  {Brooks}, {Brooks}, {Brown}, {Brunett}, {Buikema}, {Bulik}, {Bulten},
  {Buonanno}, {Buskulic}, {Buy}, {Byer}, {Cabero}, {Cadonati}, {Cagnoli},
  {Cahillane}, {Calder{\'o}n Bustillo}, {Callister}, {Calloni}, {Camp},
  {Campbell}, {Canepa}, {Cannon}, {Cao}, {Cao}, {Carapella}, {Carbognani},
  {Caride}, {Carney}, {Carullo}, {Casanueva Diaz}, {Casentini}, {Caudill},
  {Cavagli{\`a}}, {Cavalier}, {Cavalieri}, {Cella}, {Cerd{\'a}-Dur{\'a}n},
  {Cesarini}, {Chaibi}, {Chakravarti}, {Chamberlin}, {Chan}, {Chao},
  {Charlton}, {Chase}, {Chassande-Mottin}, {Chatterjee}, {Chaturvedi},
  {Chatziioannou}, {Cheeseboro}, {Chen}, {Chen}, {Chen}, {Cheng}, {Cheong},
  {Chia}, {Chiadini}, {Chincarini}, {Chiummo}, {Cho}, {Cho}, {Cho},
  {Christensen}, {Chu}, {Chua}, {Chung}, {Chung}, {Ciani}, {Cie{\'s}lar},
  {Ciobanu}, {Ciolfi}, {Cipriano}, {Cirone}, {Clara}, {Clark}, {Clearwater},
  {Cleva}, {Coccia}, {Cohadon}, {Cohen}, {Colleoni}, {Collette}, {Collins},
  {Colpi}, {Cominsky}, {Constancio}, {Conti}, {Cooper}, {Corban}, {Corbitt},
  {Cordero-Carri{\'o}n}, {Corezzi}, {Corley}, {Cornish}, {Corre}, {Corsi},
  {Cortese}, {Costa}, {Cotesta}, {Coughlin}, {Coughlin}, {Coulon},
  {Countryman}, {Couvares}, {Covas}, {Cowan}, {Coward}, {Cowart}, {Coyne},
  {Coyne}, {Creighton}, {Creighton}, {Cripe}, {Croquette}, {Crowder}, {Cullen},
  {Cumming}, {Cunningham}, {Cuoco}, {Dal Canton}, {D{\'a}lya}, {D'Angelo},
  {Danilishin}, {D'Antonio}, {Danzmann}, {Dasgupta}, {Da Silva Costa},
  {Datrier}, {Dattilo}, {Dave}, {Davier}, {Davis}, {Daw}, {DeBra},
  {Deenadayalan}, {Degallaix}, {De Laurentis}, {Del{\'e}glise}, {De Lillo},
  {Del Pozzo}, {DeMarchi}, {Demos}, {Dent}, {De Pietri}, {De Rosa}, {De Rossi},
  {DeSalvo}, {de Varona}, {Dhurandhar}, {D{\'\i}az}, {Dietrich}, {Di Fiore},
  {DiFronzo}, {Di Giorgio}, {Di Giovanni}, {Di Giovanni}, {Di Girolamo}, {Di
  Lieto}, {Ding}, {Di Pace}, {Di Palma}, {Di Renzo}, {Divakarla}, {Dmitriev},
  {Doctor}, {Donovan}, {Dooley}, {Doravari}, {Dorrington}, {Downes}, {Drago},
  {Driggers}, {Du}, {Ducoin}, {Dudi}, {Dupej}, {Durante}, {Dwyer}, {Easter},
  {Eddolls}, {Edo}, {Effler}, {Ehrens}, {Eichholz}, {Eikenberry}, {Eisenmann},
  {Eisenstein}, {Errico}, {Essick}, {Estelles}, {Estevez}, {Etienne}, {Etzel},
  {Evans}, {Evans}, {Fafone}, {Fairhurst}, {Fan}, {Farinon}, {Farr}, {Farr},
  {Fauchon-Jones}, {Favata}, {Fays}, {Fazio}, {Fee}, {Feicht}, {Fejer}, {Feng},
  {Fernandez-Galiana}, {Ferrante}, {Ferreira}, {Ferreira}, {Fidecaro}, {Fiori},
  {Fiorucci}, {Fishbach}, {Fisher}, {Fishner}, {Fittipaldi}, {Fitz-Axen},
  {Fiumara}, {Flaminio}, {Fletcher}, {Floden}, {Flynn}, {Fong}, {Font},
  {Forsyth}, {Fournier}, {Vivanco}, {Frasca}, {Frasconi}, {Frei}, {Freise},
  {Frey}, {Frey}, {Fritschel}, {Frolov}, {Fronz{\`e}}, {Fulda}, {Fyffe},
  {Gabbard}, {Gadre}, {Gaebel}, {Gair}, {Gamba}, {Gammaitoni}, {Gaonkar},
  {Garc{\'\i}a-Quir{\'o}s}, {Garufi}, {Gateley}, {Gaudio}, {Gaur}, {Gayathri},
  {Gemme}, {Genin}, {Gennai}, {George}, {George}, {George}, {Gergely},
  {Ghonge}, {Ghosh}, {Ghosh}, {Ghosh}, {Giacomazzo}, {Giaime}, {Giardina},
  {Gibson}, {Gill}, {Glover}, {Gniesmer}, {Godwin}, {Goetz}, {Goetz},
  {Goncharov}, {Gonz{\'a}lez}, {Castro}, {Gopakumar}, {Gossan}, {Gosselin},
  {Gouaty}, {Grace}, {Grado}, {Granata}, {Grant}, {Gras}, {Grassia}, {Gray},
  {Gray}, {Greco}, {Green}, {Green}, {Gretarsson}, {Grimaldi}, {Grimm},
  {Groot}, {Grote}, {Grunewald}, {Gruning}, {Guidi}, {Gulati}, {Guo}, {Gupta},
  {Gupta}, {Gupta}, {Gustafson}, {Gustafson}, {Haegel}, {Halim}, {Hall},
  {Hall}, {Hamilton}, {Hammond}, {Haney}, {Hanke}, {Hanks}, {Hanna}, {Hannam},
  {Hannuksela}, {Hansen}, {Hanson}, {Harder}, {Hardwick}, {Haris}, {Harms},
  {Harry}, {Harry}, {Hasskew}, {Haster}, {Haughian}, {Hayes}, {Healy},
  {Heidmann}, {Heintze}, {Heitmann}, {Hellman}, {Hello}, {Hemming}, {Hendry},
  {Heng}, {Hennig}, {Heurs}, {Hild}, {Hinderer}, {Ho}, {Hochheim}, {Hofman},
  {Holgado}, {Holland}, {Holt}, {Holz}, {Hopkins}, {Horst}, {Hough}, {Howell},
  {Hoy}, {Huang}, {H{\"u}bner}, {Huerta}, {Huet}, {Hughey}, {Hui}, {Husa},
  {Huttner}, {Huynh-Dinh}, {Idzkowski}, {Iess}, {Inchauspe}, {Ingram}, {Inta},
  {Intini}, {Irwin}, {Isa}, {Isac}, {Isi}, {Iyer}, {Jacqmin}, {Jadhav}, {Jani},
  {Janthalur}, {Jaranowski}, {Jariwala}, {Jenkins}, {Jiang}, {Johnson},
  {Johnson-McDaniel}, {Jones}, {Jones}, {Jones}, {Jones}, {Jonker}, {Ju},
  {Junker}, {Kalaghatgi}, {Kalogera}, {Kamai}, {Kandhasamy}, {Kang}, {Kanner},
  {Kapadia}, {Karki}, {Kashyap}, {Kasprzack}, {Kastaun}, {Katsanevas},
  {Katsavounidis}, {Katzman}, {Kaufer}, {Kawabe}, {Keerthana},
  {K{\'e}f{\'e}lian}, {Keitel}, {Kennedy}, {Key}, {Khalili}, {Khan}, {Khan},
  {Khazanov}, {Khetan}, {Khursheed}, {Kijbunchoo}, {Kim}, {Kim}, {Kim}, {Kim},
  {Kim}, {Kim}, {Kimball}, {King}, {Kinley-Hanlon}, {Kirchhoff}, {Kissel},
  {Kleybolte}, {Klika}, {Klimenko}, {Knowles}, {Koch}, {Koehlenbeck},
  {Koekoek}, {Koley}, {Kondrashov}, {Kontos}, {Koper}, {Korobko}, {Korth},
  {Kovalam}, {Kozak}, {Kr{\"a}mer}, {Kringel}, {Krishnendu}, {Kr{\'o}lak},
  {Krupinski}, {Kuehn}, {Kumar}, {Kumar}, {Kumar}, {Kumar}, {Kuo}, {Kutynia},
  {Kwang}, {Lackey}, {Laghi}, {Lai}, {Lam}, {Landry}, {Landry}, {Lane}, {Lang},
  {Lange}, {Lantz}, {Lanza}, {Lartaux-Vollard}, {Lasky}, {Laxen}, {Lazzarini},
  {Lazzaro}, {Leaci}, {Leavey}, {Lecoeuche}, {Lee}, {Lee}, {Lee}, {Lee}, {Lee},
  {Lee}, {Lehmann}, {Lenon}, {Leroy}, {Letendre}, {Levin}, {Li}, {Li}, {Li},
  {Li}, {Li}, {Lin}, {Linde}, {Linker}, {Littenberg}, {Liu}, {Liu},
  {Llorens-Monteagudo}, {Lo}, {London}, {Longo}, {Lorenzini}, {Loriette},
  {Lormand}, {Losurdo}, {Lough}, {Lousto}, {Lovelace}, {Lower}, {Lucaccioni},
  {L{\"u}ck}, {Lumaca}, {Lundgren}, {Lynch}, {Ma}, {Macas}, {Macfoy},
  {MacInnis}, {Macleod}, {Macquet}, {Maga{\~n}a Hernandez},
  {Maga{\~n}a-Sandoval}, {Magee}, {Majorana}, {Maksimovic}, {Malik}, {Man},
  {Mandic}, {Mangano}, {Mansell}, {Manske}, {Mantovani}, {Mapelli},
  {Marchesoni}, {Marion}, {M{\'a}rka}, {M{\'a}rka}, {Markakis}, {Markosyan},
  {Markowitz}, {Maros}, {Marquina}, {Marsat}, {Martelli}, {Martin}, {Martin},
  {Martinez}, {Martynov}, {Masalehdan}, {Mason}, {Massera}, {Masserot},
  {Massinger}, {Masso-Reid}, {Mastrogiovanni}, {Matas}, {Matichard}, {Matone},
  {Mavalvala}, {McCann}, {McCarthy}, {McClelland}, {McCormick}, {McCuller},
  {McGuire}, {McIsaac}, {McIver}, {McManus}, {McRae}, {McWilliams}, {Meacher},
  {Meadors}, {Mehmet}, {Mehta}, {Meidam}, {Mejuto Villa}, {Melatos}, {Mendell},
  {Mercer}, {Mereni}, {Merfeld}, {Merilh}, {Merzougui}, {Meshkov}, {Messenger},
  {Messick}, {Messina}, {Metzdorff}, {Meyers}, {Meylahn}, {Miani}, {Miao},
  {Michel}, {Middleton}, {Milano}, {Miller}, {Millhouse}, {Mills},
  {Milovich-Goff}, {Minazzoli}, {Minenkov}, {Mishkin}, {Mishra}, {Mistry},
  {Mitra}, {Mitrofanov}, {Mitselmakher}, {Mittleman}, {Mo}, {Moffa}, {Mogushi},
  {Mohapatra}, {Molina-Ruiz}, {Mondin}, {Montani}, {Moore}, {Moraru},
  {Morawski}, {Moreno}, {Morisaki}, {Mours}, {Mow-Lowry}, {Muciaccia},
  {Mukherjee}, {Mukherjee}, {Mukherjee}, {Mukherjee}, {Mukund}, {Mullavey},
  {Munch}, {Mu{\~n}iz}, {Muratore}, {Murray}, {Nagar}, {Nardecchia},
  {Naticchioni}, {Nayak}, {Neil}, {Neilson}, {Nelemans}, {Nelson}, {Nery},
  {Neunzert}, {Nevin}, {Ng}, {Ng}, {Nguyen}, {Nguyen}, {Nichols}, {Nichols},
  {Nissanke}, {Nocera}, {North}, {Nuttall}, {Obergaulinger}, {Oberling},
  {O'Brien}, {Oganesyan}, {Ogin}, {Oh}, {Oh}, {Ohme}, {Ohta}, {Okada},
  {Oliver}, {Oppermann}, {Oram}, {O'Reilly}, {Ormiston}, {Ortega},
  {O'Shaughnessy}, {Ossokine}, {Ottaway}, {Overmier}, {Owen}, {Pace}, {Pagano},
  {Page}, {Pagliaroli}, {Pai}, {Pai}, {Palamos}, {Palashov}, {Palomba}, {Pan},
  {Panda}, {Pang}, {Pankow}, {Pannarale}, {Pant}, {Paoletti}, {Paoli},
  {Parida}, {Parker}, {Pascucci}, {Pasqualetti}, {Passaquieti}, {Passuello},
  {Patil}, {Patricelli}, {Payne}, {Pearlstone}, {Pechsiri}, {Pedersen},
  {Pedraza}, {Pedurand}, {Pele}, {Penn}, {Perego}, {Perez}, {P{\'e}rigois},
  {Perreca}, {Petermann}, {Pfeiffer}, {Phelps}, {Phukon}, {Piccinni}, {Pichot},
  {Piergiovanni}, {Pierro}, {Pillant}, {Pinard}, {Pinto}, {Pirello}, {Pitkin},
  {Plastino}, {Poggiani}, {Pong}, {Ponrathnam}, {Popolizio}, {Porter},
  {Powell}, {Prajapati}, {Prasad}, {Prasai}, {Prasanna}, {Pratten},
  {Prestegard}, {Principe}, {Prodi}, {Prokhorov}, {Punturo}, {Puppo},
  {P{\"u}rrer}, {Qi}, {Quetschke}, {Quinonez}, {Raab}, {Raaijmakers},
  {Radkins}, {Radulesco}, {Raffai}, {Raja}, {Rajan}, {Rajbhandari},
  {Rakhmanov}, {Ramirez}, {Ramos-Buades}, {Rana}, {Rao}, {Rapagnani},
  {Raymond}, {Razzano}, {Read}, {Regimbau}, {Rei}, {Reid}, {Reitze},
  {Rettegno}, {Ricci}, {Richardson}, {Richardson}, {Ricker}, {Riemenschneider},
  {Riles}, {Rizzo}, {Robertson}, {Robinet}, {Rocchi}, {Rolland}, {Rollins},
  {Roma}, {Romanelli}, {Romano}, {Romel}, {Romie}, {Rose}, {Rose}, {Rose},
  {Rosell}, {Rosi{\'n}ska}, {Rosofsky}, {Ross}, {Rowan}, {Roy}, {R{\"u}diger},
  {Ruggi}, {Rutins}, {Ryan}, {Sachdev}, {Sadecki}, {Sakellariadou}, {Salafia},
  {Salconi}, {Saleem}, {Samajdar}, {Sammut}, {Sanchez}, {Sanchez},
  {Sanchis-Gual}, {Sanders}, {Santiago}, {Santos}, {Sarin}, {Sassolas},
  {Sathyaprakash}, {Sauter}, {Savage}, {Schale}, {Scheel}, {Scheuer},
  {Schmidt}, {Schnabel}, {Schofield}, {Sch{\"o}nbeck}, {Schreiber}, {Schulte},
  {Schutz}, {Scott}, {Scott}, {Seidel}, {Sellers}, {Sengupta}, {Sennett},
  {Sentenac}, {Sequino}, {Sergeev}, {Setyawati}, {Shaddock}, {Shaffer},
  {Shahriar}, {Shaner}, {Sharma}, {Sharma}, {Shawhan}, {Shen}, {Shink},
  {Shoemaker}, {Shoemaker}, {Shukla}, {ShyamSundar}, {Siellez}, {Sieniawska},
  {Sigg}, {Singer}, {Singh}, {Singh}, {Singhal}, {Sintes}, {Sitmukhambetov},
  {Skliris}, {Slagmolen}, {Slaven-Blair}, {Smith}, {Smith}, {Somala}, {Son},
  {Soni}, {Sorazu}, {Sorrentino}, {Souradeep}, {Sowell}, {Spencer}, {Spera},
  {Srivastava}, {Srivastava}, {Staats}, {Stachie}, {Standke}, {Steer},
  {Steinke}, {Steinlechner}, {Steinlechner}, {Steinmeyer}, {Stevenson},
  {Stocks}, {Stone}, {Stops}, {Strain}, {Stratta}, {Strigin}, {Strunk},
  {Sturani}, {Stuver}, {Sudhir}, {Summerscales}, {Sun}, {Sunil}, {Sur},
  {Suresh}, {Sutton}, {Swinkels}, {Szczepa{\'n}czyk}, {Tacca}, {Tait},
  {Talbot}, {Tanner}, {Tao}, {T{\'a}pai}, {Tapia}, {Tasson}, {Taylor},
  {Tenorio}, {Terkowski}, {Thomas}, {Thomas}, {Thondapu}, {Thorne}, {Thrane},
  {Tiwari}, {Tiwari}, {Tiwari}, {Toland}, {Tonelli}, {Tornasi},
  {Torres-Forn{\'e}}, {Torrie}, {T{\"o}yr{\"a}}, {Travasso}, {Traylor},
  {Tringali}, {Tripathee}, {Trovato}, {Trozzo}, {Tsang}, {Tse}, {Tso},
  {Tsukada}, {Tsuna}, {Tsutsui}, {Tuyenbayev}, {Ueno}, {Ugolini},
  {Unnikrishnan}, {Urban}, {Usman}, {Vahlbruch}, {Vajente}, {Valdes},
  {Valentini}, {van Bakel}, {van Beuzekom}, {van den Brand}, {Van Den Broeck},
  {Vander-Hyde}, {van der Schaaf}, {VanHeijningen}, {van Veggel}, {Vardaro},
  {Varma}, {Vass}, {Vas{\'u}th}, {Vecchio}, {Vedovato}, {Veitch}, {Veitch},
  {Venkateswara}, {Venugopalan}, {Verkindt}, {Vetrano}, {Vicer{\'e}}, {Viets},
  {Vinciguerra}, {Vine}, {Vinet}, {Vitale}, {Vo}, {Vocca}, {Vorvick},
  {Vyatchanin}, {Wade}, {Wade}, {Wade}, {Walet}, {Walker}, {Wallace}, {Walsh},
  {Wang}, {Wang}, {Wang}, {Wang}, {Ward}, {Warden}, {Warner}, {Was}, {Watchi},
  {Weaver}, {Wei}, {Weinert}, {Weinstein}, {Weiss}, {Wellmann}, {Wen},
  {Wessel}, {We{\ss}els}, {Westhouse}, {Wette}, {Whelan}, {White}, {Whiting},
  {Whittle}, {Wilken}, {Williams}, {Williamson}, {Willis}, {Willke}, {Winkler},
  {Wipf}, {Wittel}, {Woan}, {Woehler}, {Wofford}, {Wright}, {Wu}, {Wysocki},
  {Xiao}, {Xu}, {Yamamoto}, {Yancey}, {Yang}, {Yang}, {Yang}, {Yap}, {Yazback},
  {Yeeles}, {Yu}, {Yu}, {Yuen}, {Zadro{\.z}ny}, {Zadro{\.z}ny}, {Zanolin},
  {Zelenova}, {Zendri}, {Zevin}, {Zhang}, {Zhang}, {Zhang}, {Zhao}, {Zhao},
  {Zhou}, {Zhou}, {Zhu}, {Zimmerman}, {Zucker}, \&
  {Zweizig}}]{2020ApJ...892L...3A}
---. 2020{\natexlab{b}}, \apjl, 892, L3, \dodoi{10.3847/2041-8213/ab75f5}

\bibitem[{Adhikari {et~al.}(2021)Adhikari, Albataineh, Androic, Aniol,
  Armstrong, Averett, Ayerbe~Gayoso, Barcus, Bellini, Beminiwattha, Benesch,
  Bhatt, Bhatta~Pathak, Bhetuwal, Blaikie, Campagna, Camsonne, Cates, Chen,
  Clarke, Cornejo, Covrig~Dusa, Datta, Deshpande, Dutta, Feldman, Fuchey, Gal,
  Gaskell, Gautam, Gericke, Ghosh, Halilovic, Hansen, Hauenstein, Henry,
  Horowitz, Jantzi, Jian, Johnston, Jones, Karki, Katugampola, Keppel, King,
  King, Knauss, Kumar, Kutz, Lashley-Colthirst, Leverick, Liu, Liyange, Malace,
  Mammei, Mammei, McCaughan, McNulty, Meekins, Metts, Michaels, Mondal,
  Napolitano, Narayan, Nikolaev, Rashad, Owen, Palatchi, Pan, Pandey, Park,
  Paschke, Petrusky, Pitt, Premathilake, Puckett, Quinn, Radloff, Rahman,
  Rathnayake, Reed, Reimer, Richards, Riordan, Roblin, Seeds, Shahinyan,
  Souder, Tang, Thiel, Tian, Urciuoli, Wertz, Wojtsekhowski, Yale, Ye, Yoon,
  Zec, Zhang, Zhang, \& Zheng}]{PhysRevLett.126.172502}
Adhikari, D., Albataineh, H., Androic, D., {et~al.} 2021, Phys. Rev. Lett.,
  126, 172502, \dodoi{10.1103/PhysRevLett.126.172502}

\bibitem[{{Agazie} {et~al.}(2023){Agazie}, {Alam}, {Anumarlapudi}, {Archibald},
  {Arzoumanian}, {Baker}, {Blecha}, {Bonidie}, {Brazier}, {Brook},
  {Burke-Spolaor}, {B{\'e}csy}, {Chapman}, {Charisi}, {Chatterjee}, {Cohen},
  {Cordes}, {Cornish}, {Crawford}, {Cromartie}, {Crowter}, {Decesar},
  {Demorest}, {Dolch}, {Drachler}, {Ferrara}, {Fiore}, {Fonseca}, {Freedman},
  {Garver-Daniels}, {Gentile}, {Glaser}, {Good}, {G{\"u}ltekin}, {Hazboun},
  {Jennings}, {Jessup}, {Johnson}, {Jones}, {Kaiser}, {Kaplan}, {Kelley},
  {Kerr}, {Key}, {Kuske}, {Laal}, {Lam}, {Lamb}, {Lazio}, {Lewandowska}, {Lin},
  {Liu}, {Lorimer}, {Luo}, {Lynch}, {Ma}, {Madison}, {Maraccini}, {McEwen},
  {McKee}, {McLaughlin}, {McMann}, {Meyers}, {Mingarelli}, {Mitridate}, {Ng},
  {Nice}, {Ocker}, {Olum}, {Panciu}, {Pennucci}, {Perera}, {Pol}, {Radovan},
  {Ransom}, {Ray}, {Romano}, {Salo}, {Sardesai}, {Schmiedekamp},
  {Schmiedekamp}, {Schmitz}, {Shapiro-Albert}, {Siemens}, {Simon}, {Siwek},
  {Stairs}, {Stinebring}, {Stovall}, {Susobhanan}, {Swiggum}, {Taylor},
  {Turner}, {Unal}, {Vallisneri}, {Vigeland}, {Wahl}, {Wang}, {Witt}, {Young},
  \& {Nanograv Collaboration}}]{2023ApJ...951L...9A}
{Agazie}, G., {Alam}, M.~F., {Anumarlapudi}, A., {et~al.} 2023, \apjl, 951, L9,
  \dodoi{10.3847/2041-8213/acda9a}

\bibitem[{{Alcock} \& {Illarionov}(1980)}]{1980ApJ...235..534A}
{Alcock}, C., \& {Illarionov}, A. 1980, \apj, 235, 534, \dodoi{10.1086/157656}

\bibitem[{{AlGendy} \& {Morsink}(2014)}]{2014ApJ...791...78A}
{AlGendy}, M., \& {Morsink}, S.~M. 2014, \apj, 791, 78,
  \dodoi{10.1088/0004-637X/791/2/78}

\bibitem[{{Antoniadis} {et~al.}(2013){Antoniadis}, {Freire}, {Wex}, {Tauris},
  {Lynch}, {van Kerkwijk}, {Kramer}, {Bassa}, {Dhillon}, {Driebe}, {Hessels},
  {Kaspi}, {Kondratiev}, {Langer}, {Marsh}, {McLaughlin}, {Pennucci}, {Ransom},
  {Stairs}, {van Leeuwen}, {Verbiest}, \& {Whelan}}]{2013Sci...340..448A}
{Antoniadis}, J., {Freire}, P. C.~C., {Wex}, N., {et~al.} 2013, Science, 340,
  448, \dodoi{10.1126/science.1233232}

\bibitem[{{Arzoumanian} {et~al.}(2018){Arzoumanian}, {Baker}, {Brazier},
  {Burke-Spolaor}, {Chamberlin}, {Chatterjee}, {Christy}, {Cordes}, {Cornish},
  {Crawford}, {Thankful Cromartie}, {Crowter}, {DeCesar}, {Demorest}, {Dolch},
  {Ellis}, {Ferdman}, {Ferrara}, {Folkner}, {Fonseca}, {Garver-Daniels},
  {Gentile}, {Haas}, {Hazboun}, {Huerta}, {Islo}, {Jones}, {Jones}, {Kaplan},
  {Kaspi}, {Lam}, {Lazio}, {Levin}, {Lommen}, {Lorimer}, {Luo}, {Lynch},
  {Madison}, {McLaughlin}, {McWilliams}, {Mingarelli}, {Ng}, {Nice}, {Park},
  {Pennucci}, {Pol}, {Ransom}, {Ray}, {Rasskazov}, {Siemens}, {Simon},
  {Spiewak}, {Stairs}, {Stinebring}, {Stovall}, {Swiggum}, {Taylor},
  {Vallisneri}, {van Haasteren}, {Vigeland}, {Zhu}, \& {NANOGrav
  Collaboration}}]{2018ApJ...859...47A}
{Arzoumanian}, Z., {Baker}, P.~T., {Brazier}, A., {et~al.} 2018, \apj, 859, 47,
  \dodoi{10.3847/1538-4357/aabd3b}

\bibitem[{{Badnell} {et~al.}(2005){Badnell}, {Bautista}, {Butler}, {Delahaye},
  {Mendoza}, {Palmeri}, {Zeippen}, \& {Seaton}}]{2005MNRAS.360..458B}
{Badnell}, N.~R., {Bautista}, M.~A., {Butler}, K., {et~al.} 2005, \mnras, 360,
  458, \dodoi{10.1111/j.1365-2966.2005.08991.x}

\bibitem[{{Baub{\"o}ck} {et~al.}(2019){Baub{\"o}ck}, {Psaltis}, \&
  {{\"O}zel}}]{2019ApJ...872..162B}
{Baub{\"o}ck}, M., {Psaltis}, D., \& {{\"O}zel}, F. 2019, \apj, 872, 162,
  \dodoi{10.3847/1538-4357/aafe08}

\bibitem[{{Baym} {et~al.}(2019){Baym}, {Furusawa}, {Hatsuda}, {Kojo}, \&
  {Togashi}}]{2019ApJ...885...42B}
{Baym}, G., {Furusawa}, S., {Hatsuda}, T., {Kojo}, T., \& {Togashi}, H. 2019,
  \apj, 885, 42, \dodoi{10.3847/1538-4357/ab441e}

\bibitem[{{Bedaque} \& {Steiner}(2015)}]{2015PhRvL.114c1103B}
{Bedaque}, P., \& {Steiner}, A.~W. 2015, \prl, 114, 031103,
  \dodoi{10.1103/PhysRevLett.114.031103}

\bibitem[{{Beronya} {et~al.}(2019){Beronya}, {Karpova}, {Kirichenko},
  {Zharikov}, {Zyuzin}, {Shibanov}, \& {Cabrera-Lavers}}]{2019MNRAS.485.3715B}
{Beronya}, D.~M., {Karpova}, A.~V., {Kirichenko}, A.~Y., {et~al.} 2019, \mnras,
  485, 3715, \dodoi{10.1093/mnras/stz607}

\bibitem[{{Bhattacharya} \& {van den Heuvel}(1991)}]{1991PhR...203....1B}
{Bhattacharya}, D., \& {van den Heuvel}, E.~P.~J. 1991, \physrep, 203, 1,
  \dodoi{10.1016/0370-1573(91)90064-S}

\bibitem[{{Bhattacharyya} {et~al.}(2005){Bhattacharyya}, {Strohmayer},
  {Miller}, \& {Markwardt}}]{2005ApJ...619..483B}
{Bhattacharyya}, S., {Strohmayer}, T.~E., {Miller}, M.~C., \& {Markwardt},
  C.~B. 2005, \apj, 619, 483, \dodoi{10.1086/426383}

\bibitem[{{Bildsten} {et~al.}(1992){Bildsten}, {Salpeter}, \&
  {Wasserman}}]{1992ApJ...384..143B}
{Bildsten}, L., {Salpeter}, E.~E., \& {Wasserman}, I. 1992, \apj, 384, 143,
  \dodoi{10.1086/170860}

\bibitem[{{Blaes} {et~al.}(1992){Blaes}, {Blandford}, {Madau}, \&
  {Yan}}]{1992ApJ...399..634B}
{Blaes}, O.~M., {Blandford}, R.~D., {Madau}, P., \& {Yan}, L. 1992, \apj, 399,
  634, \dodoi{10.1086/171955}

\bibitem[{{Bogdanov} {et~al.}(2019){Bogdanov}, {Lamb}, {Mahmoodifar}, {Miller},
  {Morsink}, {Riley}, {Strohmayer}, {Tung}, {Watts}, {Dittmann}, {Chakrabarty},
  {Guillot}, {Arzoumanian}, \& {Gendreau}}]{2019ApJ...887L..26B}
{Bogdanov}, S., {Lamb}, F.~K., {Mahmoodifar}, S., {et~al.} 2019, \apjl, 887,
  L26, \dodoi{10.3847/2041-8213/ab5968}

\bibitem[{{Bogdanov} {et~al.}(2021){Bogdanov}, {Dittmann}, {Ho}, {Lamb},
  {Mahmoodifar}, {Miller}, {Morsink}, {Riley}, {Strohmayer}, {Watts},
  {Choudhury}, {Guillot}, {Harding}, {Ray}, {Wadiasingh}, {Wolff}, {Markwardt},
  {Arzoumanian}, \& {Gendreau}}]{2021ApJ...914L..15B}
{Bogdanov}, S., {Dittmann}, A.~J., {Ho}, W. C.~G., {et~al.} 2021, \apjl, 914,
  L15, \dodoi{10.3847/2041-8213/abfb79}

\bibitem[{{Buccheri} {et~al.}(1983){Buccheri}, {Bennett}, {Bignami}, {Bloemen},
  {Boriakoff}, {Caraveo}, {Hermsen}, {Kanbach}, {Manchester}, {Masnou},
  {Mayer-Hasselwander}, {{\"O}zel}, {Paul}, {Sacco}, {Scarsi}, \&
  {Strong}}]{1983A&A...128..245B}
{Buccheri}, R., {Bennett}, K., {Bignami}, G.~F., {et~al.} 1983, \aap, 128, 245

\bibitem[{{Buchner}(2016{\natexlab{a}})}]{2016ascl.soft06005B}
{Buchner}, J. 2016{\natexlab{a}}, {PyMultiNest: Python interface for
  MultiNest}, Astrophysics Source Code Library, record ascl:1606.005.
\newblock \doeprint{1606.005}

\bibitem[{{Buchner}(2016{\natexlab{b}})}]{2016S&C....26..383B}
---. 2016{\natexlab{b}}, Statistics and Computing, 26, 383,
  \dodoi{10.1007/s11222-014-9512-y}

\bibitem[{{Buchner}(2021)}]{2021JOSS....6.3001B}
---. 2021, The Journal of Open Source Software, 6, 3001,
  \dodoi{10.21105/joss.03001}

\bibitem[{{Buchner}(2023)}]{2023StSur..17..169B}
---. 2023, Statistics Surveys, 17, 169, \dodoi{10.1214/23-SS144}

\bibitem[{{Chang} \& {Bildsten}(2004)}]{2004ApJ...605..830C}
{Chang}, P., \& {Bildsten}, L. 2004, \apj, 605, 830, \dodoi{10.1086/382271}

\bibitem[{{Contopoulos} \& {Spitkovsky}(2006)}]{2006ApJ...643.1139C}
{Contopoulos}, I., \& {Spitkovsky}, A. 2006, \apj, 643, 1139,
  \dodoi{10.1086/501161}

\bibitem[{{Cromartie} {et~al.}(2020){Cromartie}, {Fonseca}, {Ransom},
  {Demorest}, {Arzoumanian}, {Blumer}, {Brook}, {DeCesar}, {Dolch}, {Ellis},
  {Ferdman}, {Ferrara}, {Garver-Daniels}, {Gentile}, {Jones}, {Lam}, {Lorimer},
  {Lynch}, {McLaughlin}, {Ng}, {Nice}, {Pennucci}, {Spiewak}, {Stairs},
  {Stovall}, {Swiggum}, \& {Zhu}}]{2020NatAs...4...72C}
{Cromartie}, H.~T., {Fonseca}, E., {Ransom}, S.~M., {et~al.} 2020, Nature
  Astronomy, 4, 72, \dodoi{10.1038/s41550-019-0880-2}

\bibitem[{{Danielewicz} {et~al.}(2002){Danielewicz}, {Lacey}, \&
  {Lynch}}]{2002Sci...298.1592D}
{Danielewicz}, P., {Lacey}, R., \& {Lynch}, W.~G. 2002, Science, 298, 1592,
  \dodoi{10.1126/science.1078070}

\bibitem[{{De} {et~al.}(2018){De}, {Finstad}, {Lattimer}, {Brown}, {Berger}, \&
  {Biwer}}]{2018PhRvL.121i1102D}
{De}, S., {Finstad}, D., {Lattimer}, J.~M., {et~al.} 2018, \prl, 121, 091102,
  \dodoi{10.1103/PhysRevLett.121.091102}

\bibitem[{De {et~al.}(2018)De, Finstad, Lattimer, Brown, Berger, \&
  Biwer}]{PhysRevLett.121.091102}
De, S., Finstad, D., Lattimer, J.~M., {et~al.} 2018, Phys. Rev. Lett., 121,
  091102, \dodoi{10.1103/PhysRevLett.121.091102}

\bibitem[{{Demorest} {et~al.}(2010){Demorest}, {Pennucci}, {Ransom}, {Roberts},
  \& {Hessels}}]{2010Natur.467.1081D}
{Demorest}, P.~B., {Pennucci}, T., {Ransom}, S.~M., {Roberts}, M.~S.~E., \&
  {Hessels}, J.~W.~T. 2010, \nat, 467, 1081, \dodoi{10.1038/nature09466}

\bibitem[{{Dittmann}(2024)}]{2024arXiv240416928D}
{Dittmann}, A.~J. 2024, arXiv e-prints, arXiv:2404.16928,
  \dodoi{10.48550/arXiv.2404.16928}

\bibitem[{Dittmann {et~al.}(2024)Dittmann, Miller, Lamb, Holt, Chirenti, Wolff,
  Bogdanov, Guillot, Ho, Morsink, Arzoumanian, \&
  Gendreau}]{dittmann_2024_10215109}
Dittmann, A.~J., Miller, M.~C., Lamb, F.~K., {et~al.} 2024, {Updated NICER PSR
  J0740+6620 Illinois-Maryland MCMC Samples},  Zenodo,
  \dodoi{10.5281/zenodo.10215109}

\bibitem[{{Echeveste} {et~al.}(2020){Echeveste}, {Novarino}, {Benvenuto}, \&
  {De Vito}}]{2020MNRAS.495.2509E}
{Echeveste}, M., {Novarino}, M.~L., {Benvenuto}, O.~G., \& {De Vito}, M.~A.
  2020, \mnras, 495, 2509, \dodoi{10.1093/mnras/staa1372}

\bibitem[{{Ecker} {et~al.}(2017){Ecker}, {Hoyos}, {Jokela}, {Fern{\'a}ndez}, \&
  {Vuorinen}}]{2017JHEP...11..031E}
{Ecker}, C., {Hoyos}, C., {Jokela}, N., {Fern{\'a}ndez}, D.~R., \& {Vuorinen},
  A. 2017, Journal of High Energy Physics, 2017, 31,
  \dodoi{10.1007/JHEP11(2017)031}

\bibitem[{{Essick}(2022)}]{2022ApJ...927..195E}
{Essick}, R. 2022, \apj, 927, 195, \dodoi{10.3847/1538-4357/ac517c}

\bibitem[{{Essick} {et~al.}(2023){Essick}, {Legred}, {Chatziioannou}, {Han}, \&
  {Landry}}]{2023PhRvD.108d3013E}
{Essick}, R., {Legred}, I., {Chatziioannou}, K., {Han}, S., \& {Landry}, P.
  2023, \prd, 108, 043013, \dodoi{10.1103/PhysRevD.108.043013}

\bibitem[{{Feroz} {et~al.}(2009){Feroz}, {Hobson}, \&
  {Bridges}}]{2009MNRAS.398.1601F}
{Feroz}, F., {Hobson}, M.~P., \& {Bridges}, M. 2009, \mnras, 398, 1601,
  \dodoi{10.1111/j.1365-2966.2009.14548.x}

\bibitem[{{Fonseca} {et~al.}(2016){Fonseca}, {Pennucci}, {Ellis}, {Stairs},
  {Nice}, {Ransom}, {Demorest}, {Arzoumanian}, {Crowter}, {Dolch}, {Ferdman},
  {Gonzalez}, {Jones}, {Jones}, {Lam}, {Levin}, {McLaughlin}, {Stovall},
  {Swiggum}, \& {Zhu}}]{2016ApJ...832..167F}
{Fonseca}, E., {Pennucci}, T.~T., {Ellis}, J.~A., {et~al.} 2016, \apj, 832,
  167, \dodoi{10.3847/0004-637X/832/2/167}

\bibitem[{{Fonseca} {et~al.}(2021){Fonseca}, {Cromartie}, {Pennucci}, {Ray},
  {Kirichenko}, {Ransom}, {Demorest}, {Stairs}, {Arzoumanian}, {Guillemot},
  {Parthasarathy}, {Kerr}, {Cognard}, {Baker}, {Blumer}, {Brook}, {DeCesar},
  {Dolch}, {Dong}, {Ferrara}, {Fiore}, {Garver-Daniels}, {Good}, {Jennings},
  {Jones}, {Kaspi}, {Lam}, {Lorimer}, {Luo}, {McEwen}, {McKee}, {McLaughlin},
  {McMann}, {Meyers}, {Naidu}, {Ng}, {Nice}, {Pol}, {Radovan},
  {Shapiro-Albert}, {Tan}, {Tendulkar}, {Swiggum}, {Wahl}, \&
  {Zhu}}]{2021ApJ...915L..12F}
{Fonseca}, E., {Cromartie}, H.~T., {Pennucci}, T.~T., {et~al.} 2021, \apjl,
  915, L12, \dodoi{10.3847/2041-8213/ac03b8}

\bibitem[{{Foreman-Mackey} {et~al.}(2013){Foreman-Mackey}, {Hogg}, {Lang}, \&
  {Goodman}}]{2013PASP..125..306F}
{Foreman-Mackey}, D., {Hogg}, D.~W., {Lang}, D., \& {Goodman}, J. 2013, \pasp,
  125, 306, \dodoi{10.1086/670067}

\bibitem[{{Gabriel} {et~al.}(2004){Gabriel}, {Denby}, {Fyfe}, {Hoar}, {Ibarra},
  {Ojero}, {Osborne}, {Saxton}, {Lammers}, \& {Vacanti}}]{2004ASPC..314..759G}
{Gabriel}, C., {Denby}, M., {Fyfe}, D.~J., {et~al.} 2004, in Astronomical
  Society of the Pacific Conference Series, Vol. 314, Astronomical Data
  Analysis Software and Systems (ADASS) XIII, ed. F.~{Ochsenbein}, M.~G.
  {Allen}, \& D.~{Egret}, 759

\bibitem[{{Goodman} \& {Weare}(2010)}]{2010CAMCS...5...65G}
{Goodman}, J., \& {Weare}, J. 2010, Communications in Applied Mathematics and
  Computational Science, 5, 65, \dodoi{10.2140/camcos.2010.5.65}

\bibitem[{{Guillot} {et~al.}(2019){Guillot}, {Kerr}, {Ray}, {Bogdanov},
  {Ransom}, {Deneva}, {Arzoumanian}, {Bult}, {Chakrabarty}, {Gendreau}, {Ho},
  {Jaisawal}, {Malacaria}, {Miller}, {Strohmayer}, {Wolff}, {Wood}, {Webb},
  {Guillemot}, {Cognard}, \& {Theureau}}]{2019ApJ...887L..27G}
{Guillot}, S., {Kerr}, M., {Ray}, P.~S., {et~al.} 2019, \apjl, 887, L27,
  \dodoi{10.3847/2041-8213/ab511b}

\bibitem[{{Harding} \& {Muslimov}(2001)}]{2001ApJ...556..987H}
{Harding}, A.~K., \& {Muslimov}, A.~G. 2001, \apj, 556, 987,
  \dodoi{10.1086/321589}

\bibitem[{{Harding} \& {Muslimov}(2002)}]{2002ApJ...568..862H}
---. 2002, \apj, 568, 862, \dodoi{10.1086/338985}

\bibitem[{{Hebeler} {et~al.}(2013){Hebeler}, {Lattimer}, {Pethick}, \&
  {Schwenk}}]{2013ApJ...773...11H}
{Hebeler}, K., {Lattimer}, J.~M., {Pethick}, C.~J., \& {Schwenk}, A. 2013,
  \apj, 773, 11, \dodoi{10.1088/0004-637X/773/1/11}

\bibitem[{{Ho} \& {Heinke}(2009)}]{2009Natur.462...71H}
{Ho}, W. C.~G., \& {Heinke}, C.~O. 2009, \nat, 462, 71,
  \dodoi{10.1038/nature08525}

\bibitem[{{Ho} \& {Lai}(2001)}]{2001MNRAS.327.1081H}
{Ho}, W. C.~G., \& {Lai}, D. 2001, \mnras, 327, 1081,
  \dodoi{10.1046/j.1365-8711.2001.04801.x}

\bibitem[{{Ho} \& {Lai}(2003)}]{2003MNRAS.338..233H}
---. 2003, \mnras, 338, 233, \dodoi{10.1046/j.1365-8711.2003.06047.x}

\bibitem[{{Ho} {et~al.}(2003){Ho}, {Lai}, {Potekhin}, \&
  {Chabrier}}]{2003ApJ...599.1293H}
{Ho}, W. C.~G., {Lai}, D., {Potekhin}, A.~Y., \& {Chabrier}, G. 2003, \apj,
  599, 1293, \dodoi{10.1086/379507}

\bibitem[{{Hoyos} {et~al.}(2016){Hoyos}, {Jokela}, {Rodr{\'\i}guez
  Fern{\'a}ndez}, \& {Vuorinen}}]{2016PhRvD..94j6008H}
{Hoyos}, C., {Jokela}, N., {Rodr{\'\i}guez Fern{\'a}ndez}, D., \& {Vuorinen},
  A. 2016, \prd, 94, 106008, \dodoi{10.1103/PhysRevD.94.106008}

\bibitem[{{Hunter}(2007)}]{4160265}
{Hunter}, J.~D. 2007, Computing in Science Engineering, 9, 90,
  \dodoi{10.1109/MCSE.2007.55}

\bibitem[{{Lai}(2001)}]{2001RvMP...73..629L}
{Lai}, D. 2001, Reviews of Modern Physics, 73, 629,
  \dodoi{10.1103/RevModPhys.73.629}

\bibitem[{{Landry} \& {Essick}(2019)}]{2019PhRvD..99h4049L}
{Landry}, P., \& {Essick}, R. 2019, \prd, 99, 084049,
  \dodoi{10.1103/PhysRevD.99.084049}

\bibitem[{{Lemos} {et~al.}(2023){Lemos}, {Weaverdyck}, {Rollins}, {Muir},
  {Fert{\'e}}, {Liddle}, {Campos}, {Huterer}, {Raveri}, {Zuntz}, {Di
  Valentino}, {Fang}, {Hartley}, {Aguena}, {Allam}, {Annis}, {Bertin},
  {Bocquet}, {Brooks}, {Burke}, {Carnero Rosell}, {Carrasco Kind}, {Carretero},
  {Castander}, {Choi}, {Costanzi}, {Crocce}, {da Costa}, {Pereira}, {Dietrich},
  {Everett}, {Ferrero}, {Frieman}, {Garc{\'\i}a-Bellido}, {Gatti}, {Gaztanaga},
  {Gerdes}, {Gruen}, {Gruendl}, {Gschwend}, {Gutierrez}, {Hinton}, {Hollowood},
  {Honscheid}, {James}, {Kuehn}, {Kuropatkin}, {Lima}, {March}, {Melchior},
  {Menanteau}, {Miquel}, {Morgan}, {Palmese}, {Paz-Chinch{\'o}n}, {Pieres},
  {Malag{\'o}n}, {Porredon}, {Sanchez}, {Scarpine}, {Schubnell}, {Serrano},
  {Sevilla-Noarbe}, {Smith}, {Suchyta}, {Swanson}, {Tarle}, {Thomas}, {To},
  {Varga}, {Weller}, \& {DES Collaboration}}]{2023MNRAS.521.1184L}
{Lemos}, P., {Weaverdyck}, N., {Rollins}, R.~P., {et~al.} 2023, \mnras, 521,
  1184, \dodoi{10.1093/mnras/stac2786}

\bibitem[{{Luo} {et~al.}(2019){Luo}, {Ransom}, {Demorest}, {van Haasteren},
  {Ray}, {Stovall}, {Bachetti}, {Archibald}, {Kerr}, {Colen}, \&
  {Jenet}}]{2019ascl.soft02007L}
{Luo}, J., {Ransom}, S., {Demorest}, P., {et~al.} 2019, {PINT: High-precision
  pulsar timing analysis package}, Astrophysics Source Code Library, record
  ascl:1902.007.
\newblock \doeprint{1902.007}

\bibitem[{{Luo} {et~al.}(2021){Luo}, {Ransom}, {Demorest}, {Ray}, {Archibald},
  {Kerr}, {Jennings}, {Bachetti}, {van Haasteren}, {Champagne}, {Colen},
  {Phillips}, {Zimmerman}, {Stovall}, {Lam}, \& {Jenet}}]{2021ApJ...911...45L}
---. 2021, \apj, 911, 45, \dodoi{10.3847/1538-4357/abe62f}

\bibitem[{{Margalit} \& {Metzger}(2017)}]{2017ApJ...850L..19M}
{Margalit}, B., \& {Metzger}, B.~D. 2017, \apjl, 850, L19,
  \dodoi{10.3847/2041-8213/aa991c}

\bibitem[{{Miller}(2021)}]{2013arXiv1312.0029M}
{Miller}, M.~C. 2021, in Astrophysics and Space Science Library, Vol. 461,
  Astrophysics and Space Science Library, ed. T.~M. {Belloni}, M.~{M{\'e}ndez},
  \& C.~{Zhang}, 1--51, \dodoi{10.1007/978-3-662-62110-3_1}

\bibitem[{{Miller} {et~al.}(2020){Miller}, {Chirenti}, \&
  {Lamb}}]{2020ApJ...888...12M}
{Miller}, M.~C., {Chirenti}, C., \& {Lamb}, F.~K. 2020, \apj, 888, 12,
  \dodoi{10.3847/1538-4357/ab4ef9}

\bibitem[{{Miller} \& {Lamb}(1998)}]{1998ApJ...499L..37M}
{Miller}, M.~C., \& {Lamb}, F.~K. 1998, \apjl, 499, L37, \dodoi{10.1086/311335}

\bibitem[{{Miller} {et~al.}(2019){Miller}, {Lamb}, {Dittmann}, {Bogdanov},
  {Arzoumanian}, {Gendreau}, {Guillot}, {Harding}, {Ho}, {Lattimer}, {Ludlam},
  {Mahmoodifar}, {Morsink}, {Ray}, {Strohmayer}, {Wood}, {Enoto}, {Foster},
  {Okajima}, {Prigozhin}, \& {Soong}}]{2019ApJ...887L..24M}
{Miller}, M.~C., {Lamb}, F.~K., {Dittmann}, A.~J., {et~al.} 2019, \apjl, 887,
  L24, \dodoi{10.3847/2041-8213/ab50c5}

\bibitem[{{Miller} {et~al.}(2021){Miller}, {Lamb}, {Dittmann}, {Bogdanov},
  {Arzoumanian}, {Gendreau}, {Guillot}, {Ho}, {Lattimer}, {Loewenstein},
  {Morsink}, {Ray}, {Wolff}, {Baker}, {Cazeau}, {Manthripragada}, {Markwardt},
  {Okajima}, {Pollard}, {Cognard}, {Cromartie}, {Fonseca}, {Guillemot}, {Kerr},
  {Parthasarathy}, {Pennucci}, {Ransom}, \& {Stairs}}]{2021ApJ...918L..28M}
---. 2021, \apjl, 918, L28, \dodoi{10.3847/2041-8213/ac089b}

\bibitem[{{Mroczek} {et~al.}(2023{\natexlab{a}}){Mroczek}, {Coleman Miller},
  {Noronha-Hostler}, \& {Yunes}}]{2023JPhCS2536a2006M}
{Mroczek}, D., {Coleman Miller}, M., {Noronha-Hostler}, J., \& {Yunes}, N.
  2023{\natexlab{a}}, in Journal of Physics Conference Series, Vol. 2536,
  Journal of Physics Conference Series, 012006,
  \dodoi{10.1088/1742-6596/2536/1/012006}

\bibitem[{{Mroczek} {et~al.}(2023{\natexlab{b}}){Mroczek}, {Miller},
  {Noronha-Hostler}, \& {Yunes}}]{2023arXiv230902345M}
{Mroczek}, D., {Miller}, M.~C., {Noronha-Hostler}, J., \& {Yunes}, N.
  2023{\natexlab{b}}, arXiv e-prints, arXiv:2309.02345,
  \dodoi{10.48550/arXiv.2309.02345}

\bibitem[{{Nasa High Energy Astrophysics Science Archive Research Center
  (Heasarc)}(2014)}]{2014ascl.soft08004N}
{Nasa High Energy Astrophysics Science Archive Research Center (Heasarc)}.
  2014, {HEAsoft: Unified Release of FTOOLS and XANADU}, Astrophysics Source
  Code Library, record ascl:1408.004.
\newblock \doeprint{1408.004}

\bibitem[{{N{\"a}ttil{\"a}} {et~al.}(2017){N{\"a}ttil{\"a}}, {Miller},
  {Steiner}, {Kajava}, {Suleimanov}, \& {Poutanen}}]{2017A&A...608A..31N}
{N{\"a}ttil{\"a}}, J., {Miller}, M.~C., {Steiner}, A.~W., {et~al.} 2017, \aap,
  608, A31, \dodoi{10.1051/0004-6361/201731082}

\bibitem[{{Nelson} {et~al.}(2020){Nelson}, {Ford}, {Buchner}, {Cloutier},
  {D{\'\i}az}, {Faria}, {Hara}, {Rajpaul}, \& {Rukdee}}]{2020AJ....159...73N}
{Nelson}, B.~E., {Ford}, E.~B., {Buchner}, J., {et~al.} 2020, \aj, 159, 73,
  \dodoi{10.3847/1538-3881/ab5190}

\bibitem[{{{\"O}zel} {et~al.}(2016{\natexlab{a}}){{\"O}zel}, {Psaltis},
  {Arzoumanian}, {Morsink}, \& {Baub{\"o}ck}}]{2016ApJ...832...92O}
{{\"O}zel}, F., {Psaltis}, D., {Arzoumanian}, Z., {Morsink}, S., \&
  {Baub{\"o}ck}, M. 2016{\natexlab{a}}, \apj, 832, 92,
  \dodoi{10.3847/0004-637X/832/1/92}

\bibitem[{{{\"O}zel} {et~al.}(2016{\natexlab{b}}){{\"O}zel}, {Psaltis},
  {G{\"u}ver}, {Baym}, {Heinke}, \& {Guillot}}]{2016ApJ...820...28O}
{{\"O}zel}, F., {Psaltis}, D., {G{\"u}ver}, T., {et~al.} 2016{\natexlab{b}},
  \apj, 820, 28, \dodoi{10.3847/0004-637X/820/1/28}

\bibitem[{{Pang} {et~al.}(2021){Pang}, {Tews}, {Coughlin}, {Bulla}, {Van Den
  Broeck}, \& {Dietrich}}]{2021ApJ...922...14P}
{Pang}, P. T.~H., {Tews}, I., {Coughlin}, M.~W., {et~al.} 2021, \apj, 922, 14,
  \dodoi{10.3847/1538-4357/ac19ab}

\bibitem[{{Pavlov} \& {Zavlin}(1997)}]{1997ApJ...490L..91P}
{Pavlov}, G.~G., \& {Zavlin}, V.~E. 1997, \apjl, 490, L91,
  \dodoi{10.1086/311007}

\bibitem[{{Potekhin}(2014)}]{2014PhyU...57..735P}
{Potekhin}, A.~Y. 2014, Physics Uspekhi, 57, 735,
  \dodoi{10.3367/UFNe.0184.201408a.0793}

\bibitem[{{Randhawa} {et~al.}(2019){Randhawa}, {Meisel}, {Giuliani}, {Schatz},
  {Meyer}, {Ebinger}, {Hood}, \& {Kanungo}}]{2019ApJ...887..100R}
{Randhawa}, J.~S., {Meisel}, Z., {Giuliani}, S.~A., {et~al.} 2019, \apj, 887,
  100, \dodoi{10.3847/1538-4357/ab4f71}

\bibitem[{{Ray} {et~al.}(2019){Ray}, {Arzoumanian}, {Ballantyne}, {Bozzo},
  {Brandt}, {Brenneman}, {Chakrabarty}, {Christophersen}, {DeRosa}, {Feroci},
  {Gendreau}, {Goldstein}, {Hartmann}, {Hernanz}, {Jenke}, {Kara}, {Maccarone},
  {McDonald}, {Nowak}, {Phlips}, {Remillard}, {Stevens}, {Tomsick}, {Watts},
  {Wilson-Hodge}, {Wood}, {Zane}, {Ajello}, {Alston}, {Altamirano}, {Antoniou},
  {Arur}, {Ashton}, {Auchettl}, {Ayres}, {Bachetti}, {Balokovic}, {Baring},
  {Baykal}, {Begelman}, {Bhat}, {Bogdanov}, {Briggs}, {Bulbul}, {Bult},
  {Burns}, {Cackett}, {Campana}, {Caspi}, {Cavecchi}, {Chenevez}, {Cherry},
  {Corbet}, {Corcoran}, {Corsi}, {Degenaar}, {Drake}, {Eikenberry}, {Enoto},
  {Fragile}, {Fuerst}, {Gandhi}, {Garcia}, {Goldstein}, {Gonzalez},
  {Grefenstette}, {Grinberg}, {Grossan}, {Guillot}, {Guver}, {Haggard},
  {Heinke}, {Heinz}, {Hemphill}, {Homan}, {Hui}, {Huppenkothen}, {Ingram},
  {Irwin}, {Jaisawal}, {Jaodand}, {Kalemci}, {Kaplan}, {Keek}, {Kennea},
  {Kerr}, {van der Klis}, {Kocevski}, {Koss}, {Kowalski}, {Lai}, {Lamb},
  {Laycock}, {Lazio}, {Lazzati}, {Longcope}, {Loewenstein}, {Maitra}, {Majid},
  {Maksym}, {Malacaria}, {Margutti}, {Martindale}, {McHardy}, {Meyer},
  {Middleton}, {Miller}, {Miller}, {Motta}, {Neilsen}, {Nelson}, {Noble},
  {O'Brien}, {Osborne}, {Osten}, {Ozel}, {Palliyaguru}, {Pasham}, {Patruno},
  {Pelassa}, {Petropoulou}, {Pilia}, {Pohl}, {Pooley}, {Prescod-Weinstein},
  {Psaltis}, {Raaijmakers}, {Reynolds}, {Riley}, {Salvesen}, {Santangelo},
  {Scaringi}, {Schanne}, {Schnittman}, {Smith}, {Smith}, {Snios}, {Steiner},
  {Steiner}, {Stella}, {Strohmayer}, {Sun}, {Tauris}, {Taylor}, {Tohuvavohu},
  {Vacchi}, {Vasilopoulos}, {Veledina}, {Walsh}, {Weinberg}, {Wilkins},
  {Willingale}, {Wilms}, {Winter}, {Wolff}, {in 't Zand}, {Zezas}, {Zhang}, \&
  {Zoghbi}}]{2019arXiv190303035R}
{Ray}, P.~S., {Arzoumanian}, Z., {Ballantyne}, D., {et~al.} 2019, arXiv
  e-prints, arXiv:1903.03035, \dodoi{10.48550/arXiv.1903.03035}

\bibitem[{{Remillard} {et~al.}(2022){Remillard}, {Loewenstein}, {Steiner},
  {Prigozhin}, {LaMarr}, {Enoto}, {Gendreau}, {Arzoumanian}, {Markwardt},
  {Basak}, {Stevens}, {Ray}, {Altamirano}, \& {Buisson}}]{2022AJ....163..130R}
{Remillard}, R.~A., {Loewenstein}, M., {Steiner}, J.~F., {et~al.} 2022, \aj,
  163, 130, \dodoi{10.3847/1538-3881/ac4ae6}

\bibitem[{{Rezzolla} {et~al.}(2018){Rezzolla}, {Most}, \&
  {Weih}}]{2018ApJ...852L..25R}
{Rezzolla}, L., {Most}, E.~R., \& {Weih}, L.~R. 2018, \apjl, 852, L25,
  \dodoi{10.3847/2041-8213/aaa401}

\bibitem[{{Riley} {et~al.}(2019){Riley}, {Watts}, {Bogdanov}, {Ray}, {Ludlam},
  {Guillot}, {Arzoumanian}, {Baker}, {Bilous}, {Chakrabarty}, {Gendreau},
  {Harding}, {Ho}, {Lattimer}, {Morsink}, \&
  {Strohmayer}}]{2019ApJ...887L..21R}
{Riley}, T.~E., {Watts}, A.~L., {Bogdanov}, S., {et~al.} 2019, \apjl, 887, L21,
  \dodoi{10.3847/2041-8213/ab481c}

\bibitem[{{Riley} {et~al.}(2021){Riley}, {Watts}, {Ray}, {Bogdanov}, {Guillot},
  {Morsink}, {Bilous}, {Arzoumanian}, {Choudhury}, {Deneva}, {Gendreau},
  {Harding}, {Ho}, {Lattimer}, {Loewenstein}, {Ludlam}, {Markwardt}, {Okajima},
  {Prescod-Weinstein}, {Remillard}, {Wolff}, {Fonseca}, {Cromartie}, {Kerr},
  {Pennucci}, {Parthasarathy}, {Ransom}, {Stairs}, {Guillemot}, \&
  {Cognard}}]{2021ApJ...918L..27R}
{Riley}, T.~E., {Watts}, A.~L., {Ray}, P.~S., {et~al.} 2021, \apjl, 918, L27,
  \dodoi{10.3847/2041-8213/ac0a81}

\bibitem[{{Salmi} {et~al.}(2020){Salmi}, {Suleimanov}, {N{\"a}ttil{\"a}}, \&
  {Poutanen}}]{2020A&A...641A..15S}
{Salmi}, T., {Suleimanov}, V.~F., {N{\"a}ttil{\"a}}, J., \& {Poutanen}, J.
  2020, \aap, 641, A15, \dodoi{10.1051/0004-6361/202037824}

\bibitem[{{Salmi} {et~al.}(2022){Salmi}, {Vinciguerra}, {Choudhury}, {Riley},
  {Watts}, {Remillard}, {Ray}, {Bogdanov}, {Guillot}, {Arzoumanian},
  {Chirenti}, {Dittmann}, {Gendreau}, {Ho}, {Miller}, {Morsink}, {Wadiasingh},
  \& {Wolff}}]{salmi2022}
{Salmi}, T., {Vinciguerra}, S., {Choudhury}, D., {et~al.} 2022, \apj, 941, 150,
  \dodoi{10.3847/1538-4357/ac983d}

\bibitem[{{Salmi} {et~al.}(2023){Salmi}, {Vinciguerra}, {Choudhury}, {Watts},
  {Ho}, {Guillot}, {Kini}, {Dorsman}, {Morsink}, \&
  {Bogdanov}}]{2023arXiv230809319S}
---. 2023, \apj, 956, 138, \dodoi{10.3847/1538-4357/acf49d}

\bibitem[{{Salmi} {et~al.}(2024){Salmi}, {Choudhury}, {Kini}, {Vinciguerra},
  {Guillot}, {Morsink}, {Bilous}, {Arzoumanian}, {Choudhury}, {Deneva},
  {Gendreau}, {Harding}, {Ho}, \& {Lattimer}}]{salmi2023filler}
{Salmi}, T., {Choudhury}, D., {Kini}, Y., {et~al.} 2024, \apjl, XXX, ZZZ

\bibitem[{{SAS development team}(2014)}]{2014ascl.soft04004S}
{SAS development team}. 2014, {SAS: Science Analysis System for XMM-Newton
  observatory}, Astrophysics Source Code Library, record ascl:1404.004.
\newblock \doeprint{1404.004}

\bibitem[{{Silva} {et~al.}(2021){Silva}, {Pappas}, {Yunes}, \&
  {Yagi}}]{2021PhRvD.103f3038S}
{Silva}, H.~O., {Pappas}, G., {Yunes}, N., \& {Yagi}, K. 2021, \prd, 103,
  063038, \dodoi{10.1103/PhysRevD.103.063038}

\bibitem[{{Skilling}(2004)}]{2004AIPC..735..395S}
{Skilling}, J. 2004, in American Institute of Physics Conference Series, Vol.
  735, Bayesian Inference and Maximum Entropy Methods in Science and
  Engineering: 24th International Workshop on Bayesian Inference and Maximum
  Entropy Methods in Science and Engineering, ed. R.~{Fischer}, R.~{Preuss}, \&
  U.~V. {Toussaint}, 395--405, \dodoi{10.1063/1.1835238}

\bibitem[{{Speagle}(2020)}]{2020MNRAS.493.3132S}
{Speagle}, J.~S. 2020, \mnras, 493, 3132, \dodoi{10.1093/mnras/staa278}

\bibitem[{{Stella} {et~al.}(1987){Stella}, {Priedhorsky}, \&
  {White}}]{1987ApJ...312L..17S}
{Stella}, L., {Priedhorsky}, W., \& {White}, N.~E. 1987, \apjl, 312, L17,
  \dodoi{10.1086/184811}

\bibitem[{{Strohmayer} \& {Brown}(2002)}]{2002ApJ...566.1045S}
{Strohmayer}, T.~E., \& {Brown}, E.~F. 2002, \apj, 566, 1045,
  \dodoi{10.1086/338337}

\bibitem[{{Str{\"u}der} {et~al.}(2001){Str{\"u}der}, {Briel}, {Dennerl},
  {Hartmann}, {Kendziorra}, {Meidinger}, {Pfeffermann}, {Reppin}, {Aschenbach},
  {Bornemann}, {Br{\"a}uninger}, {Burkert}, {Elender}, {Freyberg}, {Haberl},
  {Hartner}, {Heuschmann}, {Hippmann}, {Kastelic}, {Kemmer}, {Kettenring},
  {Kink}, {Krause}, {M{\"u}ller}, {Oppitz}, {Pietsch}, {Popp}, {Predehl},
  {Read}, {Stephan}, {St{\"o}tter}, {Tr{\"u}mper}, {Holl}, {Kemmer}, {Soltau},
  {St{\"o}tter}, {Weber}, {Weichert}, {von Zanthier}, {Carathanassis}, {Lutz},
  {Richter}, {Solc}, {B{\"o}ttcher}, {Kuster}, {Staubert}, {Abbey}, {Holland},
  {Turner}, {Balasini}, {Bignami}, {La Palombara}, {Villa}, {Buttler},
  {Gianini}, {Lain{\'e}}, {Lumb}, \& {Dhez}}]{2001A&A...365L..18S}
{Str{\"u}der}, L., {Briel}, U., {Dennerl}, K., {et~al.} 2001, \aap, 365, L18,
  \dodoi{10.1051/0004-6361:20000066}

\bibitem[{Tantau(2013)}]{tantau:2013a}
Tantau, T. 2013, The TikZ and PGF Packages.
\newblock \url{http://sourceforge.net/projects/pgf/}

\bibitem[{{Timokhin} \& {Harding}(2015)}]{2015ApJ...810..144T}
{Timokhin}, A.~N., \& {Harding}, A.~K. 2015, \apj, 810, 144,
  \dodoi{10.1088/0004-637X/810/2/144}

\bibitem[{{Tolman} {et~al.}(2022){Tolman}, {Philippov}, \&
  {Timokhin}}]{2022ApJ...933L..37T}
{Tolman}, E.~A., {Philippov}, A.~A., \& {Timokhin}, A.~N. 2022, \apjl, 933,
  L37, \dodoi{10.3847/2041-8213/ac7c71}

\bibitem[{{Tsai}(1974)}]{1974RvMP...46..815T}
{Tsai}, Y.-S. 1974, Reviews of Modern Physics, 46, 815,
  \dodoi{10.1103/RevModPhys.46.815}

\bibitem[{{Tsang} {et~al.}(2012){Tsang}, {Stone}, {Camera}, {Danielewicz},
  {Gandolfi}, {Hebeler}, {Horowitz}, {Lee}, {Lynch}, {Kohley}, {Lemmon},
  {M{\"o}ller}, {Murakami}, {Riordan}, {Roca-Maza}, {Sammarruca}, {Steiner},
  {Vida{\~n}a}, \& {Yennello}}]{2012PhRvC..86a5803T}
{Tsang}, M.~B., {Stone}, J.~R., {Camera}, F., {et~al.} 2012, \prc, 86, 015803,
  \dodoi{10.1103/PhysRevC.86.015803}

\bibitem[{{Turner} {et~al.}(2001){Turner}, {Abbey}, {Arnaud}, {Balasini},
  {Barbera}, {Belsole}, {Bennie}, {Bernard}, {Bignami}, {Boer}, {Briel},
  {Butler}, {Cara}, {Chabaud}, {Cole}, {Collura}, {Conte}, {Cros}, {Denby},
  {Dhez}, {Di Coco}, {Dowson}, {Ferrando}, {Ghizzardi}, {Gianotti}, {Goodall},
  {Gretton}, {Griffiths}, {Hainaut}, {Hochedez}, {Holland}, {Jourdain},
  {Kendziorra}, {Lagostina}, {Laine}, {La Palombara}, {Lortholary}, {Lumb},
  {Marty}, {Molendi}, {Pigot}, {Poindron}, {Pounds}, {Reeves}, {Reppin},
  {Rothenflug}, {Salvetat}, {Sauvageot}, {Schmitt}, {Sembay}, {Short},
  {Spragg}, {Stephen}, {Str{\"u}der}, {Tiengo}, {Trifoglio}, {Tr{\"u}mper},
  {Vercellone}, {Vigroux}, {Villa}, {Ward}, {Whitehead}, \&
  {Zonca}}]{2001A&A...365L..27T}
{Turner}, M.~J.~L., {Abbey}, A., {Arnaud}, M., {et~al.} 2001, \aap, 365, L27,
  \dodoi{10.1051/0004-6361:20000087}

\bibitem[{{van der Walt} {et~al.}(2011){van der Walt}, {Colbert}, \&
  {Varoquaux}}]{5725236}
{van der Walt}, S., {Colbert}, S.~C., \& {Varoquaux}, G. 2011, Computing in
  Science Engineering, 13, 22, \dodoi{10.1109/MCSE.2011.37}

\bibitem[{{Vinciguerra} {et~al.}(2024){Vinciguerra}, {Salmi}, {Watts},
  {Choudhury}, {Riley}, {Ray}, {Bogdanov}, {Kini}, {Guillot}, {Chakrabarty},
  {Ho}, {Huppenkothen}, {Morsink}, {Wadiasingh}, \&
  {Wolff}}]{2024ApJ...961...62V}
{Vinciguerra}, S., {Salmi}, T., {Watts}, A.~L., {et~al.} 2024, \apj, 961, 62,
  \dodoi{10.3847/1538-4357/acfb83}

\bibitem[{Virtanen {et~al.}(2020)Virtanen, Gommers, Oliphant, Haberland, Reddy,
  Cournapeau, Burovski, Peterson, Weckesser, Bright, {van der Walt}, Brett,
  Wilson, Millman, Mayorov, Nelson, Jones, Kern, Larson, Carey, Polat, Feng,
  Moore, {VanderPlas}, Laxalde, Perktold, Cimrman, Henriksen, Quintero, Harris,
  Archibald, Ribeiro, Pedregosa, {van Mulbregt}, \& {SciPy 1.0
  Contributors}}]{2020SciPy-NMeth}
Virtanen, P., Gommers, R., Oliphant, T.~E., {et~al.} 2020, Nature Methods, 17,
  261, \dodoi{10.1038/s41592-019-0686-2}

\bibitem[{{Watts} {et~al.}(2016){Watts}, {Andersson}, {Chakrabarty}, {Feroci},
  {Hebeler}, {Israel}, {Lamb}, {Miller}, {Morsink}, {{\"O}zel}, {Patruno},
  {Poutanen}, {Psaltis}, {Schwenk}, {Steiner}, {Stella}, {Tolos}, \& {van der
  Klis}}]{2016RvMP...88b1001W}
{Watts}, A.~L., {Andersson}, N., {Chakrabarty}, D., {et~al.} 2016, Reviews of
  Modern Physics, 88, 021001, \dodoi{10.1103/RevModPhys.88.021001}

\bibitem[{{Wijngaarden} {et~al.}(2019){Wijngaarden}, {Ho}, {Chang}, {Heinke},
  {Page}, {Beznogov}, \& {Patnaude}}]{2019MNRAS.484..974W}
{Wijngaarden}, M.~J.~P., {Ho}, W. C.~G., {Chang}, P., {et~al.} 2019, \mnras,
  484, 974, \dodoi{10.1093/mnras/stz042}

\bibitem[{{Wijngaarden} {et~al.}(2020){Wijngaarden}, {Ho}, {Chang}, {Page},
  {Wijnands}, {Ootes}, {Cumming}, {Degenaar}, \&
  {Beznogov}}]{2020MNRAS.493.4936W}
---. 2020, \mnras, 493, 4936, \dodoi{10.1093/mnras/staa595}

\bibitem[{{Wilms} {et~al.}(2000){Wilms}, {Allen}, \&
  {McCray}}]{2000ApJ...542..914W}
{Wilms}, J., {Allen}, A., \& {McCray}, R. 2000, \apj, 542, 914,
  \dodoi{10.1086/317016}

\bibitem[{{Wolfe} {et~al.}(2023){Wolfe}, {Talbot}, \&
  {Golomb}}]{2023PhRvD.107j4056W}
{Wolfe}, N.~E., {Talbot}, C., \& {Golomb}, J. 2023, \prd, 107, 104056,
  \dodoi{10.1103/PhysRevD.107.104056}

\bibitem[{{Wolff} {et~al.}(2021){Wolff}, {Guillot}, {Bogdanov}, {Ray}, {Kerr},
  {Arzoumanian}, {Gendreau}, {Miller}, {Dittmann}, {Ho}, {Guillemot},
  {Cognard}, {Theureau}, \& {Wood}}]{2021ApJ...918L..26W}
{Wolff}, M.~T., {Guillot}, S., {Bogdanov}, S., {et~al.} 2021, \apjl, 918, L26,
  \dodoi{10.3847/2041-8213/ac158e}

\bibitem[{{Zhang} {et~al.}(2019){Zhang}, {Santangelo}, {Feroci}, {Xu}, {Lu},
  {Chen}, {Feng}, {Zhang}, {Brandt}, {Hernanz}, {Baldini}, {Bozzo}, {Campana},
  {De Rosa}, {Dong}, {Evangelista}, {Karas}, {Meidinger}, {Meuris}, {Nandra},
  {Pan}, {Pareschi}, {Orleanski}, {Huang}, {Schanne}, {Sironi}, {Spiga},
  {Svoboda}, {Tagliaferri}, {Tenzer}, {Vacchi}, {Zane}, {Walton}, {Wang},
  {Winter}, {Wu}, {in't Zand}, {Ahangarianabhari}, {Ambrosi}, {Ambrosino},
  {Barbera}, {Basso}, {Bayer}, {Bellazzini}, {Bellutti}, {Bertucci},
  {Bertuccio}, {Borghi}, {Cao}, {Cadoux}, {Campana}, {Ceraudo}, {Chen}, {Chen},
  {Chevenez}, {Civitani}, {Cui}, {Cui}, {Dauser}, {Del Monte}, {Di Cosimo},
  {Diebold}, {Doroshenko}, {Dovciak}, {Du}, {Ducci}, {Fan}, {Favre},
  {Fuschino}, {G{\'a}lvez}, {Gao}, {Ge}, {Gevin}, {Grassi}, {Gu}, {Gu}, {Han},
  {Hong}, {Hu}, {Ji}, {Jia}, {Jiang}, {Kennedy}, {Kreykenbohm}, {Kuvvetli},
  {Labanti}, {Latronico}, {Li}, {Li}, {Li}, {Li}, {Li}, {Limousin}, {Liu},
  {Liu}, {Lu}, {Luo}, {Macera}, {Malcovati}, {Martindale}, {Michalska}, {Meng},
  {Minuti}, {Morbidini}, {Muleri}, {Paltani}, {Perinati}, {Picciotto},
  {Piemonte}, {Qu}, {Rachevski}, {Rashevskaya}, {Rodriguez}, {Schanz}, {Shen},
  {Sheng}, {Song}, {Song}, {Sgro}, {Sun}, {Tan}, {Uttley}, {Wang}, {Wang},
  {Wang}, {Wang}, {Wang}, {Wang}, {Watts}, {Wen}, {Wilms}, {Xiong}, {Yang},
  {Yang}, {Yang}, {Yu}, {Zhang}, {Zampa}, {Zampa}, {Zdziarski}, {Zhang},
  {Zhang}, {Zhang}, {Zhang}, {Zhang}, {Zhang}, {Zhang}, {Zhang}, {Zhao},
  {Zheng}, {Zhou}, {Zorzi}, \& {Zwart}}]{2019SCPMA..6229502Z}
{Zhang}, S., {Santangelo}, A., {Feroci}, M., {et~al.} 2019, Science China
  Physics, Mechanics, and Astronomy, 62, 29502,
  \dodoi{10.1007/s11433-018-9309-2}

\end{thebibliography}

\appendix

\section{Assessing the Thoroughness of our Nested Sampling Analyses}\label{app:sampling}

\begin{deluxetable*}{cccccccc}
\caption{The equatorial radii inferred by our various analyses.}
\tablehead{
\colhead{Dataset} & \colhead{\makecell{Atmosphere\\ Model}} & Nested Sampling &  \makecell{Nested Sampling \\$R_e<16\,\rm km$} & \makecell{ Nested Sampling \\ \makecell{$R_e<16\,\rm km $\\$\Delta\theta_i<0.4$}} & MCMC & \makecell{MCMC \\$R_e<16\,\rm km$} & \makecell{ MCMC \\ \makecell{$R_e<16\,\rm km $\\$\Delta\theta_i<0.4$}} }
\startdata
NICER+XMM&$\rm{H_{full}}$&$12.704_{-0.970}^{+1.484}$&$12.661_{-0.942}^{+1.368}$&$12.661_{-0.942}^{+1.368}$&$12.922_{-1.129}^{+2.089}$&$12.756_{-1.020}^{+1.495}$&$12.756_{-1.021}^{+1.495}$\\
NICER+XMM&$\rm{H_{partial}}$&$12.754_{-1.039}^{+1.656}$&$12.692_{-1.004}^{+1.454}$&$12.692_{-1.004}^{+1.454}$&$12.992_{-1.228}^{+2.364}$&$12.760_{-1.074}^{+1.534}$&$12.760_{-1.074}^{+1.534}$\\
NICER+XMM&$\rm He_{full}$&$12.773_{-1.048}^{+1.657}$&$12.705_{-1.008}^{+1.442}$&$12.705_{-1.007}^{+1.442}$&$13.067_{-1.248}^{+2.455}$&$12.812_{-1.076}^{+1.560}$&$12.814_{-1.073}^{+1.562}$\\
\hline
NICER&$\rm{H_{full}}$&$12.150_{-1.507}^{+2.305}$&$11.984_{-1.394}^{+1.881}$&N/A&$12.475_{-1.938}^{+4.062}$&$11.856_{-1.481}^{+2.302}$&N/A\\
NICER&$\rm{H_{partial}}$&$11.479_{-0.922}^{+1.232}$&$11.465_{-0.914}^{+1.216}$&$11.365_{-0.811}^{+0.970}$&$11.946_{-1.361}^{+2.512}$&$11.781_{-1.247}^{+1.946}$&$11.804_{-1.191}^{+1.826}$\\
NICER&$\rm He_{full}$&$12.552_{-1.799}^{+2.816}$&$12.211_{-1.567}^{+2.039}$&N/A&$12.723_{-2.159}^{+4.666}$&$11.877_{-1.514}^{+2.351}$&N/A\\
\enddata
\label{tab:allradiiCuts}\tablecomments{The median and $68\%$ credible regions for the equatorial radius (in km) derived from both our preliminary nested sampling analyses (using MultiNest), and from our MCMC analysis after convergence (using emcee). For the sake of comparison with other works, such as \citet{2021ApJ...918L..27R} and \citet{salmi2023filler}, we also report the credible regions resulting from each analysis after imposing an upper limit of $16\,\rm km$ on the radius. Because some of our analyses identified lunate spot configurations which were neither significant after incorporating XMM-Newton data nor allowed by the priors used in \citet{salmi2023filler}, we also include the radius constraints after eliminating these solutions by requiring the angular diameters of both spots to be less than 0.4 radians. As this table shows, our preliminary, MultiNest-derived results are typically biased towards smaller radii, and often have erroneously small uncertainties.} 
\end{deluxetable*}

In Section \ref{sec:modes} we found that the best-fitting parameter values from our analysis of NICER data alone, using fully ionized hydrogen and helium atmosphere models, which produce a lunate spot configuration, were drastically disfavored by differences in log likelihood of $>400$ when the XMM-Newton data were included. On the other hand, the best-fitting parameters from our NICER-only analysis using neutron star atmosphere models consisting of hydrogen that could be partially ionized, which describe a mode with separate and much smaller emitting regions, gave a log likelihood $\sim 35$ smaller than the lunate spot configuration.
While this first analysis strongly suggested that the lunate configurations that are capable of fitting the NICER data alone are disfavored when the XMM-Newton data are included, the possibility remained that comparably good fits exist for that or other spot configurations that differ from the maximum likelihood configurations identified during the analyses of NICER data alone. To check whether these modes were explored by our analyses, we re-analyzed the nested samples produced by MultiNest using a tempered likelihood function $L' = L^\beta$ for $\beta\in[1,0]$, such that $\beta=1$ generates samples from the posterior $\beta=0$ generates samples from the prior, using the procedure suggested in \citet{2023PhRvD.107j4056W} to initialize parallel-tempered MCMC runs using nested sampling analyses. 

To make things explicit, nested sampling determines a set of samples $i$ and the prior volumes $w_i$ associated with each likelihood isocontour. Given a likelihood function $L$, these prior volumes can be used to calculate the evidence as $Z\approx\sum w_iL_i$, after which the points can be used as weighted posterior samples with weights $p_i=w_iL_i/Z.$ Because the prior volume of each sample is independent of the likelihood function itself as long as the order between points is preserved, we are free to re-weight samples according to $p_{i,\beta}=L_i^\beta w_i/Z_\beta$. For example, using $\beta\sim1/100$, the differences observed earlier between points in different modes can be smoothed over. Thus, plotting effective pseudo-posteriors with $0\leq\beta\leq1$ can inform us whether certain regions of parameter space were favored over the prior, and explored during sampling, before the highest-likelihood regions dominated the sampling.\footnote{However, this does not prove that the parameter space was thoroughly explored, or that the tempered likelihood sufrace is accurately or precisely characterized by this analysis, as it would be in an MCMC analysis of the tempered posterior.} 

\begin{figure}
    \includegraphics[width=\linewidth]{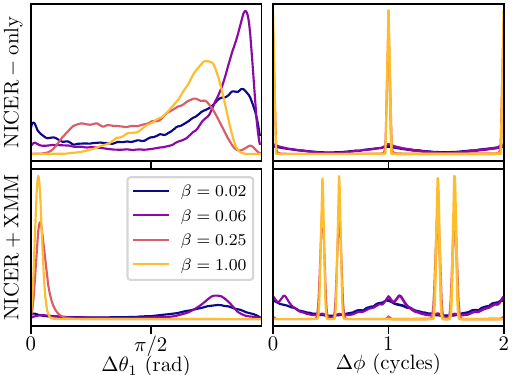}
    \caption{A demonstration, from our analyses using a fully ionized helium atmosphere model, that MultiNest explored subdominant modes in each analysis thoroughly enough to assign them appreciable mass at intermediate stages of the analysis. We accomplish this by tempering the likelihood function to place less weight on samples acquired at later times in the analysis. The top panels present effective posteriors from the analysis of NICER data alone, and the bottom panels show results from the joint analysis of NICER and XMM-Newton data. The lightest yellow lines plot the inferred posterior, while the others plot effective posteriors, tempered to varying degrees. The left panels present probability distributions for the angular radius of the first spot, and the right panels present distributions for the separation between spots; in both analyses, the subdominant mode was thoroughly explored before being abandoned for higher-likelihood ground.}
    \label{fig:betamodes}
\end{figure}

Figure \ref{fig:betamodes} illustrates the aforementioned analysis for two parameters, namely, the angular radius of the first spot and the phase offset between the two spots, for fits using fully ionized hydrogen atmospheres to the NICER data alone and to both the NICER and XMM-Newton data. Considering first the fits to only the NICER data, the results show that the mode with a large first spot, and spots separated by $\sim 0 $ rotational cycles, clearly dominates the posterior. However, after tempering the likelihood by a factor of $\beta\lesssim0.06$, we observe substantial pseudo-posterior mass at small spot angular radii and at intermediate phased offsets, suggesting that these families of modes were explored early in the analysis before the far-higher likelihood points belonging the lunate mode were identified. Similarly, while the posterior from the joint analysis of NICER and XMM data is comprised almost entirely by configurations with small, nearly antipodal spots, before honing in on these much higher-likelihood solutions the analysis thoroughly explored the crescent-shaped configurations which better suit the NICER data without the flux constraints from the data provided by XMM-Newton.

Although this analysis indicates that MultiNest was able to identity the various spot configurations capable of describing the present NICER and XMM-Newton data analyzed here, with the settings used in this work, it was unable to produce accurate posterior inferences. Table~\ref{tab:allradiiCuts} lists the $68\%$ uncertainty and median values of the equatorial radius for each of our preliminary nested sampling analyses and our converged MCMC analyses. In \emph{every} case, MultiNest estimated a posterior that is both erroneously narrow and biased towards low radii. We also list in Table~\ref{tab:allradiiCuts} the radius credible regions obtained after placing various cuts on our posterior samples to simulate the effects of imposing the priors used in \citet{salmi2023filler}. The posterior distributions resulting from these constraints agree better with the MultiNest-derived results reported in \citet{salmi2023filler}.

\section{Posterior Distributions from Analysis of Only NICER Data}\label{app:NO}
\begin{figure*}
     \centering
     \includegraphics[width=\textwidth]{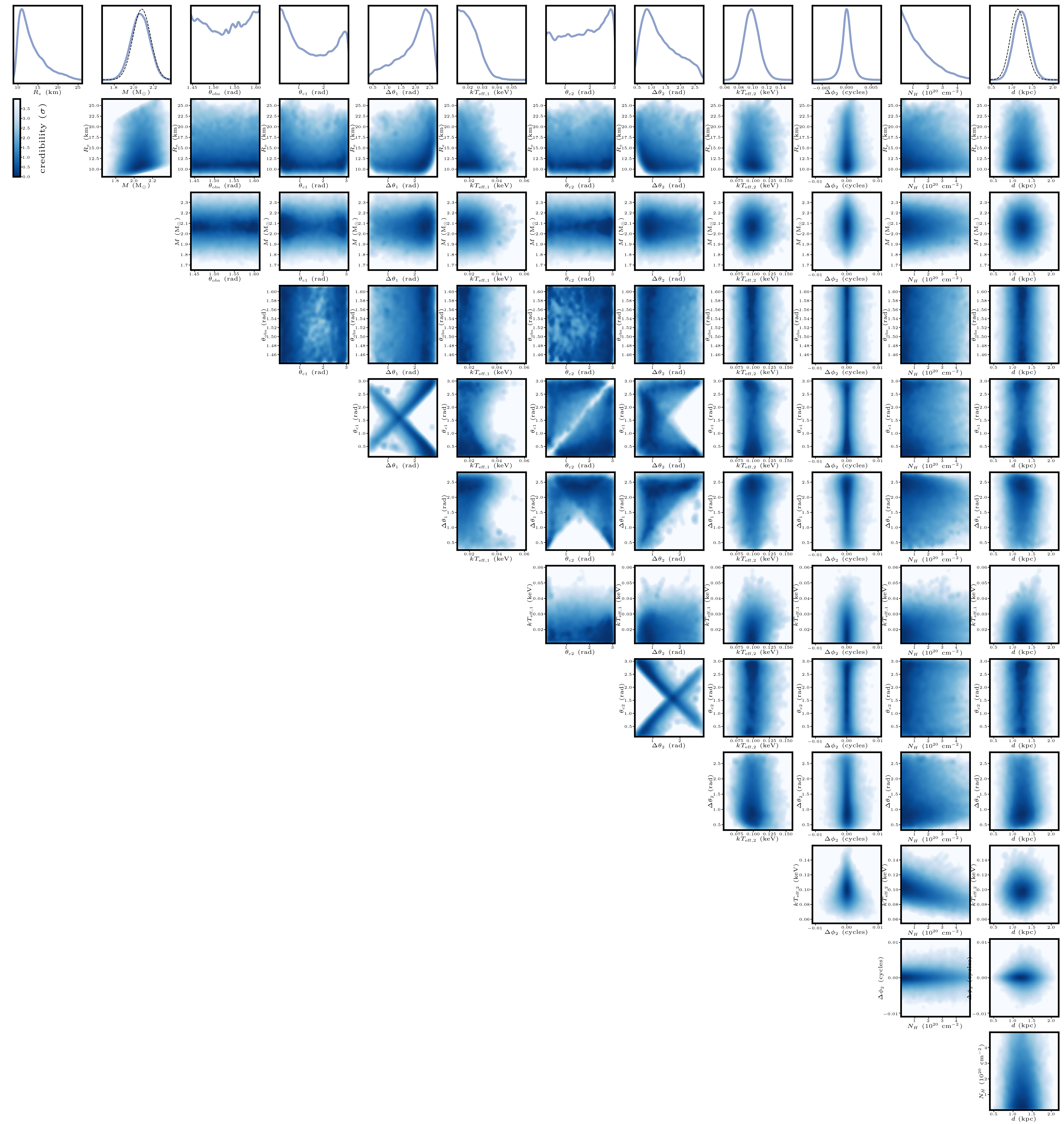}
    \caption{Posterior probability density distributions from our analysis of NICER data alone, using a fully ionized hydrogen atmosphere model. In the one-dimensional plots of the distance to and gravitational mass of the pulsar, dashed lines indicate the priors that we applied.}\label{fig:fullCornerNOf}    
\end{figure*}

Table~\ref{tab:allfullNOf} lists the median, $\pm1\sigma$, and $\pm2\sigma$ points in the posterior distributions obtained by fitting our models to only the NICER data, assuming a fully ionized hydrogen atmosphere, and the resulting maximum likelihood values for each of the parameters in these models. We display the complete corner plot of the posteriors from these same analyses in Figure~\ref{fig:fullCornerNOf}.

\renewcommand{\arraystretch}{0.875}
\setlength{\tabcolsep}{2.pt}
\begin{deluxetable}{ccccccc}
\caption{Fits to Only Nicer Data}
\tablehead{
\colhead{Parameter} & \colhead{median} & \colhead{$-1\sigma$} & \colhead{$+1\sigma$} & \colhead{$-2\sigma$} & \colhead{$+2\sigma$} & \colhead{\makecell{Maximum \\Likelihood}} 
}
\startdata
$R_e\,\rm (km)$& 12.475 & 10.537 & 16.538 & 9.602 & 21.840 & 11.731 \\
$GM/c^2R_e$& 0.245 & 0.185 & 0.289 & 0.140 & 0.308 & 0.259 \\
$M\,(M_\odot)$& 2.066 & 1.972 & 2.161 & 1.878 & 2.252 & 2.058 \\
$\theta_{\rm c1}\rm (rad)$ &1.331 & 0.384 & 2.632 & 0.149 & 2.972 & 1.680 \\
$\Delta\theta_1\,\rm (rad)$&2.096 & 1.266 & 2.485 & 0.552 & 2.671 & 1.159 \\
$kT_{\rm eff,1}\,\rm (keV)$&0.020 & 0.014 & 0.028 & 0.011 & 0.036 & 0.021 \\
$\theta_{\rm c2}\rm (rad)$&1.718 & 0.629 & 2.677 & 0.193 & 2.980 & 2.907 \\
$\Delta\theta_2\,\rm (rad)$&1.167 & 0.719 & 1.993 & 0.490 & 2.570 & 0.617 \\
$kT_{\rm eff,2}\,\rm (keV)$&0.098 & 0.088 & 0.109 & 0.078 & 0.122 & 0.107 \\
$\Delta\phi_2\,\rm (cycles)$&0.003 & 0.000 & 1.000 & 0.000 & 1.000 & 1.000 \\
$\theta_{\rm obs}\,\rm (rad)$&1.528 & 1.467 & 1.589 & 1.444 & 1.610 & 1.471 \\
$N_H\,(10^{20}\,\rm cm^{-2})$&1.056 & 0.273 & 2.524 & 0.037 & 4.182 & 0.063 \\
$d\,\rm (kpc)$&1.228 & 1.039 & 1.423 & 0.861 & 1.619 & 1.116 \\
\enddata
\label{tab:allfullNOf}
\tablecomments{A comparison of the $-2\sigma$, $-1\sigma$, median, $+1\sigma$, and $+2\sigma$, and maximum likelihood values inferred from our analysis of NICER data for PSR~J0740+6620. These results were derived under the assumption of a fully ionized pure hydrogen atmosphere.}
\end{deluxetable}
\pagebreak
\section{Posterior Distributions from Analysis of both NICER and XMM-Newton Data}\label{app:NX}
\begin{figure*}
     \centering
     \includegraphics[width=\textwidth]{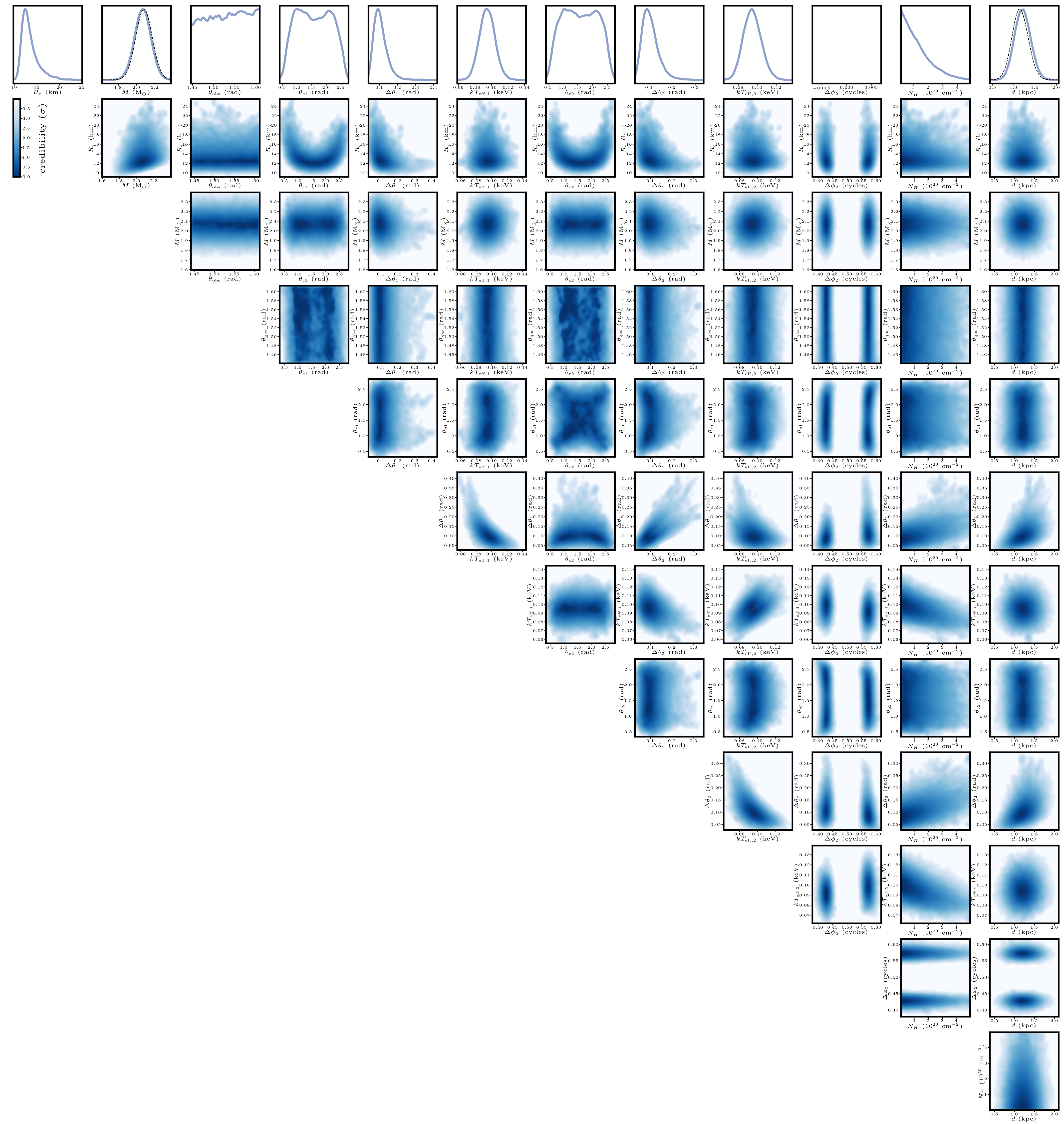}
    \caption{Posterior probability density distributions from our joint analysis of NICER and XMM-Newton data, using a fully ionized hydrogen atmosphere model. In the one-dimensional plots of the distance to and gravitational mass of the pulsar, dashed lines indicate the priors that we applied. }\label{fig:fullCornerNXf}    
\end{figure*}

Table~\ref{tab:allfullNXf} lists the median, $\pm1\sigma$, and $\pm2\sigma$ points in the posterior distributions obtained by fitting our models to both the NICER data and XMM-Newton data, assuming a fully ionized hydrogen atmosphere, and the resulting maximum likelihood values for each of the parameters in these models. We display the complete corner plot of the posteriors from these same analyses in Figure~\ref{fig:fullCornerNXf}.

\begin{deluxetable}{ccccccc}
\caption{Fits to NICER and XMM-Newton Data}
\tablehead{
\colhead{Parameter} & \colhead{median} & \colhead{$-1\sigma$} & \colhead{$+1\sigma$} & \colhead{$-2\sigma$} & \colhead{$+2\sigma$} & \colhead{\makecell{Maximum \\Likelihood}} 
}
\startdata
$R_e\,\rm (km)$ &12.922 & 11.793 & 15.010 & 10.992 & 18.574 & 11.541 \\
$GM/c^2R_e$ &0.236 & 0.204 & 0.257 & 0.165 & 0.272 & 0.250 \\
$M\,(M_\odot)$ &2.066 & 1.976 & 2.155 & 1.886 & 2.246 & 1.958 \\
$\theta_{\rm c1}\rm (rad)$ &1.573 & 0.946 & 2.222 & 0.641 & 2.546 & 1.387 \\
$\Delta\theta_1\,\rm (rad)$&0.101 & 0.072 & 0.140 & 0.051 & 0.199 & 0.092 \\
$kT_{\rm eff,1}\,\rm (keV)$&0.095 & 0.085 & 0.104 & 0.076 & 0.114 & 0.106 \\
$\theta_{\rm c2}\rm (rad)$&1.578 & 0.957 & 2.220 & 0.640 & 2.531 & 1.980 \\
$\Delta\theta_2\,\rm (rad)$&0.100 & 0.073 & 0.138 & 0.052 & 0.193 & 0.112 \\
$kT_{\rm eff,2}\,\rm (keV)$&0.095 & 0.085 & 0.104 & 0.077 & 0.114 & 0.099 \\
$\Delta\phi_2\,\rm (cycles)$&0.553 & 0.424 & 0.576 & 0.413 & 0.587 & 0.428 \\
$\theta_{\rm obs}\,\rm (rad)$&1.529 & 1.469 & 1.587 & 1.444 & 1.610 & 1.539 \\
$N_H\,(10^{20}\,\rm cm^{-2})$&0.953 & 0.255 & 2.209 & 0.034 & 3.871 & 0.023 \\
$d\,\rm (kpc)$&1.205 & 1.020 & 1.397 & 0.844 & 1.588 & 1.297 \\
\enddata
\label{tab:allfullNXf}\tablecomments{A comparison of the $-2\sigma$, $-1\sigma$, median, $+1\sigma$, and $+2\sigma$, and maximum likelihood values inferred from our joint analysis of NICER and XMM-Newton data for PSR~J0740+6620. These results were derived under the assumption of a fully ionized pure hydrogen atmosphere. }
\end{deluxetable}
\end{document}